\documentclass[a4paper,12pt]{article}
\usepackage{geometry}

\usepackage{amsmath}
\usepackage{amssymb}
\usepackage{bbm}
\usepackage{graphicx}
\usepackage{accents}
\usepackage{amsthm}
\usepackage{color}
\usepackage{graphics}
\usepackage{epsfig}
\usepackage{cancel}
\usepackage{epstopdf}
\usepackage{hyperref}
\usepackage{xcolor}
\usepackage{upgreek}

\hypersetup{
colorlinks,
linkcolor={blue},
citecolor={blue},
urlcolor={blue},
}
\usepackage{natbib}
\usepackage{stmaryrd}

\newcommand{\mbf}[1]{\boldsymbol{#1}}
\newcommand{\bsym}[1]{\mathbf{#1}}
\newcommand{\bbm}[1]{\mathbbm{#1}}
\newcommand{\mcal}[1]{\mathcal{#1}}
\newcommand{\bbbm}[1]{\bsym{\bbm{#1}}}

\newcommand{\jump}[1]{\llbracket {#1} \rrbracket}

\DeclareMathOperator{\vol}{\mcal{V}}
\DeclareMathOperator{\bvol}{\partial\mcal{V}}

\DeclareMathOperator{\defnt}{:=}
\DeclareMathOperator{\bdel}{\triangle\hspace{-1pt}}
\DeclareMathOperator{\sdel}{\delta\hspace{-1pt}}
\DeclareMathOperator{\sddel}{\delta^2\hspace{-1pt}}

\renewcommand{\div}{\text{div}}
\DeclareMathOperator{\Div}{\text{Div}}
\DeclareMathOperator{\Grad}{\text{Grad}}
\DeclareMathOperator{\grad}{\text{grad}}
\DeclareMathOperator{\Curl}{\text{Curl}}
\DeclareMathOperator{\curl}{\text{curl}}

\newcommand{\wst}[1]{{#1}^\ast}
\newcommand{\irchi}[2]{\raisebox{\depth}{$#1\chi$}}

\usepackage{mdframed} 
\newmdtheoremenv[linecolor=white,leftmargin=0,rightmargin=0,backgroundcolor=white!10,innertopmargin=0pt,ntheorem]{sidenote}{Remark}[section]

\usepackage{array}
\newcolumntype{L}[1]{>{\raggedright\let\newline\\\arraybackslash\hspace{0pt}}m{#1}}
\newcolumntype{C}[1]{>{\centering\let\newline\\\arraybackslash\hspace{0pt}}m{#1}}
\newcolumntype{R}[1]{>{\raggedleft\let\newline\\\arraybackslash\hspace{0pt}}m{#1}}

\usepackage{tabu}

\numberwithin{equation}{section}

\title{Variational principles of nonlinear magnetoelastostatics and their correspondences} 
\usepackage{fancyhdr}
\makeatletter
\let\runauthor\@author
\let\runtitle\@title
\makeatother
\author{Basant Lal Sharma$^1$, Prashant Saxena$^2$\thanks{Corresponding author email: prashant.saxena@glasgow.ac.uk}\\[2ex]
{\small $^1$Department of Mechanical Engineering, Indian Institute of Technology Kanpur}\\ 
{\small Kanpur, Uttar Pradesh 208016, India}\\
{\small $^2$Glagsow Computational Engineering Centre, James Watt School of Engineering}\\ 
{\small University of Glasgow, Glasgow G12 8LT, UK}}

\date{}

\allowdisplaybreaks 

\begin{document}

\maketitle

\begin{abstract}
We derive the equations of nonlinear magnetoelastostatics using several variational formulations involving the mechanical deformation and an independent field representing the magnetic component. 
An equivalence is also discussed, modulo certain boundary integrals or constant integrals, between these formulations using the Legendre transform and properties of Maxwell's equations. The second variation based bifurcation equations are stated for the incremental fields as well for all five variational principles. 
When the total potential energy is defined over the infinite space surrounding the body, we find that the inclusion of certain term in the energy principle, associated with the externally applied magnetic field, leads to slight changes in the Maxwell stress tensor and associated boundary conditions.
On the other hand, when the energy contained in the magnetic field is restricted to finite volumes, we find that there is a correspondence between the discussed formulations and associated expressions of physical entities.
In view of a diverse set of boundary data and nature of externally applied controls in the problems studied in the literature, along with a equally diverse  list of variational principles employed in modeling, our analysis emphasizes care in the choice of variational principle and unknown fields so that consistency with other choices is also satisfied.
\end{abstract}

\section*{Introduction}
Magnetoelastostatics concerns the analysis of suitable phenomenological models for a physical description of the equilibrium in {a} certain type of deformable solids associated with multi-functional processes involving magnetic and elastic effects. 
The main property characterizing these solids is the coupling between elastic deformation and magnetisation that they experience in the presence of externally applied mechanical as well as magnetic force fields \citep{Jolly1996b,Lokander2003,Boczkowska2010,Danas2012}.
The so called magnetoelastic coupling is known to occur due to a phenomenon involving re-configurations of small magnetic domains while a continuum vector field is borne out of an averaging of microscopic and distributed sub-fields \citep{Chatzigeorgiou2014b,Kovetz2000}. 
Thus an imposition of the magnetic field also induces a deformation of the material specimen in addition to the magnetic effects caused by the traditional mechanical forces \citep{Brown1966}.

With a history of more than five decades \citep{tiersten1964,tiersten1965,MauginandEringen1972,Maugin1988,desimone2002,kankanala2004,dorfmann2004}, the mathematical modeling of magnetoelasticity continues to be a vibrant area of research.
The presence of strong magnetoelastic coupling in some manufactured materials {such as magnetorheological elastomers (MREs) \citep{Jolly1996b}} allows the subject to be relevant for a large number of potential engineering and technological applications.
{MREs are composites made of ferromagnetic particles embedded in a polymer matrix. 
Magnetization of the ferromagnetic domains in the presence of an external magnetic field and the resulting interactions leads to a change in macroscopically observable mechanical properties.
As a result they find applications in micro-roboticss \citep{Hu2018, Ren2019}, sensors and actuators \citep{Bose2012a, Psarra2017a}, active vibration control \citep{Ginder2000}, and waveguides \citep{Saxena2018,KaramiMohammadi2019}.
Constitutive modelling of MREs has been undertaken by appropriately considering the micromechanics and derivation of coupled field equations using homogenization \citep{PonteCastaneda2011, Chatzigeorgiou2014b}; consideration of energy dissipation due to viscoelasticity of underlying matrix \citep{Saxena2013a, Saxena2014c, Ethiraj2016, Haldar2016}
}; and consideration of anisotropy due to ferromagnetic particle alignment \citep{Bustamante2010, Danas2012, Saxena2015}.

Derivation of a consistent set of partial differential equations (PDEs) and boundary conditions that describe equilibrium, analysis of the stability of equilibrium, and solution of the relevant PDEs via numerical techniques such as the finite element method requires development of appropriate variational principles. In this paper, we shall be concerned with the variational principles that have been postulated for the materials under the magnetoelastostatics assumptions and ignore any dynamic or dissipative effects. Current variational principles of magnetoelastostatics typically fall into two classes: principles based on the magnetic field or the magnetic induction as independent variable \citep{Bustamante2012, Vogel2013a} and principles based on a variant of the magnetization as an independent variable \citep{kankanala2004, Liu2014}. The typical starting point, definition of the total potential energy, is different in all these cases while it results in certain correspondence between the Euler--Lagrange equations derived.

The twofold motivation of this paper is the study of equations for the statics problem as well as the counterparts of bifurcation equations within the several variational formulations. 
Within the magnetization based principles we discuss three different formulations that utilize magnetization field per unit volume, magnetization per unit mass, and another adaption of magnetization field as an independent entity.
In fact, one of these variational principles analyzed in this paper was postulated originally, very early, by \cite[Eq.~8]{Brown1965}, while another one has been utilized in the work of \cite{kankanala2004} for a specific instability problem.
In addition to these three magnetization based principles, two more formulations are presented which are analogues of electroelastostatics {as derived in} \citep{Saxena2020}. 
For each of these variational principles, we derive the equation of equilibrium as well as the equation for the description of a state at bifurcation point.
As part of the first variation based analysis, we find that the expression for the Maxwell stress is susceptible to inclusion of certain integral terms that define suitable magnetic energy over an infinite space; the peculiar situation is however completely different from those formulations in which energy is defined over a finite domain of space.
Moreover, we present certain arguments based on Legendre transform as well as application of divergence theorem (using the properties of Maxwell fields) that suggest a direct equivalence between seemingly different formulations.

\subsubsection*{Outline}
\label{sec: outline}
This paper is organised as follows. After briefly introducing the mathematical preliminaries, we introduce the system under study and present the basic equations of nonlinear magnetoelastostatics in Section \ref{sec: basic eqns}.
In Sections \ref{sec: M formulation 1}--\ref{sec: M formulation 3}, we present the first variation of the potential energy functional corresponding to { three different magnetization vectors} ${\bbbm{M}}, {{\overline{\bbbm{M}}}}$ and ${{\bbbm{K}}}$, respectively, and then derive or state the equations for critical point by linearising the equilibrium equations.
Some auxiliary details are presented in the first appendix.
In Appendix sections \ref{sec: B formulation} and \ref{sec: H formulation}, we present the derivations of first and second variations of the potential energy functionals corresponding to the {magnetic induction} ${{\bbbm{B}}}$ and {the magnetic field} ${{\bbbm{H}}}$, respectively.

\begin{table}
\caption{Notation}
\label{table: notation}
\centering
\tabulinesep=1.2mm
\begin{tabular}{C{1cm} L{4cm} | C{1cm} L{4cm}}
\hline
$\mbf{x}$ & Position vector (spatial)& $\mbf{X}$ & Position vector (referential) \\ 
\hline
$\grad$ & Gradient (spatial) & $\Grad$ & Gradient (referential) \\ 
\hline
$\div$ & Divergence (spatial) & $\Div$ & Divergence (referential) \\ 
\hline
$\curl$ & Curl (spatial) & $\Curl$ & Curl (referential) \\ 
\hline 
$\mbf{n}$ & Unit outward normal (spatial)& $\mbf{n}_0$ & Unit outward normal (referential) \\
\hline
${\bbbm{h}}$ & Magnetic field vector (spatial) & ${{\bbbm{H}}}$ & Magnetic field vector (referential) \\
\hline
${\bbbm{b}}$ & Magnetic induction vector (spatial) & ${{\bbbm{B}}}$ & Magnetic induction vector (referential) \\
\hline
${{\bbbm{m}}}$ & Magnetisation vector per unit volume (spatial) & ${{\bbbm{M}}}$ & Magnetisation vector per unit volume (referential) \\
\hline
{${{\overline{\bbbm{m}}}}$} & { Magnetisation vector per unit mass (spatial)} & { ${{\overline{\bbbm{M}}}}$ } & {Magnetisation vector per unit mass (referential) }\\
\hline 
$\rho$ & mass density (spatial)& $\rho_0$ & mass density (referential) \\
\hline 
${\phi}$ & Magnetic scalar potential (spatial) & ${\Phi}$ & Magnetic scalar potential (referential) \\
\hline
${\bbbm{a}}$ & Magnetic vector potential (spatial) & ${\bbbm{A}}$ & Magnetic vector potential (referential) \\
\hline
${\bsym{\sigma}}$ & Cauchy stress tensor & ${\bsym{P}}$ & First Piola--Kirchhoff stress tensor \\
\hline
${\bsym{F}}$ & $\Grad{{\mathpalette\irchi\relax}}$ & ${{\bsym{P}}}_m$ & Maxwell stress tensor \\
\hline
$J$ & $\det{{\bsym{F}}}$ & ${{\bbbm{K}}}$ & $J{{\bsym{F}}}^{-\top} {{\overline{\bbbm{M}}}}$\\
\hline
$\jump{\{ \cdot \}}$ & Jump of a quantity ${\{ \cdot \}}$ across a boundary $\jump{\{ \cdot \}} =\{ \cdot \}_+ - \{ \cdot \}_- $ & $\{ \cdot \}_{,\mbf{G}}$ & Partial derivative with respect to $\mbf{G}$ \\
\hline
\end{tabular}
\end{table}

\subsubsection*{Notation}
\label{sec: notation}
We use the direct notation of tensor algebra and tensor calculus throughout the paper.
The scalar product of two vectors $\mbf{a}$ and $\mbf{b}$ is denoted as $\mbf{a} \cdot \mbf{b} =[\mbf{a}]_i [\mbf{b}]_i$ where a repeated index implies summation according to Einstein's summation convention.
{The vector (cross) product of two vectors $\mbf{a}$ and $\mbf{b}$ is denoted as $\mbf{a} \wedge \mbf{b}$ with $[\mbf{a} \wedge \mbf{b}]_i =\varepsilon_{ijk} [\mbf{a}]_j [\mbf{b}]_k $, $\varepsilon_{ijk}$ being the permutation symbol.}
The tensor product of two vectors $\mbf{a}$ and $\mbf{b}$ is a second order tensor $\bsym{H} =\mbf{a} \otimes \mbf{b}$ with $[\bsym{H}]_{ij} =[\mbf{a}]_i [\mbf{b}]_j$.
Operation of a second order tensor $\bsym{H}$ on a vector $\mbf{a}$ is given by $[\bsym{H} \mbf{a}]_i =[\bsym{H}]_{ij} [\mbf{a}]_j$.
Scalar product of two tensors $\bsym{H}$ and $\bsym{G}$ is denoted as $\bsym{H} \cdot \bsym{G} =[\bsym{H}]_{ij}[\bsym{G}]_{ij} $.
$\lVert\cdot\rVert$ represents the usual (Euclidean) norm for the mentioned vector entity.
A list of key variables {employed} throughout this manuscript is presented in Table \ref{table: notation}.

{For tensor calculus and variational method, we refer to \citep{Knowles1997, Itskov2018} and \citep{Gelfand2003}, respectively, whereas the notation and definitions of physical entities in continuum mechanics typically follow \citep{Gurtin1981}.}

\section{Nonlinear magnetoelastostatics: fundamental entities and equations}
\label{sec: basic eqns}

\begin{figure}
\begin{center}
\includegraphics[width=0.7\linewidth]{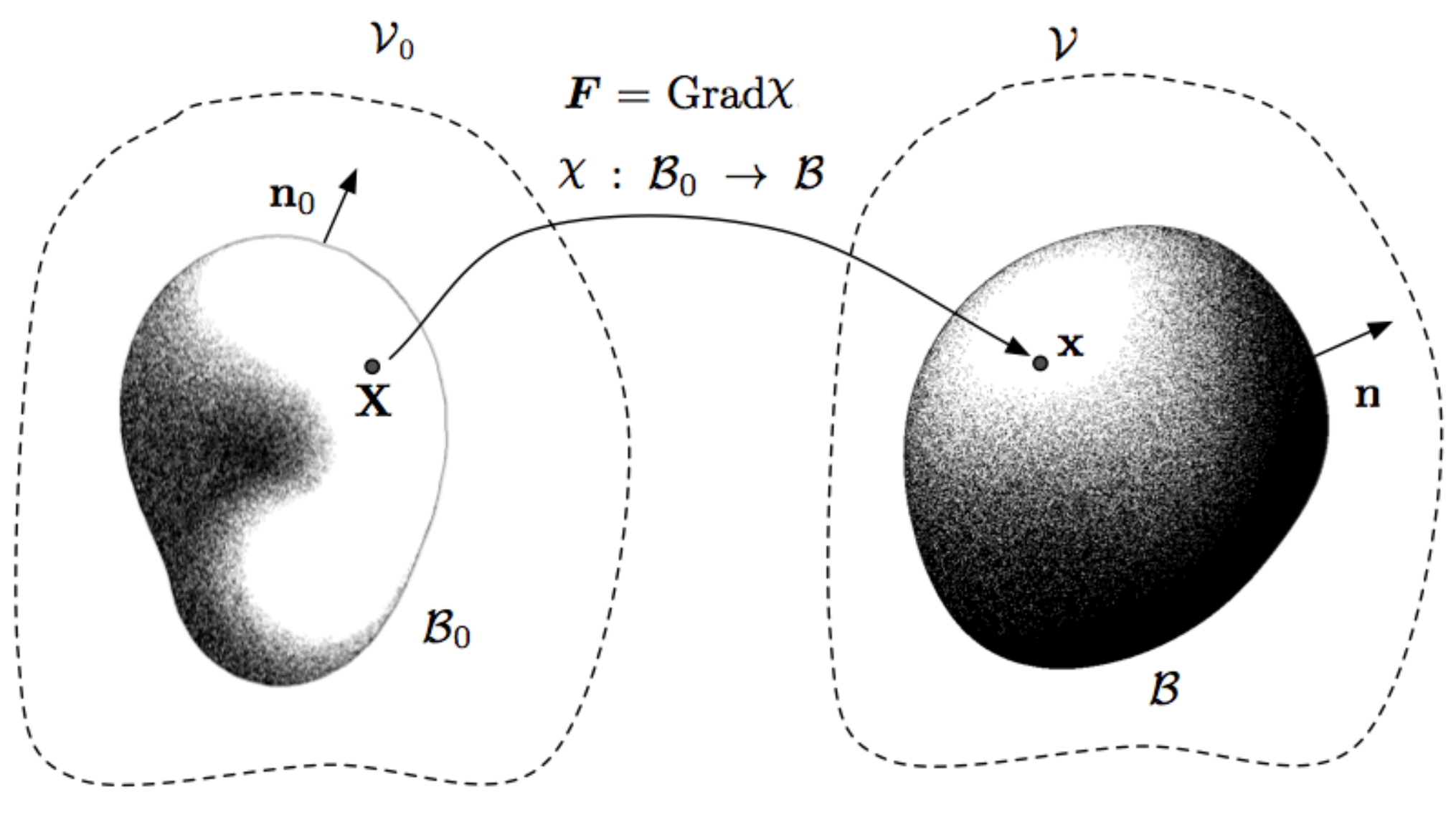}
\end{center}
\caption{A representation of the problem depicting the body in its reference and current configurations embedded in a volume $\mcal{V}$.}
\label{fig: problem cartoon}
\end{figure}

Consider a deformable body, in which its boundary or interior does not possess any distributed dipoles, occupying a three dimensional region ${\mcal{B}}$ lying inside another region $\vol$ as schematically depicted in Figure \ref{fig: problem cartoon}. We denote the {region} exterior {to} the body, relative to $\vol$, by ${\mcal{B}}'$ so that ${\mcal{B}}' =\vol \setminus ({\mcal{B}} \cup {\partial\mcal{B}} ).$ 
We assume that the body occupies a region ${\mcal{B}}_0$ in its reference configuration while {$\vol_0$ is the referential region corresponding to $\vol$, as explained below}.
The points in regions ${\mcal{B}}_0$ and ${\mcal{B}}$ corresponding to the same material point of the body are naturally mapped into each other by the deformation function
\begin{align}
{\mathpalette\irchi\relax}: {\mcal{B}}_0 \to {\mcal{B}}.
\label{eqn: deform}
\end{align}
In order to make sense of the referential (Lagrangian) description of fields in current region $\vol$, but {exterior to} the body, in a meaningful manner, we also define an extension of the deformation function ${\mathpalette\irchi\relax}$ to the part of region {exterior to} the body such that sufficient continuity requirements are maintained; {the latter region is} denoted by 
\[
{\mcal{B}}_0' =\vol_0 \setminus ({\mcal{B}}_0 \cup {\partial\mcal{B}} _0).
\]
Thus, by an abuse of notation, we assume an extension of mapping ${\mathpalette\irchi\relax}$ on a larger region, also denoted by ${\mathpalette\irchi\relax}$, i.e., 
\begin{align}
{\mathpalette\irchi\relax}: \vol_0 \to \vol.
\label{eqn: deform2}
\end{align}
In typical situations in practice, it is assumed that $\bvol_0$ and $\bvol$ coincide (for instance this is the scenario depicted in Fig. \ref{fig: problem cartoon}).

Following {the} standard notation {in continuum mechanics}, we define the deformation gradient {for points in the reference configuration ${\mcal{B}}_0$ and on its exterior relative to $\vol_0$} as 
\[
{{\bsym{F}}} \defnt \Grad {\mathpalette\irchi\relax}.
\]
The extension of {the natural definition of deformation and its gradient associated with ${\mathpalette\irchi\relax}$ on ${\mcal{B}}_0$} to $\vol_0$ {permits us later to perform some useful manipulations on the reference configuration as well as on} the exterior of the body ${\mcal{B}}_0'$ in the reference configuration.

The magnetic field vector, magnetic induction vector, and the magnetisation vector are denoted in the reference configuration as $({{\bbbm{H}}}, {{\bbbm{B}}}, {{\bbbm{M}}})$, respectively, and in the current configuration as $({\bbbm{h}}, {\bbbm{b}}, {{\bbbm{m}}})$, respectively.
These three vector fields are related by the well known constitutive relation
\begin{equation}
{\bbbm{b}} ={\mu_0} {\bbbm{h}} + {{\bbbm{m}}}.
\label{eqn: constitutive b h m}
\end{equation}
Further, the vector fields $({\bbbm{h}}, {\bbbm{b}}, {{\bbbm{m}}})$ must satisfy the Maxwell's equations
\begin{equation}
\div {\bbbm{b}} =0 \text{ and } \curl {\bbbm{h}} =\mbf{0} \text{ in } {\mcal{B}} \cup {\mcal{B}}'.
\label{eq: maxwell 1}
\end{equation}
The divergence-free and curl-free conditions \eqref{eq: maxwell 1} for ${\bbbm{b}}$ and ${\bbbm{h}}$, respectively, lead to the existence of magnetic potential (vector) field ${\bbbm{a}}$ and magnetic potential (scalar) field ${\phi}$ on ${\mcal{B}} \cup {\mcal{B}}'$; the respective expressions of ${\bbbm{b}}$ and ${\bbbm{h}}$ are given by
\begin{equation}
{\bbbm{b}} =\curl {\bbbm{a}}, \quad \quad {\bbbm{h}} =- \grad {\phi}.
\label{eq: maxwell potential}
\end{equation}
Following tradition in continuum mechanics \citep{Gurtin1981}, let $J$ denote the determinant of the deformation gradient, i.e., $J=\det{{\bsym{F}}}$ (note that $J>0$ on ${\mcal{B}}_0$ as well as ${\mcal{B}}'_0$).
The referential (Lagrangian) counterparts of ${\bbbm{b}}$ and ${\bbbm{h}}$, defined by
\begin{equation}
{{\bbbm{B}}} =J {{\bsym{F}}}^{-1} {\bbbm{b}}, \quad \quad {{\bbbm{H}}} ={{\bsym{F}}}^\top {\bbbm{h}},
\label{eqn: Bb, Hh relations}
\end{equation}
naturally satisfy the Maxwell's equations \eqref{eq: maxwell 1} in the reference configuration as
\begin{equation}
\Div {{\bbbm{B}}} =0\text{ and } \Curl {{\bbbm{H}}} =\mbf{0} \text{ in } {\mcal{B}}_0 \cup {\mcal{B}}'_0.
\label{eqn: maxwell lagrangian}
\end{equation}
Suitable referential (Lagrangian) counterparts of the magnetic vector potential and magnetic scalar potential \eqref{eq: maxwell potential} on ${\mcal{B}}_0 \cup {\mcal{B}}'_0$, based on the referential equations \eqref{eqn: maxwell lagrangian}, are given by
\begin{equation}
{{\bbbm{B}}} =\Curl {\bbbm{A}}, \quad {{\bbbm{H}}} =- \Grad {\Phi}.
\label{eqn: potentials introduction}
\end{equation}
{Concerning notational issues, a typical point in ${\mcal{B}}_0$ (as well as ${\mcal{B}}'_0$) is denoted by $\mbf{X}$, which is related (after deformation) to the point in ${\mcal{B}}$ (resp. ${\mcal{B}}'$) by the deformation function, assumed to be a sufficiently smooth mapping, ${\mathpalette\irchi\relax}$ such that $\mbf{x} ={\mathpalette\irchi\relax} (\mbf{X})$ and $\mbf{X} ={\mathpalette\irchi\relax}^{-1}(\mbf{x})$ \citep{Gurtin1981}, i.e.,
\begin{align}
\mbf{X}\mapsto\mbf{x}, \quad \mbf{x}\mapsto\mbf{X}.
\label{eqnxXXx}
\end{align}
}
{It can be shown using tensor algebra and calculus that}
\begin{equation}
{\bbbm{A}}(\mbf{X}) ={{\bsym{F}}}^\top(\mbf{X}) {\bbbm{a}}(\mbf{x}), {\Phi}(\mbf{X}) ={\phi}(\mbf{x}),
\end{equation}
for all $\mbf{X} \in {\mcal{B}}_0 \cup {\mcal{B}}'_0$.
Upon substituting the transformations \eqref{eqn: Bb, Hh relations} into the constitutive relation \eqref{eqn: constitutive b h m}, we obtain the relation
\begin{equation}
J^{-1} {{\bsym{C}}} {{\bbbm{B}}} ={\mu_0} {{\bbbm{H}}} + {{\bbbm{M}}},
\label{eqn: constitutive B H M}
\end{equation}
where ${{\bbbm{M}}}$ denotes the referential (Lagrangian) magnetisation (per unit volume) vector field.
Clearly, ${{\bbbm{M}}}$ is related to the current (spatial, Eulerian) magnetisation (per unit volume) vector field ${{\bbbm{m}}}$ by the definition (recall \eqref{eqnxXXx})
\begin{equation}
{{\bbbm{M}}}(\mbf{X}) \defnt {{\bsym{F}}}^\top(\mbf{X}) {{\bbbm{m}}}(\mbf{x}),
\label{LagrangianM}
\end{equation}
for all $\mbf{X} \in {\mcal{B}}_0 \cup {\mcal{B}}'_0$ (as ${{\bbbm{m}}}$ is zero in ${\mcal{B}}'$, we also get vanishing ${{\bbbm{M}}}$ in ${\mcal{B}}'_0$).
From the point of view of practical applications motivated by physics oriented models, it is also useful to define the magnetisation (per unit mass) ${{\overline{\bbbm{m}}}}: {\mcal{B}}\to{\bbm{R}^3}$. It is easy to see that the defining relation is
\begin{align}
{{\overline{\bbbm{m}}}}(\mbf{x})\defnt \rho(\mbf{x})^{-1} {{\bbbm{m}}}(\mbf{x}), \mbf{x}\in{\mcal{B}},
\label{eqn: defn mpm and mpv}
\end{align}
where $\rho$ stands for the mass density, i.e., a scalar field on ${\mcal{B}}$.
The referential (Lagrangian) counterpart of the spatial field ${{\overline{\bbbm{m}}}}$ is denoted by ${{\overline{\bbbm{M}}}}$, which is defined by
\begin{equation}
{{\overline{\bbbm{M}}}}(\mbf{X}) \defnt J^{-1}(\mbf{X}) {{\bsym{F}}}^\top(\mbf{X}) {{\overline{\bbbm{m}}}}(\mbf{x}), \mbf{X}\in{\mcal{B}}_0.
\label{eqn: defn mpm and mpv 2}
\end{equation}
\begin{sidenote}
When the density $\rho_0$ in the reference configuration is a constant, in particular for a homogeneous body, it is easy to see that ${{\overline{\bbbm{M}}}}$ and ${{\bbbm{M}}}$ are simply proportional (i.e., ${{\bbbm{M}}}=\rho_0{{\overline{\bbbm{M}}}}$ as $\rho_0=\rho J$).
\label{remark: M rho}
\end{sidenote}

Holding the viewpoint of several practical applications where magnetoelastic materials are involved, in certain situations it is quite convenient to distinguish the externally applied fields and the fields generated due to presence of the magnetoelastic body.
In such a typical scenario, an external magnetic field ${{\bbbm{h}}}^e$ is applied that results in the generation of a magnetic flux density ${{\bbbm{b}}}^e$ with the relation
\begin{align}
{{\bbbm{b}}}^e ={{\mu_0}} {{\bbbm{h}}}^e, 
\label{eqn: b h relation vacuum}
\end{align}
where ${{\mu_0}}$ is the (constant) magnetic permeability of vacuum. 
The presence of the magnetoelastic body creates a perturbation ({sometimes described as the} {\em self-field}) in the magnetic field that is denoted by ${\bbbm{h}}^s$ and a corresponding self-field for the magnetic flux vector denoted by ${{\bbbm{b}}}^s$ \citep[Ch.~5]{Brown1966}. 

\begin{sidenote}
{In general, in this paper the decoration with superscript `$s$' denotes the self-field or stray field while the superscript `$e$' denotes the externally applied entity.}
\label{decores}
\end{sidenote}
Thus, the total magnetic field and induction vector field 
is given by the sum
\begin{align}
{{\bbbm{h}}} ={{\bbbm{h}}}^e+{\bbbm{h}}^s, 
\quad \quad
{{\bbbm{b}}} ={{\bbbm{b}}}^e+{{\bbbm{b}}}^s.
\label{eqn: h and b field decomposition}
\end{align}
The relationship between the three magnetic vector fields ${{\bbbm{b}}}^s, {{\bbbm{h}}}^s,$ and the magnetisation per unit volume ${{\bbbm{m}}}$, is naturally given by
\begin{align}
{{\bbbm{b}}}^s ={{\mu_0}} {\bbbm{h}}^s +{{\bbbm{m}}},
\label{eqn: b h m relation total}
\end{align}
which holds on account of the relations \eqref{eqn: b h relation vacuum} and \eqref{eqn: h and b field decomposition}.

\begin{sidenote}
Concerning the units of magnetisation vector ${{\bbbm{m}}}$, we note that definition of the magnetisation vector is not standardised in literature and depending on the {choice of} units {either of} ${{\bbbm{m}}}$ and ${\mu_0} {{\bbbm{m}}}$ have been used.
Thus, the constitutive equation relating the three magnetic variables is also sometimes written as ${{\bbbm{b}}}^s ={{\mu_0}} [ {\bbbm{h}}^s +{{\bbbm{m}}} ]$ {for a different set of units}.
A detailed discussion on this topic can be found in \citep{Maugin1988}.
\end{sidenote}

\section{First formulation based on magnetisation}
\label{sec: M formulation 1}

Consider the body ${\mcal{B}}_0$ in its reference configuration (lying inside a containing space $\vol_0$).
Noting that ${{\bbbm{H}}} =-\Grad {\Phi}$ by \eqref{eqn: potentials introduction}${}_2$,
it is assumed that the total potential energy of the system is a functional of the deformation ${\mathpalette\irchi\relax}$ \eqref{eqn: deform} (and \eqref{eqn: deform2}) and the referential magnetisation ${{\bbbm{M}}}$ \eqref{LagrangianM} with the explicit expression given by \citep{Liu2014}
\begin{align}
{{\mathit{E}_{\mathtt{I}}}}[{\mathpalette\irchi\relax}, {{\bbbm{M}}}] &\defnt \int\nolimits_{{\mcal{B}}_0} {{\Omega}} ({{\bsym{F}}}, {{\bbbm{M}}}) {dv}_0 + \frac{{\mu_0}}{2} \int\nolimits_{\vol_0 } J \Vert {{\bsym{F}}}^{-\top} \Grad { {\Phi} } \rVert^2 {dv}_0 \nonumber \\
& - \int\nolimits_{{\mcal{B}}_0} \widetilde{{\mbf{f}}}^e \cdot {\mathpalette\irchi\relax} {dv}_0 - \int\nolimits_{{\partial\mcal{B}} _0}
\widetilde{\mbf{t}}^e \cdot {\mathpalette\irchi\relax} {ds}_0 + \int\nolimits_{\bvol_0} {\phi}^e \mbf{n}_0 \cdot {{\bbbm{B}}} {ds}_0,
\label{eqn: potential energy functional for M formulation 1}
\end{align}
where ${{\Omega}}$ is the (magnetoelastic) stored energy density per unit volume that depends on the deformation gradient ${{\bsym{F}}}$ and the referential magnetisation vector ${{\bbbm{M}}}$. {Integrals in equation \eqref{eqn: potential energy functional for M formulation 1} are} defined on the reference configuration and the spatial fields are mapped to the reference {configuration} by using the mapping ${\mathpalette\irchi\relax}$ as placement.
In this expression of the potential energy functional, it is assumed that ${\phi}^e$ stands for the externally applied magnetic potential on the boundary of the containing region $\mcal{V}_0$.
Note that $\widetilde{{\mbf{f}}}^e$ is the body force (vector) field per unit volume while $\widetilde{\mbf{t}}^e$ is the applied traction (vector) field {due to dead loads} at the boundary of the body in its current configuration; here also recall the notation described in Remark \ref{decores}. 

\subsection{Equilibrium: first variation}
In order to describe the state of magnetoelastic equilibrium, the particular deformation ${\mathpalette\irchi\relax}$ and magnetisation ${{\bbbm{M}}}$ at such a equilibrium corresponds to a extremum point of $E$, that is, when the first variation of the potential energy functional vanishes. In other words, it is assumed that ${\mathpalette\irchi\relax}$ and ${{\bbbm{M}}}$ satisfy
\begin{equation}
{\sdel}{{\mathit{E}_{\mathtt{I}}}}\equiv {\sdel}{{\mathit{E}_{\mathtt{I}}}}[{\mathpalette\irchi\relax}, {{\bbbm{M}}}; ({\sdel}{\mathpalette\irchi\relax}, {\sdel}{{\bbbm{M}}})] =0,
\label{eqn: first variation condition}
\end{equation}
for arbitrary but admissible variations ${\sdel}{\mathpalette\irchi\relax}$ and ${\sdel}{{\bbbm{M}}}$.
The variation of the potential energy functional ${{\mathit{E}_{\mathtt{I}}}}$ up to the first order is given by
\begin{align}
{\sdel}{{\mathit{E}_{\mathtt{I}}}} &={{\mathit{E}_{\mathtt{I}}}}[{\mathpalette\irchi\relax} + {\sdel}{\mathpalette\irchi\relax}, {{\bbbm{M}}} + {\sdel}{{\bbbm{M}}}] - {{\mathit{E}_{\mathtt{I}}}}[{\mathpalette\irchi\relax}, {{\bbbm{M}}}]\nonumber \\
&=\int\nolimits_{{\mcal{B}}_0}[ {{\Omega}}_{,{{\bsym{F}}}} \cdot {\sdel}{{\bsym{F}}} + {{\Omega}}_{,{{\bbbm{M}}}} \cdot {\sdel}{{\bbbm{M}}}] {dv}_0 - \int\nolimits_{{\mcal{B}}_0} \widetilde{{\mbf{f}}}^e \cdot {\sdel}{\mathpalette\irchi\relax} {dv}_0 - \int\nolimits_{{\partial\mcal{B}} _0} \widetilde{\mbf{t}}^e \cdot {\sdel}{\mathpalette\irchi\relax} {ds}_0 \nonumber \\
& + \int\nolimits_{\vol_0}[ - {{\widehat{{\bsym{P}}}}}_m \cdot {\sdel}{{\bsym{F}}} - J {\mu_0}[ {{\bsym{C}}}^{-1} {{\bbbm{H}}}] \cdot \Grad {\sdel}{\Phi}] {dv}_0 + \int\nolimits_{\bvol_0} {\phi}^e \mbf{n}_0 \cdot {\sdel}{{\bbbm{B}}} {ds}_0. 
\label{eqn: M form delta E 1}
\end{align}
where ${{\widehat{{\bsym{P}}}}}_m$ is a tensor field defined by
\begin{equation}
{{\widehat{{\bsym{P}}}}}_m ={\mu_0} J \bigg[ - \frac{1}{2} \big[ {{\bsym{F}}}^{-\top} {{\bbbm{H}}} \big] \cdot \big[ {{\bsym{F}}}^{-\top} {{\bbbm{H}}} \big] \bsym{I} + \big[ {{\bsym{F}}}^{-\top} {{\bbbm{H}}} \big] \otimes\big[ {{\bsym{F}}}^{-\top} {{\bbbm{H}}} \big] \bigg] {{\bsym{F}}}^{-\top},
\label{eqn: max type tensor 1}
\end{equation}
where $\bsym{I}$ is the identity tensor.
We are able to understand the physical nature of ${{\widehat{{\bsym{P}}}}}_m$ by noticing that, in the region ${\mcal{B}}'_0$ {exterior to} the body, the magnetisation ${{\bbbm{M}}} =\mbf{0}$; this results in ${{\widehat{{\bsym{P}}}}}_m ={{\bsym{P}}}_m$, where ${{\bsym{P}}}_m$ denotes the well known Maxwell stress tensor defined by
\begin{align}
{{\bsym{P}}}_m & \defnt \frac{1}{ {\mu_0} J} \bigg[ [{{\bsym{F}}} {{\bbbm{B}}} ] \otimes [{{\bsym{F}}} {{\bbbm{B}}} ] - \frac{1}{2} [{{\bsym{F}}} {{\bbbm{B}}}] \cdot [{{\bsym{F}}} {{\bbbm{B}}}] \bsym{I} \bigg] {{\bsym{F}}}^{-\top}.
\label{eqn: maxwell strPK}
\end{align}

In order to further simplify the first variation expression \eqref{eqn: M form delta E 1}, we apply the divergence theorem on the last term and use the condition from a variation of equation \eqref{eqn: maxwell lagrangian}$_1$ that $\Div({\sdel}{{\bbbm{B}}}) =0$ to get
\begin{align}
\int\nolimits_{\bvol_0} \mbf{n}_0 \cdot {\phi} {\sdel}{{\bbbm{B}}} {ds}_0 &=\int\nolimits_{\vol_0} \Div \left({\phi} {\sdel}{{\bbbm{B}}} \right) {dv}_0 =\int\nolimits_{\vol_0} \Grad({\phi}) \cdot {\sdel}{{\bbbm{B}}} {dv}_0\nonumber\\
 &=- \int\nolimits_{\vol_0} {{\bbbm{H}}} \cdot {\sdel}{{\bbbm{B}}} {dv}_0.
\label{eqn: last term M functional}
\end{align}
At this point we recall several identities for variations of ${{\bsym{C}}}, J$, etc, from {Appendix \ref{appendix: variations}}.
{Using the constitutive relation \eqref{eqn: constitutive B H M}, an increment of magnetic induction ${{\bbbm{B}}}$ up to first order can be written as}
\begin{align}
{\sdel}{{\bbbm{B}}} 
& =[ [{{\bsym{F}}}^{-\top} \cdot {\sdel}{{\bsym{F}}} ] \bsym{I} - {{\bsym{C}}}^{-1} [{\sdel}{{\bsym{F}}}]^\top {{\bsym{F}}} - {{\bsym{F}}}^{-1} [{\sdel}{{\bsym{F}}}] ] {{\bbbm{B}}} \nonumber\\
 &- {\mu_0} J {{\bsym{C}}}^{-1} \Grad {\sdel}{\Phi} + J {{\bsym{C}}}^{-1} {\sdel}{{\bbbm{M}}}.
\label{eqn: Delta B}
\end{align}
{Upon substituting \eqref{eqn: last term M functional} and \eqref{eqn: Delta B} } in the last term of equation \eqref{eqn: M form delta E 1}, we thus obtain
\begin{align}
{\sdel}{{\mathit{E}_{\mathtt{I}}}} &=\int\nolimits_{{\mcal{B}}_0}[ {{\Omega}}_{,{{\bsym{F}}}} \cdot {\sdel}{{\bsym{F}}} + {{\Omega}}_{,{{\bbbm{M}}}} \cdot {\sdel}{{\bbbm{M}}} - \widetilde{{\mbf{f}}}^e \cdot {\sdel}{\mathpalette\irchi\relax}] {dv}_0 - \int\nolimits_{{\partial\mcal{B}} _0} \widetilde{\mbf{t}}^e \cdot {\sdel}{\mathpalette\irchi\relax} {ds}_0 \nonumber \\
& + \int\nolimits_{\vol_0}[[ {{\mathring{{\bsym{P}}}}}_m- {{\widehat{{\bsym{P}}}}}_m] \cdot {\sdel}{{\bsym{F}}} - J {{\bsym{C}}}^{-1} {{\bbbm{H}}} \cdot {\sdel}{{\bbbm{M}}}] {dv}_0,
\label{eqn: delta E for M formulation 1}
\end{align}
where we have defined the tensor
\begin{align}
{{\mathring{{\bsym{P}}}}}_m &\defnt [ - [{{\bbbm{B}}} \cdot {{\bbbm{H}}}] \bsym{I} + [{{\bsym{F}}} {{\bbbm{B}}}] \otimes [{{\bsym{F}}}^{-\top} {{\bbbm{H}}}] + [{{\bsym{F}}}^{-\top} {{\bbbm{H}}}] \otimes [{{\bsym{F}}} {{\bbbm{B}}}] ] {{\bsym{F}}}^{-\top} \nonumber\\
&=2 {{\widehat{{\bsym{P}}}}}_m + J [ - [{{\bsym{C}}}^{-1} {{\bbbm{M}}} \cdot {{\bbbm{H}}}] \bsym{I} + [{{\bsym{F}}}^{-\top} {{\bbbm{M}}}] \otimes [{{\bsym{F}}}^{-\top} {{\bbbm{H}}}] \nonumber\\
 &+ [{{\bsym{F}}}^{-\top} {{\bbbm{H}}}] \otimes [{{\bsym{F}}}^{-\top} {{\bbbm{M}}}] ] {{\bsym{F}}}^{-\top}. 
\label{eqn: max type tensor 2}
\end{align}
As observed above, ${{\bbbm{M}}} =\mbf{0}$ in {the region ${\mcal{B}}_0'$}, which leads to ${{\mathring{{\bsym{P}}}}}_m =2 {{\bsym{P}}}_m$.

{Upon splitting the (third term) integral over $\vol_0$ in \eqref{eqn: delta E for M formulation 1} to a sum of the integrals on disjoint regions ${\mcal{B}}_0$ and ${\mcal{B}}_0'$, we obtain }
\begin{align}
{\sdel}{{\mathit{E}_{\mathtt{I}}}} &=\int\nolimits_{{\mcal{B}}_0} [[{{\Omega}}_{{{\bsym{F}}}} + {{\mathring{{\bsym{P}}}}}_m- {{\widehat{{\bsym{P}}}}}_m] \cdot {\sdel}{{\bsym{F}}} - \widetilde{{\mbf{f}}}^e \cdot {\sdel}{\mathpalette\irchi\relax} +[ {{\Omega}}_{,{{\bbbm{M}}}} - J {{\bsym{C}}}^{-1} {{\bbbm{H}}}] \cdot {\sdel}{{\bbbm{M}}}] {dv}_0\nonumber \\
& - \int\nolimits_{{\partial\mcal{B}} _0} \widetilde{\mbf{t}}^e \cdot {\sdel}{\mathpalette\irchi\relax} {ds}_0 + \int\nolimits_{{\mcal{B}}_0'} {{\bsym{P}}}_m \cdot {\sdel}{{\bsym{F}}} {dv}_0.\nonumber
\end{align}
This is rewritten with the use of divergence theorem as
\begin{align}
{\sdel}{{\mathit{E}_{\mathtt{I}}}} &=\int\nolimits_{{\mcal{B}}_0}[ -[ \Div ({{\Omega}}_{,{{\bsym{F}}}} + {{\mathring{{\bsym{P}}}}}_m- {{\widehat{{\bsym{P}}}}}_m ) + \widetilde{{\mbf{f}}}^e] \cdot {\sdel}{\mathpalette\irchi\relax} +[ {{\Omega}}_{,{{\bbbm{M}}}} - J {{\bsym{C}}}^{-1} {{\bbbm{H}}}] \cdot {\sdel}{{\bbbm{M}}}] {dv}_0 \nonumber \\
& + \int\nolimits_{{\partial\mcal{B}} _0}[[[ {{\Omega}}_{,{{\bsym{F}}}} + {{\mathring{{\bsym{P}}}}}_m- {{\widehat{{\bsym{P}}}}}_m]|_- - {{\bsym{P}}}_m|_+] \mbf{n}_0 - \widetilde{\mbf{t}}^e] \cdot {\sdel}{\mathpalette\irchi\relax} {ds}_0 \nonumber \\
&- \int\nolimits_{{\mcal{B}}_0'} \Div {{\bsym{P}}}_m 
\cdot {\sdel}{\mathpalette\irchi\relax} {dv}_0 + \int\nolimits_{\bvol_0} {{\bsym{P}}}_m \mbf{n}_0 \cdot {\sdel}{\mathpalette\irchi\relax} {ds}_0.\nonumber
\end{align}
Following the traditional definition, at this point, by virtue of inspection of the above form of the first variation of the potential energy functional, we define the first Piola--Kirchhoff stress in the body as
\begin{equation}
{\bsym{P}} \defnt {{\Omega}}_{,{{\bsym{F}}}} + {{\mathring{{\bsym{P}}}}}_m- {{\widehat{{\bsym{P}}}}}_m, \quad \text{ in } \quad {\mcal{B}}_0, 
\label{eqn: first PK for M formulation 1}
\end{equation}
while we have the natural stress tensor, i.e. Maxwell stress, ${\bsym{P}} ={{\bsym{P}}}_m$ defined by equation \eqref{eqn: maxwell strPK} {exterior to} the body, i.e., in ${\mcal{B}}_0'$.

\begin{sidenote}
The Cauchy stress ${\bsym{\sigma}}$ in the body is related to the first Piola--Kirchhoff {stress}
${{\bsym{P}}}$ by the Piola transform as
${\bsym{\sigma}} \mathrm{cof}({{\bsym{F}}}) ={{\bsym{P}}};$ also sometimes referred as the Nanson's relation.
Upon using the relation \eqref{eqn: Bb, Hh relations}$_1$ and the tensor field stated as \eqref{eqn: maxwell strPK}, the counterpart ${\bsym{\sigma}}_m$ of the Cauchy stress ${\bsym{\sigma}}$ in ${\mcal{B}}'$ (vacuum) is given by the expression
${\bsym{\sigma}} ={\bsym{\sigma}}_m =\frac{1}{ {\mu_0}} \left[ {\bbbm{b}} \otimes {\bbbm{b}} - \frac{1}{2} [{\bbbm{b}} \cdot {\bbbm{b}}] \bsym{I} \right] \text{ in } {\mcal{B}}'.$
\label{cauchystr}
\end{sidenote}

Upon applying the condition \eqref{eqn: first variation condition} to the first variation calculated above, the coefficients appearing with the arbitrary variations ${\sdel}{\mathpalette\irchi\relax}$ and ${\sdel}{{\bbbm{M}}}$ also should vanish for the requirement that ${\sdel}{{\mathit{E}_{\mathtt{I}}}}$ must be zero at equilibrium (i.e., ${\mathpalette\irchi\relax}, {{\bbbm{M}}}$ corresponding to a extremum point of $E$).
Vanishing of the coefficients of ${\sdel}{{\bbbm{M}}}$ results in the following constitutive relation between ${{\bbbm{H}}}$ and ${{\bbbm{M}}}$
\begin{equation}
{{\bbbm{H}}} =J^{-1} {{\bsym{C}}} {{\Omega}}_{, {{\bbbm{M}}}} \quad \quad \text{ in } \; {\mcal{B}}_0.
\label{eqn: constitutive magnetisation magnetic field}
\end{equation}
{Upon substituting the above expression for ${{\bbbm{H}}}$ in equations \eqref{eqn: max type tensor 1}, \eqref{eqn: max type tensor 2}, and \eqref{eqn: first PK for M formulation 1}} 
the total first Piola--Kirchhoff stress can be rewritten in terms of the independent quantities ${{\bsym{F}}}$ and ${{\bbbm{M}}}$ as
\begin{align}
{\bsym{P}} &={{\Omega}}_{,{{\bsym{F}}}} + {{\mathring{{\bsym{P}}}}}_m- {{\widehat{{\bsym{P}}}}}_m\\
&={{\Omega}}_{,{{\bsym{F}}}} + {\mu_0} J^{-1}[ - \frac{1}{2} {{\Omega}}_{,{{\bbbm{M}}}} \cdot \big[ {{\bsym{C}}} {{\Omega}}_{,{{\bbbm{M}}}} \big] \bsym{I} + {{{\bsym{F}}}}{{\Omega}}_{,{{\bbbm{M}}}} \otimes \big[ {{{\bsym{F}}}} {{\Omega}}_{,{{\bbbm{M}}}} \big]] {{\bsym{F}}}^{-\top} \nonumber \\
&+[ - \big[ {{\bbbm{M}}} \cdot {{\Omega}}_{,{{\bbbm{M}}}} \big] \bsym{I} + {{{\bsym{F}}}^{-\top}}{{\bbbm{M}}} \otimes {{{\bsym{F}}}} {{\Omega}}_{,{{\bbbm{M}}}} + {{{\bsym{F}}}} {{\Omega}}_{,{{\bbbm{M}}}} \otimes {{{\bsym{F}}}^{-\top}}{{\bbbm{M}}}] {{\bsym{F}}}^{-\top}.
\label{eqn: first PK in F and M terms}
\end{align}
Also ${\bsym{P}}=J(J^{-1}{{\Omega}}_{,{{\bsym{F}}}}{{\bsym{F}}}^{\top}+ {\bbbm{h}} \otimes {\bbbm{b}}-\frac{{\mu_0}}{2} ({\bbbm{h}} \cdot {\bbbm{h}}) \bsym{I} + \{{{\bbbm{m}}} \otimes {\bbbm{h}}- ({{\bbbm{m}}} \cdot {\bbbm{h}}) \bsym{I}\}){{\bsym{F}}}^{-\top},$ which differs from that given by \citep{kankanala2004} (see their eq. (2.26)) due to the presence of terms in the curly brackets. Vanishing of the coefficients of ${\sdel}{\mathpalette\irchi\relax}$ results in the following equations
\begin{subequations}
\begin{align}
\Div {{\bsym{P}}} + \widetilde{{\mbf{f}}}^e =\mbf{0} \quad &\text{ in } \quad {\mcal{B}}_0,
\label{eqn: gov 1 M formulation} \\
\Div {{\bsym{P}}} =\mbf{0} \quad &\text{ in } \quad {\mcal{B}}_0',\\
\jump{{{\bsym{P}}}} \mbf{n}_0 + \widetilde{\mbf{t}}^e=\mbf{0} \quad &\text{ on } \quad {\partial\mcal{B}} _0, \\
{{\bsym{P}}} \mbf{n}_0 =\mbf{0} \quad &\text{ on } \quad \bvol_0.
\end{align}
\label{eqn: Euler based M}
\end{subequations}
Here
$\jump{\{ \cdot \}} =\{ \cdot \}_+ - \{ \cdot \}_- $ with plus sign representing that side of the boundary (surface) which is reached along the unit {\em outward} normal vector.

\begin{sidenote}
We note that in this formulation based on the magnetisation vector, we have to apriori use both the Maxwell's equations \eqref{eqn: maxwell lagrangian} to impose conditions on ${{\bbbm{B}}}$ and ${{\bbbm{H}}}$ unlike the two formulations based on ${\bbbm{B}}$ and ${\bbbm{H}}$ presented in Appendix sections \ref{sec: B formulation} and \ref{sec: H formulation}. in which one condition was imposed and the other was derived.
Also unlike {those} two formulations, stress does not have a simple expression of being a derivative of the stored energy density with respect to the deformation gradient tensor.
The procedure implies the constitutive relation \eqref{eqn: constitutive magnetisation magnetic field} between ${{\bbbm{H}}}$ and ${{\bbbm{M}}}$.
\end{sidenote}

\subsection{Perturbation of equilibrium equation at critical point}
In terms of the variations ${\bdel}{\mathpalette\irchi\relax}$ and ${\bdel}{{\bbbm{M}}}$, we find the perturbation in the first Piola--Kirchhoff stress using equation \eqref{eqn: first PK in F and M terms} as
\begin{align}
{\bdel}{\bsym{P}}& ={{\Omega}}_{,{{\bsym{F}}} {{\bsym{F}}}} {\bdel}{{\bsym{F}}} + \frac{1}{2} [ {{\Omega}}_{,{{\bsym{F}}} {{\bbbm{M}}}} + \wst{{\Omega}}_{{{\bsym{F}}} {{\bbbm{M}}}} ] {\bdel}{{\bbbm{M}}} \nonumber \\
& - {\mu_0} J^{-1} [ {{\bsym{F}}}^{-\top} \cdot {\bdel}{{\bsym{F}}} ] [ - \frac{1}{2} {{\Omega}}_{,{{\bbbm{M}}}} \cdot \big[ {{\bsym{C}}} {{\Omega}}_{,{{\bbbm{M}}}} \big] \bsym{I} + {{\Omega}}_{,{{\bbbm{M}}}} \otimes \big[ {{\bsym{C}}} {{\Omega}}_{,{{\bbbm{M}}}} \big] ] {{\bsym{F}}}^{-\top} \nonumber \\
& - {\mu_0} J^{-1} [ - \frac{1}{2} {{\Omega}}_{,{{\bbbm{M}}}} \cdot \big[ {{\bsym{C}}} {{\Omega}}_{,{{\bbbm{M}}}} \big] \bsym{I} + {{\Omega}}_{,{{\bbbm{M}}}} \otimes \big[ {{\bsym{C}}} {{\Omega}}_{,{{\bbbm{M}}}} \big] ] {{\bsym{F}}}^{-\top} [{\bdel}{{\bsym{F}}}]^\top {{\bsym{F}}}^{-\top} \nonumber \\
& + {\mu_0} J^{-1} [ - \bigg[ {{\bsym{F}}} {{\Omega}}_{,{{\bbbm{M}}}} \cdot [ {\bdel}{{\bsym{F}}} {{\Omega}}_{,{{\bbbm{M}}}} + {{\bsym{F}}} \big[{{\Omega}}_{,{{\bbbm{M}}} {{\bbbm{M}}}} {\bdel}{{\bbbm{M}}} + \frac{1}{2} {{\Omega}}_{,{{\bbbm{M}}} {{\bsym{F}}}} {\bdel}{{\bsym{F}}} + \frac{1}{2} \wst{{\Omega}}_{{{\bbbm{M}}} {{\bsym{F}}}} {\bdel}{{\bsym{F}}} \big] ] \bigg] \bsym{I} \nonumber \\
& + \big[{{\Omega}}_{,{{\bbbm{M}}} {{\bbbm{M}}}} {\bdel}{{\bbbm{M}}} + \frac{1}{2} {{\Omega}}_{,{{\bbbm{M}}} {{\bsym{F}}}} {\bdel}{{\bsym{F}}} + \frac{1}{2} \wst{{\Omega}}_{{{\bbbm{M}}} {{\bsym{F}}}} {\bdel}{{\bsym{F}}} \big] \otimes \big[ {{\bsym{C}}} {{\Omega}}_{,{{\bbbm{M}}}} \big] \nonumber \\
& + {{\Omega}}_{,{{\bbbm{M}}}} \otimes \bigg[ {{\bsym{C}}} \big[{{\Omega}}_{,{{\bbbm{M}}} {{\bbbm{M}}}} {\bdel}{{\bbbm{M}}} + \frac{1}{2} {{\Omega}}_{,{{\bbbm{M}}} {{\bsym{F}}}} {\bdel}{{\bsym{F}}} + \frac{1}{2} \wst{{\Omega}}_{{{\bbbm{M}}} {{\bsym{F}}}} {\bdel}{{\bsym{F}}} \big] + [ [{\bdel}{{\bsym{F}}}]^\top {{\bsym{F}}} + {{\bsym{F}}}^\top {\bdel}{{\bsym{F}}} ] {{\Omega}}_{,{{\bbbm{M}}}} \bigg] ] {{\bsym{F}}}^{-\top} \nonumber \\
& - \bigg[ - \big[ {{\bbbm{M}}} \cdot {{\Omega}}_{,{{\bbbm{M}}}} \big] \bsym{I} + {{\bbbm{M}}} \otimes {{\Omega}}_{,{{\bbbm{M}}}} + {{\Omega}}_{,{{\bbbm{M}}}} \otimes {{\bbbm{M}}} \bigg] {{\bsym{F}}}^{-\top} [{\bdel}{{\bsym{F}}}]^\top {{\bsym{F}}}^{-\top} \nonumber \\
& + \bigg( - \bigg[ {\bdel}{{\bbbm{M}}} \cdot {{\Omega}}_{, {{\bbbm{M}}}} + {{\bbbm{M}}} \cdot [ {{\Omega}}_{,{{\bbbm{M}}} {{\bbbm{M}}}} {\bdel}{{\bbbm{M}}} + \frac{1}{2} {{\Omega}}_{,{{\bbbm{M}}} {{\bsym{F}}}} {\bdel}{{\bsym{F}}} + \frac{1}{2} \wst{{\Omega}}_{{{\bbbm{M}}} {{\bsym{F}}}} {\bdel}{{\bsym{F}}} ] \bigg] \bsym{I} \nonumber \\
& + {\bdel}{{\bbbm{M}}} \otimes {{\Omega}}_{, {{\bbbm{M}}}} + {{\bbbm{M}}} \otimes [ {{\Omega}}_{,{{\bbbm{M}}} {{\bbbm{M}}}} {\bdel}{{\bbbm{M}}} + \frac{1}{2} {{\Omega}}_{,{{\bbbm{M}}} {{\bsym{F}}}} {\bdel}{{\bsym{F}}} + \frac{1}{2} \wst{{\Omega}}_{{{\bbbm{M}}} {{\bsym{F}}}} {\bdel}{{\bsym{F}}} ] \nonumber \\
& + [ {{\Omega}}_{,{{\bbbm{M}}} {{\bbbm{M}}}} {\bdel}{{\bbbm{M}}} + \frac{1}{2} {{\Omega}}_{,{{\bbbm{M}}} {{\bsym{F}}}} {\bdel}{{\bsym{F}}} + \frac{1}{2} \wst{{\Omega}}_{{{\bbbm{M}}} {{\bsym{F}}}} {\bdel}{{\bsym{F}}} ] \otimes {{\bbbm{M}}} + {{\Omega}}_{,{{\bbbm{M}}}} \otimes {{\bbbm{M}}} \bigg) {{\bsym{F}}}^{-\top},
\label{eqn: delta P for M formulation}
\end{align}
where we have defined two third order tensors $\wst{{\Omega}}_{ {{\bsym{F}}} {{\bbbm{M}}}}$ and $\wst{{\Omega}}_{ {{\bbbm{M}}} {{\bsym{F}}} }$ which have the following property
\begin{equation}
[ \wst{{\Omega}}_{{{\bsym{F}}} {{\bbbm{M}}}} \mbf{u}] \cdot \bsym{U} =[ {{\Omega}}_{, {{\bbbm{M}}} {{\bsym{F}}}} \bsym{U} ] \cdot \mbf{u}, \quad [ \wst{{\Omega}}_{ {{\bbbm{M}}} {{\bsym{F}}}} \bsym{U}] \cdot \mbf{u} =[ {{\Omega}}_{, {{\bsym{F}}} {{\bbbm{M}}}} \mbf{u}] \cdot \bsym{U},
\end{equation}
$\mbf{u}$ being an arbitrary vector and $\bsym{U}$ being an arbitrary second order tensor.
For the bifurcation analysis of critical point $({\mathpalette\irchi\relax}, {{\bbbm{M}}})$, using \eqref{eqn: Euler based M},
the perturbations ${\bdel}{\mathpalette\irchi\relax}$ and ${\bdel}{{\bbbm{M}}}$ in the equilibrium state need to satisfy the following following partial differential equations and boundary conditions:
\begin{subequations}
\begin{align}
\Div {\bdel}{{\bsym{P}}} =\mbf{0} \quad &\text{ in } \quad {\mcal{B}}_0,
\label{eqn: incr gov 1 M formulation} \\
\Div {\bdel}{{\bsym{P}}} =\mbf{0} \quad &\text{ in } \quad {\mcal{B}}_0',\\
\jump{{\bdel}{{\bsym{P}}}} \mbf{n}_0 =\mbf{0} \quad &\text{ on } \quad {\partial\mcal{B}} _0,
\label{eqn: Perturb based M BC} \\
{\bdel}{{\bsym{P}}} \mbf{n}_0 =\mbf{0} \quad &\text{ on } \quad \bvol_0.
\end{align}
\label{eqn: Perturb based M}
\end{subequations}
The set of equations \eqref{eqn: Perturb based M} need to be solved for the non-trivial unknown functions $({\bdel}{\mathpalette\irchi\relax}, {\bdel}{\bbbm{M}})$ describing the onset of bifurcation.

\begin{sidenote}
Perturbation in the Maxwell stress ${\bdel}{{\bsym{P}}}_m$ in ${\mcal{B}}_0'$ in terms of ${\bdel}{{\bsym{F}}}$ and ${\bdel}{{\bbbm{H}}}$ is given by Equation \eqref{eqn: delta Pm expression 2}.
The boundary condition \eqref{eqn: Perturb based M BC} connects ${\bdel}{\bsym{P}}$ \eqref{eqn: delta P for M formulation} and ${\bdel}{{\bsym{P}}}_m$ \eqref{eqn: delta Pm expression 2} through the constitutive relation \eqref{eqn: constitutive magnetisation magnetic field} for ${{\bbbm{H}}}$.
\end{sidenote}

\begin{sidenote}
In the context of the first variation as well as the critical point perturbation, above expressions and equations are similar to those obtained in two other formulations based on ${{\bbbm{B}}}$ and ${{\bbbm{H}}}$. These are summarized in Appendix \ref{sec: B formulation} and \ref{sec: H formulation}, where the derivations provided in \citep{Saxena2020} for the case of electroelastic materials are closely followed.
\end{sidenote}

\section{Second formulation based on magnetisation}
\label{sec: M formulation 2}
Suppose that the physical space {exterior to} ${\mcal{B}}$ is the entire space outside; in other words, we assume that 
\begin{align}
\vol={\bbm{R}^3}.
\end{align}
We consider that scenario when the potential energy functional depends on the {magnetic} energy stored in the entire space, due to the so called stray field ${{\bbbm{h}}}^s$, and also includes a contribution of the work done by an {external} magnetic field ${{\bbbm{h}}}^e$ on the {magnetisation induced in the} body. 
As a consequence of this, unlike the formulation presented {in Section \ref{sec: M formulation 1}} and in Appendix \ref{sec: B formulation} and \ref{sec: H formulation}, we do not have any contribution due to those terms that involve an integral on the boundary of the region exterior to the body, i.e., on $\bvol$.
In particular, the total (magnetoelastic) {stored} energy $\overline{\mcal{E}}$ in {the considered system} is the sum of the energy stored in the body and the {stray magnetic field energy} of the entire space. The explicit mathematical expression of the energy, as a functional of the deformation ${\mathpalette\irchi\relax}$ \eqref{eqn: deform} (and \eqref{eqn: deform2}) and the spatial magnetisation ${{\overline{\bbbm{m}}}}$ \eqref{eqn: defn mpm and mpv}, \eqref{eqn: defn mpm and mpv 2} (per unit mass), is given by
\begin{align}
\overline{\mcal{E}}({\mathpalette\irchi\relax}, {{\overline{\bbbm{m}}}}) \defnt \int\nolimits_{{\mcal{B}}} \rho \overline{{\widehat{\Omega}}}({{\bsym{F}}}, {{\overline{\bbbm{m}}}}) {dv}+\int\nolimits_{\bbm{R}^3} \frac{1}{2}{{\mu_0}} {{\bbbm{h}}}^s \cdot {{\bbbm{h}}}^s {dv},
\label{eqn: total energy 1}
\end{align}
where we have defined $\overline{{\widehat{\Omega}}}$ as the Helmholtz energy (per unit mass). 
Following the physical nature of the stray fields, also by convention, it is assumed that the stray magnetic field ${\bbbm{h}}^s$ decays (in a suitable manner) far away from the body, that is $\Vert {\bbbm{h}}^s \Vert \to 0$ as $\Vert \mbf{x} \Vert \to \infty$ (recall that $\mbf{x}$ denotes the position vector in current configuration).

The work done on the magnetoelastic body (same as the negative of the potential energy of the applied dead loading) by externally applied mechanical and magnetic forces is given by \citep{kankanala2004}
\begin{align}
\int\nolimits_{{\mcal{B}}} \rho {{\bbbm{h}}}^e \cdot {{\overline{\bbbm{m}}}} {dv}+\int\nolimits_{{\mcal{B}}} \rho {{\mbf{f}}^e} \cdot {\mathpalette\irchi\relax} {dv}+\int\nolimits_{{\partial\mcal{B}} } \mbf{t}^e \cdot {\mathpalette\irchi\relax} {ds},
\label{eqn: defn W}
\end{align}
where ${{\mbf{f}}^e}$ denotes the body force (per unit mass) and $\mbf{t}^e$ denotes the {mechanical} traction {(per unit area of the current configuration)} while
{and the first term is identified as the Zeeman energy} \citep{Feynmann1965}. It is emphasized {that ${{\mbf{f}}^e}, \mbf{t}^e$ and ${{\bbbm{h}}}^e$ are external {\em dead loads}}.

Using \eqref{eqn: total energy 1} and \eqref{eqn: defn W}, the potential energy $\overline{{\mathit{E}_{\mathtt{I\hspace{-2pt}I}}}}$ of the system comprising of the body and the surrounding space is then given by $\overline{\mcal{E}}$ minus magnetoelastic work done, i.e.,
\begin{align}
\overline{{\mathit{E}_{\mathtt{I\hspace{-2pt}I}}}} ({\mathpalette\irchi\relax}, {{\overline{\bbbm{m}}}}) & \defnt \int\nolimits_{{\mcal{B}}} \big[\rho \overline{{\widehat{\Omega}}}({{\bsym{F}}}, {{\overline{\bbbm{m}}}})- {{\bbbm{h}}}^e \cdot (\rho{{\overline{\bbbm{m}}}})-\rho{{\mbf{f}}^e} \cdot {\mathpalette\irchi\relax} \big] {dv} \nonumber\\
& {- \int\nolimits_{{\partial\mcal{B}} } \mbf{t}^e \cdot {\mathpalette\irchi\relax} {ds}+ \frac{1}{2}{{\mu_0}} \int\nolimits_{{\mcal{B}}}{\bbbm{h}}^s \cdot {\bbbm{h}}^s {dv}+\frac{1}{2}{{\mu_0}} \int\nolimits_{{\mcal{B}}'} {\bbbm{h}}^s \cdot {\bbbm{h}}^s} {dv}. 
\label{eqn: first definition of Pi}
\end{align}

\begin{sidenote}
We emphasize that {even though the {Eulerian} expression of} the potential energy $\overline{{\mathit{E}_{\mathtt{I\hspace{-2pt}I}}}}$ {is the same as that provided by \cite{kankanala2004}, our formulation is} markedly different from theirs; since we consider the mechanical {deformation} ${\mathpalette\irchi\relax}$ and the magnetisation in the body ${{\overline{\bbbm{m}}}}$ as the only two unknown fields of the problem. 
 {Moreover, our referential formulation is quite different from that of \cite{kankanala2004} as discussed below.}
In terms of ${\mathpalette\irchi\relax}$ and ${{\overline{\bbbm{m}}}}$, the 
magnetic vector field ${{\bbbm{b}}}^s $ can be found by employing the Maxwell's equations stated in Section \ref{sec: basic eqns}.
As 
\begin{align}
{{\bbbm{b}}}^s ={{\mu_0}} {\bbbm{h}}^s + \rho{{\overline{\bbbm{m}}}}
\label{eqn: b h m relation total1}
\end{align}
and 
${\bbbm{h}}^s=-\grad{\phi}^s$
by \eqref{eqn: b h m relation total}${}_2$, \eqref{eqn: defn mpm and mpv}, 
while ${\phi}^s$ is found from the condition \eqref{eq: maxwell 1}, i.e. ${\div} {{\bbbm{b}}}^s =0$, that ${{\bbbm{b}}}^s$ satisfies. 
\end{sidenote}

It is preferable to write the potential energy in equation \eqref{eqn: first definition of Pi} in the reference (Lagrangian) configuration, i.e. all field variables are functions of the reference position vector $\mbf{X}$ instead of the current position vector $\mbf{x}$ ($={\mathpalette\irchi\relax}(\mbf{X})$).

The Helmholtz energy function $\overline{{\widehat{\Omega}}}({{\bsym{F}}}, {{\overline{\bbbm{m}}}})$ in \eqref{eqn: first definition of Pi} is mapped to ${{\widehat{\Omega}}}({{\bsym{F}}}, {{\overline{\bbbm{M}}}})$ (recall \eqref{eqnxXXx}),
i.e.,
\[
\overline{{\widehat{\Omega}}}({{\bsym{F}}}({\mathpalette\irchi\relax}^{-1}(\mbf{x})), {{\overline{\bbbm{m}}}}(\mbf{x}))={{\widehat{\Omega}}}({{\bsym{F}}}(\mbf{X}), {{\overline{\bbbm{M}}}}(\mbf{X})), \mbf{X}\in{\mcal{B}}_0.
\]
It is emphasized that the field ${{\bbbm{h}}}^e$ depends directly on the spatial location $\mbf{x}$ (unlike ${{\mbf{f}}}^e$) and is therefore explicitly mentioned as such.

Recall {equation \eqref{eqn: defn mpm and mpv} and Remark \ref{remark: M rho}}, and in particular the relations
\[
{{\bbbm{m}}}=\rho{{\overline{\bbbm{m}}}}=\rho J{{\bsym{F}}}^{-\top}{{\overline{\bbbm{M}}}}=\rho_0{{\bsym{F}}}^{-\top}{{\overline{\bbbm{M}}}}={{\bsym{F}}}^{-\top}{{\bbbm{M}}},\text{ and }{{\bbbm{M}}}=\rho_0{{\overline{\bbbm{M}}}}.
\]
Using the transformations stated above, we can redefine the {expression of the} potential energy {functional} \eqref{eqn: first definition of Pi} in a referential description as
\begin{align}
{{\mathit{E}_{\mathtt{I\hspace{-2pt}I}}}}({\mathpalette\irchi\relax}, {{\overline{\bbbm{M}}}}) 
& \defnt \int\nolimits_{{\mcal{B}}_0} \rho_0 \big[ {{\widehat{\Omega}}}({{\bsym{F}}}, {{\overline{\bbbm{M}}}})- J{{\bbbm{h}}}^e({\mathpalette\irchi\relax}(\mbf{X})) \cdot {{\bsym{F}}}^{-\top}{{\overline{\bbbm{M}}}}\big]{dv}_0 \nonumber \\
&
- \int\nolimits_{{\mcal{B}}_0} \widetilde{{\mbf{f}}}^e \cdot {\mathpalette\irchi\relax} {dv}_0 -\int\nolimits_{{\partial\mcal{B}} _0} \widetilde{\mbf{t}}^e \cdot {\mathpalette\irchi\relax} ds_0 \nonumber \\
&+\frac{1}{2}{{\mu_0}} \int\nolimits_{ {\mcal{B}}_0}J{{\bsym{F}}}^{-\top}{{\bbbm{H}}}^s \cdot {{\bsym{F}}}^{-\top}{{\bbbm{H}}}^s {dv}_0
+ \frac{1}{2}{{\mu_0}} \int\nolimits_{{\mcal{B}}'} {{\bbbm{h}}}^s \cdot {{\bbbm{h}}}^s {dv},
\label{eqn: potential in deformed configuration}
\end{align}
where $\widetilde{\mbf{t}}^e$ is the force per unit area of the current configuration placed on the reference configuration, i.e., $\widetilde{\mbf{t}}^e(\mbf{X})ds_0=\mbf{t}^e(\mbf{x})ds, \mbf{X}\in{\partial\mcal{B}} _0$. Also the body force per unit volume $\widetilde{\mbf{f}}^e$ on the reference configuration is related to ${\mbf{f}}^e$ by $\widetilde{\mbf{f}}^e(\mbf{X}){dv}_0=\rho\mbf{f}^e(\mbf{x}){dv}, \mbf{X}\in{\mcal{B}}_0$. In the first term of \eqref{eqn: potential in deformed configuration}, we have highlighted the dependence on $\mbf{X}$ for additional clarity.

Recall the extension of ${\mathpalette\irchi\relax}$ to ${\mcal{B}}'_0$ is also denoted by ${\mathpalette\irchi\relax}$ and the mapping ${\mathpalette\irchi\relax}$ is sufficiently smooth and it maps ${\partial\mcal{B}} _0$ to ${\partial\mcal{B}} $ such that it identifies with ${\mathpalette\irchi\relax}$ in that region and its gradient ${{\bsym{F}}}$ identifies with the deformation gradient ${{\bsym{F}}}$ of ${\mathpalette\irchi\relax}$ on the common boundary ${\partial\mcal{B}} _0$. 
In vacuum far from ${\partial\mcal{B}} _0$ the deformation gradient ${{\bsym{F}}}$ can very well be assumed to be identity for convenience. We can rewrite the last term of the potential energy in equation \eqref{eqn: potential in deformed configuration} so that the entire expressions becomes
\begin{align}
{{\mathit{E}_{\mathtt{I\hspace{-2pt}I}}}}({\mathpalette\irchi\relax}, {{\overline{\bbbm{M}}}}) 
& =\int\nolimits_{{\mcal{B}}_0} \rho_0 \big[ {{\widehat{\Omega}}}({{\bsym{F}}}, {{\overline{\bbbm{M}}}})- J{{\bbbm{h}}}^e({\mathpalette\irchi\relax}(\mbf{X})) \cdot {{\bsym{F}}}^{-\top}{{\overline{\bbbm{M}}}}\big]{dv}_0 \nonumber \\
&
- \int\nolimits_{{\mcal{B}}_0} \widetilde{{\mbf{f}}}^e \cdot {\mathpalette\irchi\relax} {dv}_0-\int\nolimits_{{\partial\mcal{B}} _0} \widetilde{\mbf{t}}^e \cdot {\mathpalette\irchi\relax} ds_0\nonumber \\
&+ \frac{1}{2}{{\mu_0}} \int\nolimits_{ {\mcal{B}}_0}J{{\bsym{C}}}^{-1}{{\bbbm{H}}}^s \cdot {{\bbbm{H}}}^s {dv}_0+\frac{1}{2}{{\mu_0}} \int\nolimits_{{\mcal{B}}'_0} J{{\bsym{C}}}^{-1}{{\bbbm{H}}}^s \cdot {{\bbbm{H}}}^s {dv}_0. 
\label{eqn: potential in reference configuration E2}
\end{align}

\subsection{Equilibrium: first variation}
Upon using the expressions for increments,
\begin{align}
{{\mathit{E}_{\mathtt{I\hspace{-2pt}I}}}} ({\mathpalette\irchi\relax}+{\sdel}{\mathpalette\irchi\relax}, {{\overline{\bbbm{M}}}}+{\sdel}{{\overline{\bbbm{M}}}})-{{\mathit{E}_{\mathtt{I\hspace{-2pt}I}}}} ({\mathpalette\irchi\relax}, {{\overline{\bbbm{M}}}}) ={\sdel}{{\mathit{E}_{\mathtt{I\hspace{-2pt}I}}}}({\mathpalette\irchi\relax}, {{\overline{\bbbm{M}}}})[{\sdel}{\mathpalette\irchi\relax}, {\sdel}{{\overline{\bbbm{M}}}}] \nonumber\\
+\frac{1}{2} {\sddel} {{\mathit{E}_{\mathtt{I\hspace{-2pt}I}}}} ({\mathpalette\irchi\relax}, {{\overline{\bbbm{M}}}})[{\sdel}{\mathpalette\irchi\relax}, {\sdel}{{\overline{\bbbm{M}}}}]+ o_2[{\sdel}{\mathpalette\irchi\relax}, {\sdel}{{\overline{\bbbm{M}}}}],
\end{align}
where $o_2[{\sdel}{\mathpalette\irchi\relax}, {\sdel}{{\overline{\bbbm{M}}}}]$ are the terms of order higher than two in ${\sdel}{\mathpalette\irchi\relax}$ and ${\sdel}{{\overline{\bbbm{M}}}}$; ${\sdel}{{\mathit{E}_{\mathtt{I\hspace{-2pt}I}}}}$ and ${\sddel} {{\mathit{E}_{\mathtt{I\hspace{-2pt}I}}}}$ are the first and the second variations of ${{\mathit{E}_{\mathtt{I\hspace{-2pt}I}}}}$, respectively.

The first variation ${\sdel}{{\mathit{E}_{\mathtt{I\hspace{-2pt}I}}}}[{\sdel}{\mathpalette\irchi\relax}, {\sdel}{{\overline{\bbbm{M}}}}]$, written simply as ${\sdel}{{\mathit{E}_{\mathtt{I\hspace{-2pt}I}}}}$, is given by
\begin{align}
&{\sdel}{{\mathit{E}_{\mathtt{I\hspace{-2pt}I}}}} =\int\nolimits_{{\mcal{B}}_0} \rho_0 \Big[ {{\widehat{\Omega}}}_{,{{\bsym{F}}}} \cdot {\sdel}{{\bsym{F}}}+{{\widehat{\Omega}}}_{, {{\overline{\bbbm{M}}}}} \cdot {\sdel}{{\overline{\bbbm{M}}}}- J(\grad^{\top} {{\bbbm{h}}}^e) {{\bsym{F}}}^{-\top}{{\overline{\bbbm{M}}}} \cdot {\sdel}{\mathpalette\irchi\relax} \nonumber\\
&- J{{\bbbm{h}}}^e \cdot{{\bsym{F}}}^{-\top} {\sdel}{{\overline{\bbbm{M}}}}- J{{\bbbm{h}}}^e \cdot{\sdel}{{\bsym{F}}}^{-\top} {{\overline{\bbbm{M}}}}- {\sdel}J{{\bbbm{h}}}^e \cdot{{\bsym{F}}}^{-\top} {{\overline{\bbbm{M}}}}\big]{dv}_0 \nonumber \\
&
- \int\nolimits_{{\mcal{B}}_0} \widetilde{{\mbf{f}}}^e \cdot {\sdel}{\mathpalette\irchi\relax} {dv}_0-\int\nolimits_{{\partial\mcal{B}} _0} \widetilde{\mbf{t}}^e \cdot {\sdel}{\mathpalette\irchi\relax} ds_0\nonumber\\
& +{{\mu_0}} \int\nolimits_{{\mcal{B}}_0} J {{\bsym{C}}}^{-1} {{\bbbm{H}}}^s \cdot {\sdel}{{\bbbm{H}}}^s {dv}_0+\frac{1}{2}{{\mu_0}} \int\nolimits_{{\mcal{B}}_0} \big[ J {\sdel}{{\bsym{C}}}^{-1} {{\bbbm{H}}}^s \cdot {{\bbbm{H}}}^s + {\sdel}J {{\bsym{C}}}^{-1} {{\bbbm{H}}}^s \cdot {{\bbbm{H}}}^s \big] {dv}_0 \nonumber\\
& +{{\mu_0}} \int\nolimits_{{\mcal{B}}'_0} J {{\bsym{C}}}^{-1} {{\bbbm{H}}}^s \cdot {\sdel}{{\bbbm{H}}}^s {dv}_0+ \frac{1}{2}{{\mu_0}} \int\nolimits_{{\mcal{B}}'_0} \big[ J {\sdel}{{\bsym{C}}}^{-1} {{\bbbm{H}}}^s \cdot {{\bbbm{H}}}^s + {\sdel}J {{\bsym{C}}}^{-1} {{\bbbm{H}}}^s \cdot {{\bbbm{H}}}^s \big] {dv}_0.
\label{delta Pi eq}
\end{align}
Using the identities for variations of ${{\bsym{C}}}, J$ from {Appendix \ref{appendix: variations}},
\begin{align}
\frac{1}{2}{{\mu_0}} \int\nolimits_{{\mcal{B}}_0} \big[ J {\sdel}{{\bsym{C}}}^{-1} {{\bbbm{H}}}^s \cdot {{\bbbm{H}}}^s + {\sdel}J {{\bsym{C}}}^{-1} {{\bbbm{H}}}^s \cdot {{\bbbm{H}}}^s \big] {dv}_0=\int\nolimits_{{\mcal{B}}_0} [- {{\check{{\bsym{P}}}}}_m \cdot {\sdel}{{\bsym{F}}} ]{dv}_0,\label{twoterms}\\
\text{where }
{{\check{{\bsym{P}}}}}_m\defnt {{\mu_0}}J\bigg[ {\bbbm{h}}^s \otimes {{\bbbm{h}}}^s -\frac{1}{2} [ {\bbbm{h}}^s \cdot {\bbbm{h}}^s ] \bsym{I} \bigg] {{\bsym{F}}}^{-\top}.
\label{eqn: definintion of first Piola in vacuum} 
\end{align}
\begin{sidenote}
{${{\check{{\bsym{P}}}}}_m$ resembles the tensor ${{\bsym{P}}}_m$} as defined in \eqref{eqn: maxwell strPK} {in the region exterior to the body ${\mcal{B}}_0$}.
{Indeed}, 
\[
\frac{1}{2}{{\mu_0}} \int\nolimits_{{\mcal{B}}_0'} \big[ J {\sdel}{{\bsym{C}}}^{-1} {{\bbbm{H}}}^s \cdot {{\bbbm{H}}}^s +{\sdel}J {{\bsym{C}}}^{-1} {{\bbbm{H}}}^s \cdot {{\bbbm{H}}}^s \big] {dv}_0=\int\nolimits_{{\mcal{B}}'_0} \big[ - {{\check{{\bsym{P}}}}}_m \cdot {\sdel}{{\bsym{F}}}\big] {dv}_0,
\]
which leads to {${{\check{{\bsym{P}}}}}_m ={{\mu_0}} J({\bbbm{h}}^s \otimes {\bbbm{h}}^s-\frac{1}{2} ({\bbbm{h}}^s \cdot {\bbbm{h}}^s )\bsym{I}){{\bsym{F}}}^{-\top},$
that can be compared with the Maxwell stress tensor ${{\bsym{P}}}_m={{\mu_0}} J({\bbbm{h}} \otimes {\bbbm{h}}-\frac{1}{2} ({\bbbm{h}} \cdot {\bbbm{h}} )\bsym{I}){{\bsym{F}}}^{-\top}$} from equation \eqref{eqn: maxwell strPK}, {exterior to} the body ${\mcal{B}}_0$.
Thus, it is not same as that obtained by the other three formulations; in particular, $ {{\check{{\bsym{P}}}}}_m$ decays as $\lVert\mbf{X}\rVert \to \infty$. This anomaly is due to the presence of an applied external field in infinite space which corresponds to a non-vanishing `external' Maxwell stress.
\end{sidenote}

We write a first order variation of the magnetic induction vector using the constitutive relation \eqref{eqn: constitutive B H M} 
{(with ${{\bbbm{M}}}=\rho_0{{\overline{\bbbm{M}}}}, {\sdel}{{\bbbm{H}}}^s=-\Grad{\sdel}{\Phi}^s$)} as
\begin{align}
{\sdel}{{\bbbm{B}}}^s 
&={\sdel}(J {{\bsym{C}}}^{-1})({\mu_0} {{\bbbm{H}}}^s + \rho_0{{\overline{\bbbm{M}}}})+J {{\bsym{C}}}^{-1}{\sdel}({\mu_0} {{\bbbm{H}}}^s + \rho_0{{\overline{\bbbm{M}}}})\nonumber\\
& =\Big[ [{{\bsym{F}}}^{-\top} \cdot {\sdel}{{\bsym{F}}} ] \bsym{I}-{{\bsym{C}}}^{-1} {\sdel}{{\bsym{F}}}^\top {{\bsym{F}}}-{{\bsym{F}}}^{-1} {\sdel}{{\bsym{F}}} \Big] {{\bbbm{B}}}^s \nonumber \\
&- {{\mu_0}} J {{\bsym{C}}}^{-1} \Grad {\sdel}{\Phi}^s+\rho_0 J {{\bsym{C}}}^{-1} {\sdel}{{\overline{\bbbm{M}}}}. 
\label{eqn: constitutive delB}
\end{align}
We use the divergence theorem and use the condition from a variation of equation 
\eqref{eqn: maxwell lagrangian}$_1$
that $\Div({\sdel}{{\bbbm{B}}}^s) =0$ to get
\begin{align}
-\int\nolimits_{{\mcal{B}}_0}{{\bbbm{H}}}^s \cdot{\sdel}{{\bbbm{B}}}^s {dv}_0&=\int\nolimits_{{\partial\mcal{B}} _0} \mbf{n}_0 \cdot {\Phi}^s {\sdel}{{\bbbm{B}}}^s ds_0,\\
-\int\nolimits_{{\mcal{B}}_0'} {{\bbbm{H}}}^s \cdot {\sdel}{{\bbbm{B}}}^s {dv}_0&=\int\nolimits_{{\partial\mcal{B}} _0'} \mbf{n}_0 \cdot {\Phi}^s {\sdel}{{\bbbm{B}}}^s ds_0=-\int\nolimits_{{\partial\mcal{B}} _0} \mbf{n}_0 \cdot {\Phi}^s {\sdel}{{\bbbm{B}}}^s ds_0.
\label{HBiden1}
\end{align}
Also, due to \eqref{eqn: constitutive delB},
\begin{align}
-{{\mu_0}}\int\nolimits_{{\mcal{B}}_0} J {{\bsym{C}}}^{-1} {{\bbbm{H}}}^s \cdot {\sdel}{{\bbbm{H}}}^s {dv}_0
&=\int\nolimits_{{\mcal{B}}_0} -{{\mathring{{\bsym{P}}}}}_m \cdot{\sdel}{{\bsym{F}}} {dv}_0\nonumber \\
&+ \int\nolimits_{{\mcal{B}}_0}{{\bbbm{H}}}^s \cdot(\rho_0 J {{\bsym{C}}}^{-1} {\sdel}{{\overline{\bbbm{M}}}}- {\sdel}{{\bbbm{B}}}^s) {dv}_0,
\end{align}
where ${{\mathring{{\bsym{P}}}}}_m$ is defined by
\begin{align}
{{\mathring{{\bsym{P}}}}}_m &\defnt 2 {{\check{{\bsym{P}}}}}_m +\rho_0 J \big(-({{\bsym{C}}}^{-1} {{\overline{\bbbm{M}}}} \cdot {{\bbbm{H}}}^s) \bsym{I}+({{\bsym{F}}}^{-\top} {{\overline{\bbbm{M}}}}) \otimes ({{\bsym{F}}}^{-\top} {{\bbbm{H}}}^s)\nonumber\\
&+({{\bsym{F}}}^{-\top} {{\bbbm{H}}}^s) \otimes ({{\bsym{F}}}^{-\top} {{\overline{\bbbm{M}}}}) \big) {{\bsym{F}}}^{-\top}. 
\label{defTtilde1}
\end{align}
Similarly,
\begin{align}
-{{\mu_0}}\int\nolimits_{{\mcal{B}}_0'} J {{\bsym{C}}}^{-1} {{\bbbm{H}}}^s \cdot {\sdel}{{\bbbm{H}}}^s {dv}_0&=
\int\nolimits_{{\mcal{B}}_0'} -2 {{\check{{\bsym{P}}}}}_m \cdot{\sdel}{{\bsym{F}}} {dv}_0
+\int\nolimits_{{\partial\mcal{B}} _0'} \mbf{n}_0 \cdot {\Phi}^s {\sdel}{{\bbbm{B}}}^s ds_0,
\end{align}
where $\int\nolimits_{{\partial\mcal{B}} _0'} \mbf{n}_0 \cdot {\Phi}^s {\sdel}{{\bbbm{B}}}^s ds_0=-\int\nolimits_{{\partial\mcal{B}} _0} \mbf{n}_0 \cdot {\Phi}^s {\sdel}{{\bbbm{B}}}^s ds_0.$

Upon changing the derivatives from current to reference configuration, we get
\begin{align}
\grad {{\bbbm{h}}}^e =\big[ \Grad {{\bbbm{h}}}^e \big] {{\bsym{F}}}^{-1}.
\end{align}
Using these expressions,
the first variation ${\sdel}{{\mathit{E}_{\mathtt{I\hspace{-2pt}I}}}}$ can, therefore, be rewritten as
\begin{align}
&{\sdel}{{\mathit{E}_{\mathtt{I\hspace{-2pt}I}}}} =\int\nolimits_{{\mcal{B}}_0} \rho_0 \big({{\widehat{\Omega}}}_{,{{\bsym{F}}}} \cdot {\sdel}{{\bsym{F}}}+{{\widehat{\Omega}}}_{, {{\overline{\bbbm{M}}}}} \cdot {\sdel}{{\overline{\bbbm{M}}}}- J{{\bsym{F}}}^{-\top}(\Grad^{\top} {{\bbbm{h}}}^e) {{\bsym{F}}}^{-\top}{{\overline{\bbbm{M}}}} \cdot {\sdel}{\mathpalette\irchi\relax} \nonumber\\
&- J{{\bbbm{h}}}^e \cdot{{\bsym{F}}}^{-\top} {\sdel}{{\overline{\bbbm{M}}}}- J{{\bbbm{h}}}^e \cdot{\sdel}{{\bsym{F}}}^{-\top} {{\overline{\bbbm{M}}}}- {\sdel}J{{\bbbm{h}}}^e \cdot{{\bsym{F}}}^{-\top} {{\overline{\bbbm{M}}}}\big]{dv}_0 \nonumber \\
&
- \int\nolimits_{{\mcal{B}}_0} \widetilde{{\mbf{f}}}^e \cdot {\sdel}{\mathpalette\irchi\relax} {dv}_0-\int\nolimits_{{\partial\mcal{B}} _0} \widetilde{\mbf{t}}^e \cdot {\sdel}{\mathpalette\irchi\relax} ds_0\nonumber\\
&+\int\nolimits_{{\mcal{B}}_0} (- {{\check{{\bsym{P}}}}}_m \cdot {\sdel}{{\bsym{F}}}){dv}_0+\int\nolimits_{{\mcal{B}}'_0} (- {{\check{{\bsym{P}}}}}_m \cdot {\sdel}{{\bsym{F}}}){dv}_0\nonumber\\
&+\int\nolimits_{{\mcal{B}}_0} {{\mathring{{\bsym{P}}}}}_m \cdot{\sdel}{{\bsym{F}}} {dv}_0 - \int\nolimits_{{\mcal{B}}_0}{{\bbbm{H}}}^s \cdot(\rho_0 J {{\bsym{C}}}^{-1} {\sdel}{{\overline{\bbbm{M}}}} ) {dv}_0-\int\nolimits_{{\partial\mcal{B}} _0} \mbf{n}_0 \cdot {\Phi}^s {\sdel}{{\bbbm{B}}}^s ds_0\nonumber\\
&+\int\nolimits_{{\mcal{B}}_0'} {{\mathring{{\bsym{P}}}}}_m \cdot{\sdel}{{\bsym{F}}} {dv}_0+\int\nolimits_{{\partial\mcal{B}} _0} \mbf{n}_0 \cdot {\Phi}^s {\sdel}{{\bbbm{B}}}^s ds_0.
\end{align}

Assuming the continuity of ${\Phi}^s$, i.e., ${\Phi}^s|_+-{\Phi}^s|_-$ on the boundary ${\partial\mcal{B}} _0$, the two terms involving ${\Phi}^s$ and $ {\sdel}{{\bbbm{B}}}^s$ cancel; the latter is obtained by using the variation of the condition
\begin{align}
\jump{{{\bbbm{B}}}^s}\cdot\mbf{n}_0=0.
\label{jumpBeq}
\end{align}

Apply the divergence theorem on the terms containing gradients of ${\sdel}{\mathpalette\irchi\relax}$ to get
\begin{align}
{\sdel}{{\mathit{E}_{\mathtt{I\hspace{-2pt}I}}}} &=\int\nolimits_{{\mcal{B}}_0} \bigg(-\big(\Div(\rho_0 {{\widehat{\Omega}}}_{,{{\bsym{F}}}}+\overline{{\bsym{P}}}_m+{{\mathring{{\bsym{P}}}}}_m - {{\check{{\bsym{P}}}}}_m)+ \widetilde{{\mbf{f}}}^e\nonumber\\
&+\rho_0 J{{\bsym{F}}}^{-\top} (\Grad^{\top} {{\bbbm{h}}}^e) {{\bsym{F}}}^{-\top}{{\overline{\bbbm{M}}}}\big)\cdot {\sdel}{\mathpalette\irchi\relax} \nonumber\\
&+\rho_0 ({{\widehat{\Omega}}}_{, {{\overline{\bbbm{M}}}}} - J {{\bsym{F}}}^{-1}{{\bbbm{h}}}^e- J {{\bsym{F}}}^{-1}{\bbbm{h}}^s) \cdot {\sdel}{{\overline{\bbbm{M}}}} \bigg) {dv}_0 \nonumber \\
&+\int\nolimits_{{\partial\mcal{B}} _0} \big((\rho_0 {{\widehat{\Omega}}}_{,{{\bsym{F}}}}+\overline{{\bsym{P}}}_m+{{\mathring{{\bsym{P}}}}}_m - {{\check{{\bsym{P}}}}}_m)_- \mbf{n}_0
-({{\mathring{{\bsym{P}}}}}_m - {{\check{{\bsym{P}}}}}_m)_+ \mbf{n}_0
-\widetilde{\mbf{t}}^e \big) \cdot {\sdel}{\mathpalette\irchi\relax} ds_0 \nonumber\\
&-\int\nolimits_{{\mcal{B}}'_0} \Div({{\mathring{{\bsym{P}}}}}_m - {{\check{{\bsym{P}}}}}_m) \cdot {\sdel}{\mathpalette\irchi\relax} {dv}_0,
\end{align}
where $\overline{{\bsym{P}}}_m$ is defined by
\begin{align}
\overline{{\bsym{P}}}_m&\defnt\rho_0 (J{{\bsym{F}}}^{-\top}{{\overline{\bbbm{M}}}} \otimes{{\bbbm{h}}}^e-(J{{\bsym{F}}}^{-\top}{{\overline{\bbbm{M}}}}\cdot{{\bbbm{h}}}^e)\bsym{I}){{\bsym{F}}}^{-\top}\nonumber\\
& {=
\rho_0 (J{{\bsym{F}}}^{-\top}{{\overline{\bbbm{M}}}} \otimes {{\bsym{F}}}^{-\top}{{\bsym{F}}}^{\top}{{\bbbm{h}}}^e-(J{{\bsym{F}}}^{-1}{{\bsym{F}}}^{-\top}{{\overline{\bbbm{M}}}}\cdot{{\bsym{F}}}^{\top}{{\bbbm{h}}}^e)\bsym{I}){{\bsym{F}}}^{-\top}
},
\end{align}
and we have used the assumptions that ${\mathpalette\irchi\relax}$ and ${\sdel}{\mathpalette\irchi\relax}$ are continuous across ${\partial\mcal{B}} _0$; and ${\bbbm{h}}^s \to \mbf{0}$ as $\lVert\mbf{X}\rVert \to \infty$.
Note that
\begin{align}
{{\mathring{{\bsym{P}}}}}_m +\overline{{\bsym{P}}}_m&=2 {{\check{{\bsym{P}}}}}_m +\rho_0 J \big(-({{\bsym{C}}}^{-1} {{\overline{\bbbm{M}}}} \cdot {{{\bsym{F}}}^{\top}}({\bbbm{h}}^s {+{\bbbm{h}}^{e}})) \bsym{I}\nonumber \\
&+({{\bsym{F}}}^{-\top} {{\overline{\bbbm{M}}}}) \otimes {{{\bsym{F}}}^{-\top}{{\bsym{F}}}^{\top}}({\bbbm{h}}^s {+{\bbbm{h}}^{e}})\nonumber\\
&+({{\bsym{F}}}^{-\top} {{\bbbm{H}}}^s) \otimes ({{\bsym{F}}}^{-\top} {{\overline{\bbbm{M}}}}) \big) {{\bsym{F}}}^{-\top}. 
\end{align}
In vacuum ${{\overline{\bbbm{M}}}} =\mbf{0}$ which leads to 
\[
{{\mathring{{\bsym{P}}}}}_m+\overline{{\bsym{P}}}_m={{\mathring{{\bsym{P}}}}}_m =2 {{\check{{\bsym{P}}}}}_m.
\]
With the defining expression
\begin{align}
{{\bsym{P}}}=\rho_0 {{\widehat{\Omega}}}_{,{{\bsym{F}}}}+\overline{{\bsym{P}}}_m+{{\mathring{{\bsym{P}}}}}_m - {{\check{{\bsym{P}}}}}_m,
\end{align}
the tensor ${\bsym{P}}$ can be identified as the total first Piola--Kirchhoff stress tensor and ${{{{\bsym{P}}}^{\star}}}={{\bsym{P}}}={{\check{{\bsym{P}}}}}_m$ can be identified as the Maxwell stress tensor in vacuum.

Corresponding to the equilibrium condition of vanishing of the first variation ${\sdel}{{\mathit{E}_{\mathtt{I\hspace{-2pt}I}}}}$ of the potential energy ${{\mathit{E}_{\mathtt{I\hspace{-2pt}I}}}}$, using the classical methods in the calculus of variations \citep{Gelfand2003}, i.e.,
\begin{align}
{\sdel}{{\mathit{E}_{\mathtt{I\hspace{-2pt}I}}}}({\mathpalette\irchi\relax}, {{\overline{\bbbm{M}}}})[{\sdel}{\mathpalette\irchi\relax}, {\sdel}{{\overline{\bbbm{M}}}}]=0,
\end{align}
since the increment ${\sdel}{{\overline{\bbbm{M}}}}$ is arbitrary, 
we arrive at the constitutive relation
\begin{align}
{ {{\widehat{\Omega}}}_{, {{\overline{\bbbm{M}}}}} 
=J {{\bsym{F}}}^{-1}[ {{\bbbm{h}}}^s + {\bbbm{h}}^e]
=J{{\bsym{C}}}^{-1} [{{\bbbm{H}}}^s + {{\bbbm{H}}}^e], }
\label{eqn: sec5 const relation of hs, he and omega} 
\end{align}
which is, { remarkably,} same as \eqref{eqn: constitutive magnetisation magnetic field}.

\begin{sidenote}
In particular, inside the body ${{\bsym{P}}}$ is given by (as ${{\bbbm{h}}}=J^{-1}{{\bsym{F}}}{{\widehat{\Omega}}}_{, {{\overline{\bbbm{M}}}}} $)
\begin{align}
{{\bsym{P}}}&=\rho_0 {{\widehat{\Omega}}}_{,{{\bsym{F}}}}+\rho_0 J \big(-({{\bsym{C}}}^{-1} {{\overline{\bbbm{M}}}} \cdot {{\bsym{F}}}^{\top}{\bbbm{h}}) \bsym{I}+({{\bsym{F}}}^{-\top} {{\overline{\bbbm{M}}}}) \otimes {{\bsym{F}}}^{-\top}{{\bsym{F}}}^{\top}{\bbbm{h}}\nonumber\\
&+{\bbbm{h}}^s\otimes ({{\bsym{F}}}^{-\top} {{\overline{\bbbm{M}}}}) \big) {{\bsym{F}}}^{-\top}+\bigg[ {{\mu_0}}J {\bbbm{h}}^s \otimes {{\bbbm{h}}}^s -\frac{{{\mu_0}} J}{2} [ {\bbbm{h}}^s \cdot {\bbbm{h}}^s ] \bsym{I} \bigg] {{\bsym{F}}}^{-\top}\nonumber\\
&=\rho_0 {{\widehat{\Omega}}}_{,{{\bsym{F}}}}+{{\mu_0}}J^{-1}\bigg[-\frac{ J^2}{2} [ {\bbbm{h}}^s \cdot {\bbbm{h}}^s ] \bsym{I} + J^2 {\bbbm{h}}^s \otimes {{\bbbm{h}}}^s \bigg] {{\bsym{F}}}^{-\top}\nonumber\\
&+\rho_0 [-({{\overline{\bbbm{M}}}} \cdot {{\widehat{\Omega}}}_{, {{\overline{\bbbm{M}}}}}) \bsym{I}+({{\bsym{F}}}^{-\top} {{\overline{\bbbm{M}}}}) \otimes {{\bsym{F}}}{{\widehat{\Omega}}}_{, {{\overline{\bbbm{M}}}}}+J{\bbbm{h}}^s\otimes ({{\bsym{F}}}^{-\top} {{\overline{\bbbm{M}}}}) ] {{\bsym{F}}}^{-\top},
\label{eqn: sec5 first PK full expression}
\end{align}
which differs from the expression \eqref{eqn: first PK in F and M terms} by the following term:
\begin{align}
{{\bsym{P}}}^N&={{\mu_0}}J\bigg[-\frac{ 1}{2} [ {\bbbm{h}}^e \cdot {\bbbm{h}}^e ] \bsym{I} + {\bbbm{h}}^e \otimes {{\bbbm{h}}}^e \bigg] {{\bsym{F}}}^{-\top}\nonumber\\
&+{{\mu_0}}J\bigg[-[ {\bbbm{h}}^s \cdot {\bbbm{h}}^e ] \bsym{I} + {\bbbm{h}}^s \otimes {{\bbbm{h}}}^e+ {\bbbm{h}}^e \otimes {{\bbbm{h}}}^s \bigg] {{\bsym{F}}}^{-\top}\nonumber\\
&+\rho_0 J{\bbbm{h}}^e\otimes ({{\bsym{F}}}^{-\top} {{\overline{\bbbm{M}}}}) {{\bsym{F}}}^{-\top}.
\end{align}
With $\psi(a)=-\frac{1}{2}(a\cdot a)\bsym{I} + a\otimes a$,
the first and second line in ${{\bsym{P}}}^N$ can be written as $\psi({\bbbm{h}}^e+{\bbbm{h}}^s)-\psi({\bbbm{h}}^s)$.
The difference between the two definitions of the stress tensor is not surprising. It is known that these could be different expressions, yet physically equivalent, as they depend on the formulation, see for example \citep{Hutter1978} who presented this aspect of the Maxwell stress tensor while analyzing several formulations of electromagnetism in the theory of deformable media.

\end{sidenote}

Since the increment ${\sdel}{\mathpalette\irchi\relax}$ is arbitrary, we arrive at the following equation of equilibrium in magnetoelastostatics (for a system of magnetoelastic body and its surrounding vacuum)
\begin{subequations}
\begin{align}
\Div{\bsym{P}}+ \widetilde{{\mbf{f}}}^e +\rho_0 J {{\bsym{F}}}^{-\top} [ \Grad^{\top} {{\bbbm{h}}}^e ] {{\bsym{F}}}^{-\top}{{\overline{\bbbm{M}}}}=\mbf{0}&
\text{ in } {\mcal{B}}_0,
\label{eqn: sec5 reference balance lin momentum} \\
\Div {{\check{{\bsym{P}}}}}_m =\mbf{0} & \text{ in } {\mcal{B}}'_0
\label{eqn: sec5 reference balance maxwell stress} \\
[{\bsym{P}}- {{\check{{\bsym{P}}}}}_m ]
\mbf{n}_0 =\widetilde{\mbf{t}}^e & \text{ on } {\partial\mcal{B}} _0.
\label{eqn: sec5 reference boundary condition traction}
\end{align}
\label{eqn: mag form 4 Euler eqns}
\end{subequations}

\subsection{Perturbation of equilibrium equation at critical point}
For the analysis of the critical point $({\mathpalette\irchi\relax}, {{\overline{\bbbm{M}}}})$, the perturbations ${\bdel}{\mathpalette\irchi\relax} $ and ${\sdel}{{\overline{\bbbm{M}}}}$ in the equilibrium state need to satisfy certain incremental equations and boundary conditions.
They are derived by a perturbation of \eqref{eqn: mag form 4 Euler eqns} and are stated below.
Recalling from equation \eqref{eqn: sec5 const relation of hs, he and omega} that $ {{\bbbm{h}}}^s =J^{-1} {{\bsym{F}}} {{\widehat{\Omega}}}_{, {{\overline{\bbbm{M}}}}} - {\bbbm{h}}^e $,
perturbation in the first Piola--Kirchhoff stress can be written using the equation \eqref{eqn: sec5 first PK full expression} as
\begin{align}
 {\bdel}{\bsym{P}} &=\rho_0 \bigg[ {{\widehat{\Omega}}}_{, {{\bsym{F}}} {{\bsym{F}}}} {\bdel}{{\bsym{F}}} + \frac{1}{2} \big[ {{\widehat{\Omega}}}_{, {{\bsym{F}}} {{\overline{\bbbm{M}}}}} + \wst{{\widehat{\Omega}}}_{ {{\bsym{F}}} {{\overline{\bbbm{M}}}} } \big] {\bdel}{{\overline{\bbbm{M}}}} \bigg]\nonumber \\
& - \mu_0 J^{-1} [ {{\bsym{F}}}^{-\top} \cdot {\bdel}{{\bsym{F}}} ] \bigg[-\frac{ J^2}{2} [ {\bbbm{h}}^s \cdot {\bbbm{h}}^s ] \bsym{I} + J^2 {\bbbm{h}}^s \otimes {{\bbbm{h}}}^s \bigg] {{\bsym{F}}}^{-\top} \nonumber \\
 &+ {{\mu_0}}J^{-1} \Bigg[ - J^2 \bigg[ [{{\bsym{F}}}^{-\top} \cdot {\bdel}{{\bsym{F}}} ] [ {\bbbm{h}}^s \cdot {\bbbm{h}}^s ] + {\bbbm{h}}^s \cdot {\bdel}{\bbbm{h}}^s \bigg] \bsym{I} + 2 J^2 [{{\bsym{F}}}^{-\top} \cdot {\bdel}{{\bsym{F}}} ] {\bbbm{h}}^s \otimes {{\bbbm{h}}}^s \nonumber \\
 & + J^2 \big[ {\bdel}{\bbbm{h}}^s \otimes {{\bbbm{h}}}^s + {\bbbm{h}}^s \otimes {\bdel}{{\bbbm{h}}}^s \big] \Bigg] {{\bsym{F}}}^{-\top} - {\mu_0} J^{-1} \bigg[-\frac{ J^2}{2} [ {\bbbm{h}}^s \cdot {\bbbm{h}}^s ] \bsym{I} + J^2 {\bbbm{h}}^s \otimes {{\bbbm{h}}}^s \bigg] {{\bsym{F}}}^{-\top} [{\bdel}{{\bsym{F}}}]^\top {{\bsym{F}}}^{-\top }\nonumber \\
 &+ \rho_0 \Bigg[ -\Big[{\bdel}{{\overline{\bbbm{M}}}} \cdot {{\widehat{\Omega}}}_{, {{\overline{\bbbm{M}}}}} + {{\overline{\bbbm{M}}}} \cdot \bigg[ \frac{1}{2} \big[ {{\widehat{\Omega}}}_{, {{\overline{\bbbm{M}}}} {{\bsym{F}}}} + \wst{{\widehat{\Omega}}}_{ {{\overline{\bbbm{M}}}} {{\bsym{F}}}} \big] {\bdel}{{\bsym{F}}} + {{\widehat{\Omega}}}_{, {{\overline{\bbbm{M}}}} {{\overline{\bbbm{M}}}}} {\bdel}{{\overline{\bbbm{M}}}} \bigg] \Big] \bsym{I} \nonumber\\
 & + [ - {{\bsym{F}}}^{-\top} {\bdel}{{\bsym{F}}} {{\bsym{F}}}^{-\top} {{\overline{\bbbm{M}}}} + {{\bsym{F}}}^{-\top} {\bdel}{{\overline{\bbbm{M}}}} ] \otimes {{\bsym{F}}}{{\widehat{\Omega}}}_{, {{\overline{\bbbm{M}}}}} \nonumber \\
 & + {{\bsym{F}}}^{-\top} {{\overline{\bbbm{M}}}} \otimes \bigg[ {\bdel}{{\bsym{F}}} {{\widehat{\Omega}}}_{, {{\overline{\bbbm{M}}}}} + \frac{1}{2}{{\bsym{F}}} \big[ {{\widehat{\Omega}}}_{, {{\overline{\bbbm{M}}}} {{\bsym{F}}}} + \wst{{\widehat{\Omega}}}_{ {{\overline{\bbbm{M}}}} {{\bsym{F}}}} \big] {\bdel}{{\bsym{F}}} + {{\bsym{F}}} {{\widehat{\Omega}}}_{, {{\overline{\bbbm{M}}}} {{\overline{\bbbm{M}}}}} {\bdel}{{\overline{\bbbm{M}}}} \bigg] \nonumber \\
 & + \big[ J [{{\bsym{F}}} ^{-\top} \cdot {\bdel}{{\bsym{F}}}] {\bbbm{h}}^s + J {\bdel}{\bbbm{h}}^s \big]\otimes [ {{\bsym{F}}}^{-\top} {{\overline{\bbbm{M}}}} ] + J {\bbbm{h}}^s \otimes \big[ {{\bsym{F}}}^{-\top} {\bdel}{{\overline{\bbbm{M}}}} - {{\bsym{F}}}^{-\top} [{\bdel}{{\bsym{F}}} ]^\top {{\bsym{F}}}^{-\top} \big] \Bigg] {{\bsym{F}}}^{-\top} \nonumber \\
 & -\rho_0 \bigg[-({{\overline{\bbbm{M}}}} \cdot {{\widehat{\Omega}}}_{, {{\overline{\bbbm{M}}}}}) \bsym{I}+({{\bsym{F}}}^{-\top} {{\overline{\bbbm{M}}}}) \otimes {{\bsym{F}}}{{\widehat{\Omega}}}_{, {{\overline{\bbbm{M}}}}}+J{\bbbm{h}}^s\otimes ({{\bsym{F}}}^{-\top} {{\overline{\bbbm{M}}}}) \bigg] {{\bsym{F}}}^{-\top} [{\bdel}{{\bsym{F}}}]^\top {{\bsym{F}}}^{-\top},
\end{align}
where we can obtain the expression for ${\bdel}{\bbbm{h}}^s $ from equation \eqref{eqn: sec5 const relation of hs, he and omega} as
\begin{align}
 {\bdel}{\bbbm{h}}^s &=J^{-1} {{\bsym{F}}} \big[ {{\widehat{\Omega}}}_{, {{\overline{\bbbm{M}}}} {{\bsym{F}}}} {\bdel}{{\bsym{F}}} + {{\widehat{\Omega}}}_{, {{\overline{\bbbm{M}}}} {{\overline{\bbbm{M}}}}} {\bdel}{{\overline{\bbbm{M}}}} \big] \nonumber \\
&- J^{-1} [ {{\bsym{F}}}^{-\top} \cdot {\bdel}{{\bsym{F}}} ] {{\bsym{F}}} {{\widehat{\Omega}}}_{, {{\overline{\bbbm{M}}}}} + J^{-1} {\bdel}{{\bsym{F}}} {{\widehat{\Omega}}}_{, {{\overline{\bbbm{M}}}}} .
\end{align}
We have also introduced two second order tensors $\wst{{\widehat{\Omega}}}_{{{\overline{\bbbm{M}}}} {{\bsym{F}}}}$ and $\wst{{\widehat{\Omega}}}_{{{\bsym{F}}} {{\overline{\bbbm{M}}}} }$ with the property 
\begin{equation}
 \Big[ \wst{{\widehat{\Omega}}}_{{{\overline{\bbbm{M}}}} {{\bsym{F}}}} \bsym{U} \Big] \mbf{u} =\Big[ {{\widehat{\Omega}}}_{, {{\bsym{F}}} {{\overline{\bbbm{M}}}}} \mbf{u} \Big] \cdot \bsym{U}, \quad \quad \Big[ \wst{{\widehat{\Omega}}}_{{{\bsym{F}}} {{\overline{\bbbm{M}}}} } \mbf{u} \Big] \cdot \bsym{U} =\Big[ {{\widehat{\Omega}}}_{, {{\overline{\bbbm{M}}}} {{\bsym{F}}}} \bsym{U} \Big] \cdot \mbf{u} .
\end{equation}
for arbitrary vector $\mbf{u}$ and arbitrary second order tensor $\bsym{U}$. The expression for ${\bdel}{{\check{{\bsym{P}}}}}_m$ is obtained from equation \eqref{eqn: definintion of first Piola in vacuum} as
\begin{align}
 {\bdel}{{\check{{\bsym{P}}}}}_m & ={{\mu_0}}J \big[ {{\bsym{F}}}^{-\top} \cdot {\bdel}{{\bsym{F}}} \big] \bigg[ {\bbbm{h}}^s \otimes {{\bbbm{h}}}^s -\frac{1}{2} [ {\bbbm{h}}^s \cdot {\bbbm{h}}^s ] \bsym{I} \bigg] {{\bsym{F}}}^{-\top} \nonumber \\
 & + {{\mu_0}}J \bigg[ {\bbbm{h}}^s \otimes {\bdel}{\bbbm{h}}^s + {\bdel}{\bbbm{h}}^s \otimes {\bbbm{h}}^s - \big[ {\bbbm{h}}^s \cdot {\bdel}{\bbbm{h}}^s \big] \bsym{I} \bigg] {{\bsym{F}}}^{-\top} \nonumber \\
 & - {{\mu_0}}J \bigg[ {\bbbm{h}}^s \otimes {{\bbbm{h}}}^s -\frac{1}{2} [ {\bbbm{h}}^s \cdot {\bbbm{h}}^s ] \bsym{I} \bigg] {{\bsym{F}}}^{-\top} [{\bdel}{{\bsym{F}}}]^\top {{\bsym{F}}}^{-\top}.
\end{align}
Finally above leads to the following partial differential equations and boundary conditions
\begin{subequations}
 \begin{align}
 \Div {\bdel}{\bsym{P}} +\rho_0 J {{\bsym{F}}}^{-\top} \big[ \Grad^{\top} {{\bbbm{h}}}^e \big] {{\bsym{F}}}^{-\top} {\bdel}{{\overline{\bbbm{M}}}} & \nonumber \\
 +\rho_0 J \big[ {{\bsym{F}}}^{-\top} \cdot {\bdel}{{\bsym{F}}} \big] {{\bsym{F}}}^{-\top} \big[ \Grad^{\top} {{\bbbm{h}}}^e \big] {{\bsym{F}}}^{-\top}{{\overline{\bbbm{M}}}} & \nonumber \\
 - \rho_0 J {{\bsym{F}}}^{-\top} [{\bdel}{{\bsym{F}}} ]^\top {{\bsym{F}}}^{-\top} \big[ \Grad^{\top} {{\bbbm{h}}}^e \big] {{\bsym{F}}}^{-\top}{{\overline{\bbbm{M}}}} &\nonumber \\
 - \rho_0 J {{\bsym{F}}}^{-\top} \big[ \Grad^{\top} {{\bbbm{h}}}^e \big] {{\bsym{F}}}^{-\top} [{\bdel}{{\bsym{F}}} ]^\top {{\bsym{F}}}^{-\top} {{\overline{\bbbm{M}}}} &=0 
\text{ in }{\mcal{B}}_0,\\
 \Div {\bdel}{{\check{{\bsym{P}}}}}_m &=0 \text{ in } {\mcal{B}}'_0 \\
 \big[{\bdel}{\bsym{P}}- {\bdel}{{\check{{\bsym{P}}}}}_m \big]
\mbf{n}_0 &=0 \text{ on }{\partial\mcal{B}} _0.
 \end{align}
\end{subequations}

\section{Third formulation based on magnetisation}
\label{sec: M formulation 3}
In the back drop of the two formulations provided thus far based on the magnetisation, we investigate in this section the expressions provided by \cite{kankanala2004} which also assume that the stored energy density depends on the magnetisation as the additional field besides the deformation gradient.
Following \cite{kankanala2004}, in this case, the magnetisation per unit mass pulled back to the reference configuration (recall \eqref{eqnxXXx}), i.e., 
\begin{align}
{{\bbbm{K}}}(\mbf{X}) \defnt {{\overline{\bbbm{m}}}}(\mbf{x}) =J(\mbf{X}){{\bsym{F}}}^{-\top}(\mbf{X}) {{\overline{\bbbm{M}}}}(\mbf{X}), \mbf{X}\in{\mcal{B}}_0.
\label{defKvec}
\end{align}
is itself treated as a material field. In particular, note that the direction of the referential vector field ${{\bbbm{K}}}$ on ${\mcal{B}}_0$ is same as that of the spatial vector field ${{\overline{\bbbm{m}}}}$ on ${\mcal{B}}$, while it differs from the choice of the referential field ${{\overline{\bbbm{M}}}}$ due to the presence of cofactor map for ${{\bsym{F}}}$ (Nanson's relation).
The total potential energy of the system is written as
\begin{align}
{{\mathit{E}_{\mathtt{I\hspace{-2pt}I\hspace{-2pt}I}}}}({\mathpalette\irchi\relax}, {{\bbbm{K}}}) 
& \defnt \int\nolimits_{{\mcal{B}}_0} \rho_0 \big[ {{\widetilde{\Omega}}}({{\bsym{F}}}, {{\bbbm{K}}})- {{\bbbm{h}}}^e({\mathpalette\irchi\relax}(\mbf{X})) \cdot {{\bbbm{K}}}\big]{dv}_0 \nonumber \\
&
- \int\nolimits_{{\mcal{B}}_0} \widetilde{{\mbf{f}}}^e \cdot {\mathpalette\irchi\relax} {dv}_0-\int\nolimits_{{\partial\mcal{B}} _0} \widetilde{\mbf{t}}^e \cdot {\mathpalette\irchi\relax} ds_0\nonumber \\
&+ \frac{1}{2}{{\mu_0}} \int\nolimits_{ {\mcal{B}}_0}J{{\bsym{C}}}^{-1}{{\bbbm{H}}}^s \cdot {{\bbbm{H}}}^s {dv}_0+\frac{1}{2}{{\mu_0}} \int\nolimits_{{\mcal{B}}'_0} J{{\bsym{C}}}^{-1}{{\bbbm{H}}}^s \cdot {{\bbbm{H}}}^s {dv}_0.
\label{PifunctionalKR}
\end{align}
Here $\widetilde{{\mbf{f}}^e}$ represents the body force (per unit volume) and $\widetilde{\mbf{t}}^e$ denotes the {mechanical} traction. In contrast to \eqref{eqn: defn W}, the term corresponding to the Zeeman energy is written differently.
Note {from equation \eqref{eqn: b h m relation total}} that $J^{-1}{{\bsym{F}}}{{\bbbm{B}}}^s={{\mu_0}}{{\bsym{F}}}^{-\top}{{\bbbm{H}}}^s+\rho {{\bbbm{K}}},$ i.e., ${{\bbbm{B}}}^s={{\mu_0}}J{{\bsym{C}}}^{-1}{{\bbbm{H}}}^s+\rho_0 {{\bsym{F}}}^{-1}{{\bbbm{K}}}$, so that
\begin{align}
{\sdel}{{\bbbm{B}}}^s 
&={\sdel}(J {{\bsym{C}}}^{-1}) [{\mu_0} {{\bbbm{H}}}^s] + \rho_0{\sdel}({{\bsym{F}}}^{-1}{{\bbbm{K}}}), \nonumber\\
& =\Big[ [{{\bsym{F}}}^{-\top} \cdot {\sdel}{{\bsym{F}}} ] \bsym{I}-{{\bsym{C}}}^{-1} {\sdel}{{\bsym{F}}}^\top {{\bsym{F}}}-{{\bsym{F}}}^{-1} {\sdel}{{\bsym{F}}} \Big]{{\mu_0}}J{{\bsym{C}}}^{-1}{{\bbbm{H}}}^s-\rho_0 {{\bsym{F}}}^{-1}{\sdel}{{\bsym{F}}}{{\bsym{F}}}^{-1} {{\bbbm{K}}} \nonumber \\
&- {{\mu_0}} J {{\bsym{C}}}^{-1} \Grad {\sdel}{\Phi}^s+\rho_0 {{\bsym{F}}}^{-1} {\sdel}{{\bbbm{K}}},\nonumber\\
& =\Big[ [{{\bsym{F}}}^{-\top} \cdot {\sdel}{{\bsym{F}}} ] \bsym{I}-{{\bsym{C}}}^{-1} {\sdel}{{\bsym{F}}}^\top {{\bsym{F}}}\Big]{{\mu_0}}J{{\bsym{C}}}^{-1}{{\bbbm{H}}}^s
-{{\bsym{F}}}^{-1} {\sdel}{{\bsym{F}}} {{\bbbm{B}}}^s\nonumber\\
&- {{\mu_0}} J {{\bsym{C}}}^{-1} \Grad {\sdel}{\Phi}^s+\rho_0 {{\bsym{F}}}^{-1} {\sdel}{{\bbbm{K}}}. 
\label{KTeq1term}
\end{align}
Using this relation, we can rewrite the following integral that occurs in the first variation of potential energy as
\begin{align}
-{{\mu_0}}\int\nolimits_{{\mcal{B}}_0} J {{\bsym{C}}}^{-1} {{\bbbm{H}}}^s \cdot {\sdel}{{\bbbm{H}}}^s {dv}_0
&=\int\nolimits_{{\mcal{B}}_0} -{{\widetilde{{\bsym{P}}}}}_m \cdot{\sdel}{{\bsym{F}}} {dv}_0\nonumber \\
&+ \int\nolimits_{{\mcal{B}}_0}{{\bbbm{H}}}^s \cdot(\rho_0 {{\bsym{F}}}^{-1} {\sdel}{{\bbbm{K}}}- {\sdel}{{\bbbm{B}}}^s) {dv}_0,
\label{KTeq2term}
\end{align}
where the integrand of the first term on the right hand side, i.e., $-{{\widetilde{{\bsym{P}}}}}_m\cdot {\sdel}{{\bsym{F}}}$, can be expanded as
\begin{align}
-{{\widetilde{{\bsym{P}}}}}_m\cdot {\sdel}{{\bsym{F}}}
&={{\mu_0}}J({{\bsym{F}}}^{-\top} \cdot {\sdel}{{\bsym{F}}}) {{\bsym{C}}}^{-1}{{\bbbm{H}}}^s\cdot{{\bbbm{H}}}^s-{{\mu_0}}J{{\bsym{C}}}^{-1} {\sdel}{{\bsym{F}}}^\top {{\bsym{F}}} {{\bsym{C}}}^{-1}{{\bbbm{H}}}^s\cdot{{\bbbm{H}}}^s\nonumber\\
&
-{{\bsym{F}}}^{-\top}{{\bbbm{H}}}^s\otimes{{\bbbm{B}}}^s\cdot {\sdel}{{\bsym{F}}}\nonumber\\
&=(-2{{\check{{\bsym{P}}}}}_m-\rho_0 {\bbbm{h}}^s\otimes{{\bbbm{K}}}{{\bsym{F}}}^{-\top})\cdot {\sdel}{{\bsym{F}}}.
\end{align}
Thus, 
\begin{equation}
 {{\widetilde{{\bsym{P}}}}}_m-{{\check{{\bsym{P}}}}}_m={{\check{{\bsym{P}}}}}_m+\rho_0 {\bbbm{h}}^s\otimes{{\bbbm{K}}}{{\bsym{F}}}^{-\top}={{\check{{\bsym{P}}}}}_m+J {\bbbm{h}}^s\otimes{{\bbbm{m}}}{{\bsym{F}}}^{-\top}. 
 \label{eqn: sec6 strPKent expression}
\end{equation}

From \eqref{twoterms} we already know a part of the expression of the first variation of stray field energy term.
{Therefore, we write the first variation of the potential energy \eqref{PifunctionalKR} as}
\begin{align}
{\sdel}{{\mathit{E}_{\mathtt{I\hspace{-2pt}I\hspace{-2pt}I}}}}({\mathpalette\irchi\relax}, {{\bbbm{K}}}) 
& =\int\nolimits_{{\mcal{B}}_0} \rho_0 \big[{{\widetilde{\Omega}}}_{,{{\bsym{F}}}} \cdot {\sdel}{{\bsym{F}}}+{{\widetilde{\Omega}}}_{, {{\bbbm{K}}}} \cdot {\sdel}{{\bbbm{K}}}- {{\bsym{F}}}^{-\top}(\Grad^{\top} {{\bbbm{h}}}^e) {{\bbbm{K}}}\cdot {\sdel}{\mathpalette\irchi\relax}-{{\bbbm{h}}}^e\cdot {\sdel}{{\bbbm{K}}}\big]{dv}_0 \nonumber \\
&
- \int\nolimits_{{\mcal{B}}_0} \widetilde{{\mbf{f}}}^e \cdot {\sdel}{\mathpalette\irchi\relax} {dv}_0-\int\nolimits_{{\partial\mcal{B}} _0} \widetilde{\mbf{t}}^e \cdot{\sdel}{\mathpalette\irchi\relax} ds_0\nonumber \\
&+\int\nolimits_{{\mcal{B}}_0} (- {{\check{{\bsym{P}}}}}_m \cdot {\sdel}{{\bsym{F}}}){dv}_0+\int\nolimits_{{\mcal{B}}'_0} (- {{\check{{\bsym{P}}}}}_m \cdot {\sdel}{{\bsym{F}}}){dv}_0\nonumber\\
&+\int\nolimits_{{\mcal{B}}_0} {{\widetilde{{\bsym{P}}}}}_m \cdot{\sdel}{{\bsym{F}}} {dv}_0 - \int\nolimits_{{\mcal{B}}_0}{{\bbbm{H}}}^s \cdot(\rho_0 {{\bsym{F}}}^{-1} {\sdel}{{\bbbm{K}}} ) {dv}_0-\int\nolimits_{{\partial\mcal{B}} _0} \mbf{n}_0 \cdot {\Phi}^s {\sdel}{{\bbbm{B}}}^s ds_0\nonumber\\
&+\int\nolimits_{{\mcal{B}}_0'} {{\widetilde{{\bsym{P}}}}}_m \cdot{\sdel}{{\bsym{F}}} {dv}_0+\int\nolimits_{{\partial\mcal{B}} _0} \mbf{n}_0 \cdot {\Phi}^s {\sdel}{{\bbbm{B}}}^s ds_0.
\end{align}
{On applying} the divergence theorem on the terms containing gradient of ${\sdel}{\mathpalette\irchi\relax}$, we get
\begin{align}
{\sdel}{{\mathit{E}_{\mathtt{I\hspace{-2pt}I\hspace{-2pt}I}}}} &=\int\nolimits_{{\mcal{B}}_0} \bigg(-\big[ \Div(\rho_0 {{\widetilde{\Omega}}}_{,{{\bsym{F}}}}+{{\widetilde{{\bsym{P}}}}}_m - {{\check{{\bsym{P}}}}}_m)+ \widetilde{{\mbf{f}}}^e+\rho_0 {{\bsym{F}}}^{-\top} (\Grad^{\top} {{\bbbm{h}}}^e) {{\bbbm{K}}} \big] \cdot {\sdel}{\mathpalette\irchi\relax} \nonumber\\
&+\rho_0 \big[ {{\widetilde{\Omega}}}_{, {{\bbbm{K}}}} - {{\bbbm{h}}}^e- {\bbbm{h}}^s \big] \cdot {\sdel}{{\bbbm{K}}} \bigg) {dv}_0 \nonumber \\
&+\int\nolimits_{{\partial\mcal{B}} _0} \big((\rho_0 {{\widetilde{\Omega}}}_{,{{\bsym{F}}}}+{{\widetilde{{\bsym{P}}}}}_m - {{\check{{\bsym{P}}}}}_m)_- \mbf{n}_0
-({{\widetilde{{\bsym{P}}}}}_m - {{\check{{\bsym{P}}}}}_m)_+ \mbf{n}_0
-\widetilde{\mbf{t}}^e \big) \cdot {\sdel}{\mathpalette\irchi\relax} ds_0 \nonumber\\
&-\int\nolimits_{{\mcal{B}}'_0} \Div({{\widetilde{{\bsym{P}}}}}_m - {{\check{{\bsym{P}}}}}_m) \cdot {\sdel}{\mathpalette\irchi\relax} {dv}_0.
\end{align}

{Since the increments ${\sdel}{\mathpalette\irchi\relax}$ and ${\sdel}{{\bbbm{K}}}$ are arbitrary, we arrive at the following Euler--Lagrange equations for this variational problem}
\begin{subequations}
\begin{align}
 \Div{\bsym{P}}+ \widetilde{{\mbf{f}}}^e+\rho_0 {{\bsym{F}}}^{-\top} (\Grad^{\top} {{\bbbm{h}}}^e) {{\bbbm{K}}} =\mbf{0} \quad & \text{ in } \quad {\mcal{B}}_0, \\
 \jump{{{\bsym{P}}}} \mbf{n}_0 + \widetilde{\mbf{t}}^e=\mbf{0} \quad &\text{ on } \quad {\partial\mcal{B}} _0, \\
 \Div{\bsym{P}} =\mbf{0} \quad & \text{ in } \quad {\mcal{B}}_0' \\
 {\bbbm{h}} ={{\widetilde{\Omega}}}_{, {{\bbbm{K}}}} \quad & \text{ in } \quad {\mcal{B}}_0,
\end{align}
\label{eqn: sec6 EL equations 1}
\end{subequations}
where we have recognised the total first Piola--Kirchhoff stress tensor in the body and in vacuum as
\begin{subequations}
\begin{align}
 {\bsym{P}} =\rho_0 {{\widetilde{\Omega}}}_{,{{\bsym{F}}}}+{{\widetilde{{\bsym{P}}}}}_m - {{\check{{\bsym{P}}}}}_m \quad & \text{ in } \quad {\mcal{B}}_0,\\
 {\bsym{P}} ={{\widetilde{{\bsym{P}}}}}_m - {{\check{{\bsym{P}}}}}_m \quad & \text{ in } \quad {\mcal{B}}_0' .
\end{align}
\label{eqn: Sec6 stress expression}
\end{subequations}

\begin{sidenote}
From \eqref{eqn: definintion of first Piola in vacuum} and \eqref{eqn: sec6 strPKent expression} (recall Remark \ref{cauchystr}), we can write the total Cauchy stress on ${\mcal{B}}$ as
\begin{equation}
{\bsym{\sigma}}=J^{-1}{\bsym{P}} {{\bsym{F}}}^{\top}
=\rho {{\widetilde{\Omega}}}_{,{{\bsym{F}}}}{{\bsym{F}}}^{\top}+{\bbbm{h}}^s \otimes {{\bbbm{b}}}^s -\frac{1}{2}{{\mu_0}}({\bbbm{h}}^s \cdot {\bbbm{h}}^s) \bsym{I} .
\end{equation}
\end{sidenote}

\section{Correspondence between variational principles}
\label{sec: comparison}
Thus far, we have presented three different magnetisation based formulations where the difference between these variational principles occurs due the choice of particular magnetisation field. In addition to these, we have also presented two other formulations in Appendix \ref{sec: B formulation} and \ref{sec: H formulation} where in place of the magnetisation field the stored energy density depends on ${{\bbbm{B}}}$ and ${{\bbbm{H}}}$, respectively. 
Since the mechanical work terms involving the body force and the surface traction are same in all these formulations (in the referential description), i.e.,
\[
\mcal{W}_M \defnt- \int\nolimits_{{\mcal{B}}_0} \widetilde{{\mbf{f}}}^e \cdot {\mathpalette\irchi\relax} {dv}_0 - \int\nolimits_{{\partial\mcal{B}} _0} \widetilde{\mbf{t}}^e \cdot {\mathpalette\irchi\relax} ds_0,
\]
(which also equals its spatial description $- \int\nolimits_{{\mcal{B}}} \rho{{\mbf{f}}}^e \cdot {\mathpalette\irchi\relax} {dv} - \int\nolimits_{{\partial\mcal{B}} } {\mbf{t}}^e \cdot {\mathpalette\irchi\relax} ds$) so we sometimes compare only the remaining terms.
Using the constitutive relation \eqref{eqn: constitutive B H M} and the fact that ${{\bbbm{M}}}$ vanishes outside the body ${\mcal{B}}_0$, we get
\begin{align}
&\frac{1}{2}{{\mu_0}} \int\nolimits_{\vol_0} J \lVert {{\bsym{F}}}^{-\top} {{\bbbm{H}}} \rVert^2 {dv}_0-\int\nolimits_{\vol_0} {{\bbbm{H}}} \cdot {{\bbbm{B}}} {dv}_0\nonumber \\
&=-\frac{1}{2}{{\mu_0}} \int\nolimits_{\vol_0} J \lVert {{\bsym{F}}}^{-\top} {{\bbbm{H}}} \rVert^2 {dv}_0-\int\nolimits_{{\mcal{B}}_0}J{{\bsym{C}}}^{-1}{{\bbbm{M}}}\cdot {{\bbbm{H}}} {dv}_0.
\end{align}
In a similar manner, we find that
\begin{align}
\frac{1}{2}{{\mu_0}} \int\nolimits_{{\mcal{B}}'_0} J \lVert {{\bsym{F}}}^{-\top} {{\bbbm{H}}} \rVert^2 {dv}_0-\int\nolimits_{{\mcal{B}}'_0} {{\bbbm{H}}} \cdot {{\bbbm{B}}} {dv}_0
&=-\frac{1}{2}{{\mu_0}} \int\nolimits_{{\mcal{B}}'_0} J \lVert {{\bsym{F}}}^{-\top} {{\bbbm{H}}} \rVert^2 {dv}_0.
\end{align}
At this point, it is useful to recall Remark \ref{decores}.
Using \eqref{eqn: maxwell lagrangian}${}_1$ and \eqref{eqn: potentials introduction}${}_2$,
\begin{align}
-\int\nolimits_{\bvol_0} {\Phi} {{\bbbm{B}}}^e \cdot \mbf{n}_0 ds_0&=-\int\nolimits_{\vol_0}\Div({\Phi} {{\bbbm{B}}}^e) {dv}_0
=-\int\nolimits_{\vol_0}[{\Phi}\Div{{\bbbm{B}}}^e+\Grad {\Phi}\cdot {{\bbbm{B}}}^e]{dv}_0\nonumber \\
&
=\int\nolimits_{\vol_0} {{\bbbm{H}}}\cdot {{\bbbm{B}}}^e{dv}_0.
\end{align}
In general, we have
\begin{align}
\int\nolimits_{\vol_0} {{\bbbm{H}}}\cdot {{\bbbm{B}}} {dv}_0 &=
-\int\nolimits_{\bvol_0} \mbf{n}_0 \cdot {\Phi} {{\bbbm{B}}} ds_0,\nonumber\\
\int\nolimits_{\vol_0} {{\bbbm{H}}}\cdot {{\bbbm{B}}}^e{dv}_0&=-\int\nolimits_{\bvol_0} \mbf{n}_0 \cdot {\Phi} {{\bbbm{B}}}^e ds_0,\nonumber\\
\int\nolimits_{\vol_0} {{\bbbm{H}}}^e \cdot {{\bbbm{B}}} {dv}_0&=-\int\nolimits_{\bvol_0} \mbf{n}_0 \cdot {\Phi}^e {{\bbbm{B}}} ds_0.
\label{somerelations}
\end{align}
Also, these relations can be re-written further, for example, $\int\nolimits_{\vol_0} {{\bbbm{H}}}\cdot {{\bbbm{B}}}^e{dv}_0=-\int\nolimits_{\bvol_0} \mbf{n}_0 \cdot {\Phi} {{\bbbm{B}}}^e ds_0=-{{\mu_0}}\int\nolimits_{\bvol_0} \mbf{n}_0 \cdot {\Phi} J {{\bsym{C}}}^{-1}{{\bbbm{H}}}^e ds_0.$

\subsection{Potential energy functionals based on ${\bbbm{M}}$, ${{\bbbm{B}}}$, and ${{\bbbm{H}}}$}
The variational formulations based on ${{\bbbm{B}}}$ and ${{\bbbm{H}}}$ can be related by applying a Legendre-type transform on the energy functions ${{\mathring{\Omega}}}$ and ${\check{\Omega}}$ as ${{\mathring{\Omega}}}({{\bsym{F}}}, {{\bbbm{B}}}) ={\check{\Omega}}({{\bsym{F}}}, {{\bbbm{H}}}) + {{\bbbm{B}}} \cdot {{\bbbm{H}}}$ \citep{dorfmann2004}.
Moreover, we note that the three variational formulations based on ${\bbbm{M}}$, ${{\bbbm{B}}}$, and ${{\bbbm{H}}}$ can be mutually related by a set of Legendre-type transform on the stored energy density functions ${{\Omega}}, {{\mathring{\Omega}}},$ and ${{\check{\Omega}}}$, respectively, so that
\begin{align}
{{\Omega}}({{\bsym{F}}}, {{\bbbm{M}}}) &={{\mathring{\Omega}}}({{\bsym{F}}}, {{\bbbm{B}}}) -\frac{1}{2}{{\mu_0}}J{{\bsym{C}}}^{-1}{{\bbbm{H}}}\cdot{{\bbbm{H}}}\nonumber\\
&={{\mathring{\Omega}}}({{\bsym{F}}}, {{\bbbm{B}}})+\frac{1}{{\mu_0}}{{\bbbm{M}}}\cdot{{\bbbm{B}}} -\frac{1}{2{{\mu_0}}}J{{\bsym{C}}}^{-1}{{\bbbm{M}}}\cdot{{\bbbm{M}}}-\frac{1}{2{{\mu_0}}}J^{-1}{{\bsym{C}}}{{\bbbm{B}}}\cdot{{\bbbm{B}}},\\
{{\Omega}}({{\bsym{F}}}, {{\bbbm{M}}}) &={{\check{\Omega}}}({{\bsym{F}}}, {{\bbbm{H}}}) + {{\bbbm{B}}} \cdot {{\bbbm{H}}}-\frac{1}{2}{{\mu_0}}J{{\bsym{C}}}^{-1}{{\bbbm{H}}}\cdot{{\bbbm{H}}}\nonumber\\
&={{\check{\Omega}}}({{\bsym{F}}}, {{\bbbm{H}}}) + J{{\bsym{C}}}^{-1}{{\bbbm{H}}}\cdot{{\bbbm{M}}}+\frac{1}{2}{{\mu_0}}J{{\bsym{C}}}^{-1}{{\bbbm{H}}}\cdot{{\bbbm{H}}},\\
{{\mathring{\Omega}}}({{\bsym{F}}}, {{\bbbm{B}}}) &={{\check{\Omega}}}({{\bsym{F}}}, {{\bbbm{H}}}) + {{\bbbm{B}}} \cdot {{\bbbm{H}}}.
\end{align}
By a direct calculation, it can be verified that the above relations result in the magnetic constitutive relations \eqref{eqn: H =omegaB}, \eqref{eq: PK H formulation}${}_2$, and \eqref{eqn: constitutive magnetisation magnetic field};
in particular,
\[
{{\mathring{\Omega}}}_{, {{\bbbm{B}}}}={{\bbbm{H}}}, \quad
{{\check{\Omega}}}_{, {{\bbbm{H}}}}=-{{\bbbm{B}}}, \quad
{{\Omega}}_{, {{\bbbm{M}}}}=J{{\bsym{C}}}^{-1}{{\bbbm{H}}} \quad \text{ in } {\mcal{B}}_0.
\]
As such, these relations lead to different convexity properties for ${{\mathring{\Omega}}}({{\bsym{F}}}, {{\bbbm{B}}})$, ${\check{\Omega}}({{\bsym{F}}}, {{\bbbm{H}}})$, and ${{\Omega}}({{\bsym{F}}}, {{\bbbm{M}}})$ in general.

As a consequence of above, it is natural to establish the relationship between the three variational formulations based on ${{\bbbm{B}}}$, ${{\bbbm{H}}}$ and ${\bbbm{M}}$. Recall that the total potential energy \eqref{eqn: potential energy functional for M formulation 1} is a functional of the deformation ${\mathpalette\irchi\relax}$ \eqref{eqn: deform} (and \eqref{eqn: deform2}) and the referential magnetisation ${{\bbbm{M}}}$ \eqref{LagrangianM}.
Indeed, the variational formulation \eqref{eqn: potential energy functional for M formulation 1} can be expressed as
\begin{align}
{{\mathit{E}_{\mathtt{I}}}}[{\mathpalette\irchi\relax}, {{\bbbm{M}}}]+\mcal{W}_M 
&=\int\nolimits_{{\mcal{B}}_0} {{\Omega}} ({{\bsym{F}}}, {{\bbbm{M}}}) {dv}_0+\frac{1}{2}{{\mu_0}} \int\nolimits_{\vol_0} J \lVert {{\bsym{F}}}^{-\top} \Grad {\Phi} \rVert^2 {dv}_0\nonumber\\
&+\int\nolimits_{\bvol_0} {\Phi}^e \mbf{n}_0 \cdot {{\bbbm{B}}} ds_0\nonumber\\
&=\int\nolimits_{{\mcal{B}}_0} {{\Omega}} ({{\bsym{F}}}, {{\bbbm{M}}}) {dv}_0+\frac{1}{2}{{\mu_0}} \int\nolimits_{\vol_0} J \lVert {{\bsym{F}}}^{-\top} {{\bbbm{H}}} \rVert^2 {dv}_0
-\int\nolimits_{\vol_0} {{\bbbm{H}}}^e \cdot {{\bbbm{B}}} {dv}_0\nonumber\\
&=\left[\int\nolimits_{{\mcal{B}}_0} {{\Omega}} ({{\bsym{F}}}, {{\bbbm{M}}}) {dv}_0+\frac{1}{2}{{\mu_0}} \int\nolimits_{{\mcal{B}}_0} J \lVert {{\bsym{F}}}^{-\top} {{\bbbm{H}}} \rVert^2 {dv}_0\right]\nonumber\\
&
+\frac{1}{2}{{\mu_0}} \int\nolimits_{{\mcal{B}}'_0} J \lVert {{\bsym{F}}}^{-\top} {{\bbbm{H}}} \rVert^2 {dv}_0-\int\nolimits_{\vol_0} {{\bbbm{H}}}^e \cdot {{\bbbm{B}}} {dv}_0,
\label{neweqn: E 3}
\end{align}
which can be written as
\begin{align}
{{\mathit{E}_{\mathtt{I}}}}[{\mathpalette\irchi\relax}, {{\bbbm{M}}}]+\mcal{W}_M 
&={{\mathit{E}_{\mathtt{I\hspace{-1pt}V}}}}[{\mathpalette\irchi\relax}, {{\bbbm{A}}}]+\mcal{W}_M.
\end{align}
Above is the exact relationship between the variational principles analyzed in Section \ref{sec: M formulation 1} and Appendix \ref{sec: B formulation}.
Recall
that the total potential energy \eqref{eqn: potential energy functional for B formulation} is a functional of the deformation ${\mathpalette\irchi\relax}$ \eqref{eqn: deform} (and \eqref{eqn: deform2}) and the referential counterpart ${\bbbm{B}}$ of ${\bbbm{b}}$ (via the referential magnetic vector potential ${\bbbm{A}}$ \eqref{eqn: potentials introduction}${}_1$).
Also,
\begin{align}
\frac{1}{2}{{\mu_0}} \int\nolimits_{\vol_0} J \lVert {{\bsym{F}}}^{-\top} {{\bbbm{H}}} \rVert^2 {dv}_0 &-\int\nolimits_{\vol_0} {{\bbbm{H}}} \cdot {{\bbbm{B}}} {dv}_0 \nonumber \\
&=-\frac{1}{2}{{\mu_0}} \int\nolimits_{\vol_0} J \lVert {{\bsym{F}}}^{-\top} {{\bbbm{H}}} \rVert^2 {dv}_0-\int\nolimits_{{\mcal{B}}_0}J{{\bsym{C}}}^{-1}{{\bbbm{M}}}\cdot {{\bbbm{H}}} {dv}_0,
\end{align}
so that
\begin{align}
{{\mathit{E}_{\mathtt{I}}}}[{\mathpalette\irchi\relax}, {{\bbbm{M}}}]+\mcal{W}_M
&=\int\nolimits_{{\mcal{B}}_0} {{\Omega}} ({{\bsym{F}}}, {{\bbbm{M}}}) {dv}_0-\frac{1}{2}{{\mu_0}} \int\nolimits_{\vol_0}J \lVert {{\bsym{F}}}^{-\top} {{\bbbm{H}}} \rVert^2 {dv}_0+\int\nolimits_{\vol_0} ({{\bbbm{H}}}-{{\bbbm{H}}}^e) \cdot {{\bbbm{B}}} {dv}_0\nonumber\\
&-\int\nolimits_{{\mcal{B}}_0}J{{\bsym{C}}}^{-1}{{\bbbm{M}}}\cdot {{\bbbm{H}}} {dv}_0\nonumber\\
&=\int\nolimits_{{\mcal{B}}_0} ({{\Omega}} ({{\bsym{F}}}, {{\bbbm{M}}})-J{{\bsym{C}}}^{-1}{{\bbbm{H}}}\cdot {{\bbbm{M}}}-\frac{1}{2}{{\mu_0}} J \lVert {{\bsym{F}}}^{-\top} {{\bbbm{H}}} \rVert^2) {dv}_0\nonumber\\
&
-\frac{1}{2}{{\mu_0}} \int\nolimits_{{\mcal{B}}'_0} J \lVert {{\bsym{F}}}^{-\top} {{\bbbm{H}}} \rVert^2 {dv}_0+\int\nolimits_{\vol_0} ({{\bbbm{H}}}-{{\bbbm{H}}}^e) \cdot {{\bbbm{B}}} {dv}_0.
\end{align}
Hence,
\eqref{neweqn: E 3} can be written as
\begin{align}
{{\mathit{E}_{\mathtt{I}}}}[{\mathpalette\irchi\relax}, {{\bbbm{M}}}]+\mcal{W}_M 
&={\mathit{E}_{\mathtt{V}}}[{\mathpalette\irchi\relax}, {{\Phi}}]+\mcal{W}_M
-\int\nolimits_{\vol_0} {{\bbbm{H}}}\cdot{{\bbbm{B}}}^e {dv}_0+\int\nolimits_{\vol_0} ({{\bbbm{H}}}-{{\bbbm{H}}}^e) \cdot {{\bbbm{B}}} {dv}_0,
\end{align}
which is the relationship between the variational principles analyzed in Section \ref{sec: M formulation 1} and Appendix \ref{sec: H formulation}.
Here we recall that the total potential energy \eqref{eqn: potential energy functional for H formulation} is a functional of the deformation ${\mathpalette\irchi\relax}$ \eqref{eqn: deform} (and \eqref{eqn: deform2}) and {the referential magnetic field vector ${{\bbbm{H}}}$} (via the referential magnetic scalar potential ${\Phi}$ \eqref{eqn: potentials introduction}${}_2$).

\begin{sidenote}
{Upon using equations} \eqref{eqn: curl identity}, \eqref{eqn: maxwell lagrangian}${}_2$ and \eqref{eqn: potentials introduction}${}_1$, {we can write}
\begin{align}
\int\nolimits_{\bvol_0} \left[{{\bbbm{H}}}^e\wedge {\bbbm{A}} \right] \cdot \mbf{n}_0 ds_0
&=\int\nolimits_{\vol_0} \Div\left[{{\bbbm{H}}}^e\wedge {\bbbm{A}} \right]{dv}_0
\nonumber\\
&=\int\nolimits_{\vol_0} \left[\Curl {{\bbbm{H}}}^e \cdot {\bbbm{A}}-[\Curl {\bbbm{A}}] \cdot {{\bbbm{H}}}^e \right]{dv}_0
=-\int\nolimits_{\vol_0} {{\bbbm{B}}} \cdot {{\bbbm{H}}}^e {dv}_0.
\end{align}
{Hence,} the total potential energy functional \eqref{eqn: potential energy functional for B formulation} can be re-written as
\begin{align}
{{\mathit{E}_{\mathtt{I\hspace{-1pt}V}}}}[{\mathpalette\irchi\relax}, {{\bbbm{A}}}]+\mcal{W}_M
&=\int\nolimits_{{\mcal{B}}_0} {{\mathring{\Omega}}} ({{\bsym{F}}}, {{\bbbm{B}}}) {dv}_0+\frac{1}{2 {{\mu_0}}} \int\nolimits_{{\mcal{B}}'_0} J^{-1} \lVert{{\bsym{F}}} {{\bbbm{B}}}\rVert^2{dv}_0-\int\nolimits_{\vol_0} {{\bbbm{H}}}^e\cdot{{\bbbm{B}}} {dv}_0\nonumber\\
&=\int\nolimits_{{\mcal{B}}_0} {{\mathring{\Omega}}} ({{\bsym{F}}}, {{\bbbm{B}}}) {dv}_0+\frac{1}{2 {{\mu_0}}} \int\nolimits_{{\mcal{B}}'_0} J^{-1} \lVert{{\bsym{F}}} {{\bbbm{B}}}\rVert^2{dv}_0+\int\nolimits_{\bvol_0} \mbf{n}_0 \cdot {\Phi}^e {{\bbbm{B}}} ds_0.
\label{neweqn: E 1}
\end{align}
In the variational formulation for the total potential energy functional \eqref{eqn: potential energy functional for H formulation} we have
\begin{align}
{\mathit{E}_{\mathtt{V}}}[{\mathpalette\irchi\relax}, {{\Phi}}]+\mcal{W}_M
&=\int\nolimits_{{\mcal{B}}_0} {{\check{\Omega}}} ({{\bsym{F}}}, {{\bbbm{H}}}) {dv}_0-\frac{1}{2}{{\mu_0}} \int\nolimits_{{\mcal{B}}'_0} J \lVert{{\bsym{F}}}^{-\top} {{\bbbm{H}}}\rVert^2 {dv}_0+\int\nolimits_{\vol_0} {{\bbbm{H}}}\cdot {{\bbbm{B}}}^e{dv}_0\nonumber\\
&=\int\nolimits_{{\mcal{B}}_0} {{\check{\Omega}}} ({{\bsym{F}}}, {{\bbbm{H}}}) {dv}_0-\frac{1}{2}{{\mu_0}} \int\nolimits_{{\mcal{B}}'_0} J \lVert{{\bsym{F}}}^{-\top} {{\bbbm{H}}}\rVert^2 {dv}_0
-\int\nolimits_{\bvol_0} \mbf{n}_0 \cdot {\Phi} {{\bbbm{B}}}^e ds_0.
\label{neweqn: E 2}
\end{align}
Hence,
\begin{align}
{\mathit{E}_{\mathtt{V}}}[{\mathpalette\irchi\relax}, {{\Phi}}]+\mcal{W}_M
&=\int\nolimits_{{\mcal{B}}_0} {\check{\Omega}} ({{\bsym{F}}}, {{\bbbm{H}}}) {dv}_0+\frac{1}{2}{{\mu_0}} \int\nolimits_{{\mcal{B}}'_0} J \lVert{{\bsym{F}}}^{-\top} {{\bbbm{H}}}\rVert^2 {dv}_0\nonumber\\
&-\int\nolimits_{{\mcal{B}}'_0} {{\bbbm{H}}} \cdot {{\bbbm{B}}} {dv}_0+\int\nolimits_{\vol_0} {{\bbbm{H}}}\cdot {{\bbbm{B}}}^e{dv}_0\nonumber\\
&=\int\nolimits_{{\mcal{B}}_0} ({\check{\Omega}} ({{\bsym{F}}}, {{\bbbm{H}}}) + {{\bbbm{B}}} \cdot {{\bbbm{H}}}){dv}_0+\frac{1}{2 {{\mu_0}}} \int\nolimits_{{\mcal{B}}'_0} J^{-1} \lVert{{\bsym{F}}} {{\bbbm{B}}}\rVert^2{dv}_0\nonumber\\
&-\int\nolimits_{\vol_0} {{\bbbm{H}}}\cdot ({{\bbbm{B}}}-{{\bbbm{B}}}^e){dv}_0,
\end{align}
which can be written as
\begin{align}
{\mathit{E}_{\mathtt{V}}}[{\mathpalette\irchi\relax}, {{\Phi}}]+\mcal{W}_M 
&={{\mathit{E}_{\mathtt{I\hspace{-1pt}V}}}}[{\mathpalette\irchi\relax}, {{\bbbm{A}}}]+\mcal{W}_M 
+\int\nolimits_{\vol_0} {{\bbbm{H}}}^e\cdot{{\bbbm{B}}} {dv}_0-\int\nolimits_{\vol_0} {{\bbbm{H}}}\cdot ({{\bbbm{B}}}-{{\bbbm{B}}}^e){dv}_0\nonumber\\
&={{\mathit{E}_{\mathtt{I\hspace{-1pt}V}}}}[{\mathpalette\irchi\relax}, {{\bbbm{A}}}]+\mcal{W}_M 
+\int\nolimits_{\vol_0} {{\bbbm{H}}}\cdot {{\bbbm{B}}}^e {dv}_0-\int\nolimits_{\vol_0} ({{\bbbm{H}}}-{{\bbbm{H}}}^e)\cdot{{\bbbm{B}}} {dv}_0.
\end{align}
Also
\begin{align}
\int\nolimits_{\vol_0} {{\bbbm{H}}}\cdot {{\bbbm{B}}}^e{dv}_0&={{\mu_0}}\int\nolimits_{\vol_0} {{\bbbm{H}}}\cdot J {{\bsym{C}}}^{-1}{{\bbbm{H}}}^e{dv}_0
=\int\nolimits_{\vol_0} (J^{-1} {{\bsym{C}}}{{\bbbm{B}}}-{{\bbbm{M}}}\bbm{1}_{{\mcal{B}}_0})\cdot J {{\bsym{C}}}^{-1}{{\bbbm{H}}}^e{dv}_0\nonumber\\
&=\int\nolimits_{\vol_0} ({{\bbbm{B}}}-J {{\bsym{C}}}^{-1}{{\bbbm{M}}}\bbm{1}_{{\mcal{B}}_0})\cdot {{\bbbm{H}}}^e{dv}_0\nonumber\\
&=\int\nolimits_{\vol_0} {{\bbbm{H}}}^e\cdot {{\bbbm{B}}} {dv}_0
-\int\nolimits_{{\mcal{B}}_0} J {{\bsym{F}}}^{-\top}{{\bbbm{M}}}\cdot{{\bsym{F}}}^{-\top} {{\bbbm{H}}}^e{dv}_0.
\end{align}
Hence,
\begin{align}
{\mathit{E}_{\mathtt{V}}}[{\mathpalette\irchi\relax}, {{\Phi}}]+\mcal{W}_M
&=\int\nolimits_{{\mcal{B}}_0} ({\check{\Omega}} ({{\bsym{F}}}, {{\bbbm{H}}}) + {{\bbbm{B}}} \cdot {{\bbbm{H}}}){dv}_0+\frac{1}{2 {{\mu_0}}} \int\nolimits_{{\mcal{B}}'_0} J^{-1} \lVert{{\bsym{F}}} {{\bbbm{B}}}\rVert^2{dv}_0\nonumber\\
&-\int\nolimits_{\vol_0} ({{\bbbm{H}}}-{{\bbbm{H}}}^e)\cdot {{\bbbm{B}}} {dv}_0-\int\nolimits_{{\mcal{B}}_0} {{\bsym{F}}}^{-\top} {{\bbbm{H}}}^e \cdot J {{\bsym{F}}}^{-\top}{{\bbbm{M}}} {dv}_0.
\end{align}
\end{sidenote}

\subsection{Potential energy functionals based on ${\bbbm{M}}, {\overline{\bbbm{M}}},$ and ${\bbbm{K}}$}
Following the arguments in Section \ref{sec: M formulation 2}, we assumed that
\begin{align}
\vol=\vol_0={\bbm{R}^3},
\end{align}
for the formulation presented in Section \ref{sec: M formulation 1}. This needed some changes in the expression \eqref{eqn: potential energy functional for M formulation 1}. Clearly, the only term that needs to be re-written is the last term in \eqref{eqn: potential energy functional for M formulation 1}. Using the nature of magnetic field in vacuum we have by \eqref{somerelations}${}_3$,
$\int\nolimits_{\bvol_0} {\phi}^e \mbf{n}_0 \cdot {{\bbbm{B}}} {ds}_0=-\int\nolimits_{\vol_0} {{\bbbm{H}}}^e \cdot {{\bbbm{B}}} {dv}_0.$
Hence, based on the definitions \eqref{eqn: potential energy functional for M formulation 1} and \eqref{eqn: potential in reference configuration E2}, we get
\begin{align}
 {\mathit{E}_{\mathtt{I}}} - {\mathit{E}_{\mathtt{I\hspace{-2pt}I}}} &= {\mathit{E}_{\mathtt{I}}}[{\mathpalette\irchi\relax}, {{\bbbm{M}}}] - {\mathit{E}_{\mathtt{I\hspace{-2pt}I}}}[{\mathpalette\irchi\relax}, {{\overline{\bbbm{M}}}}] \nonumber\\
 &= \int\nolimits_{{\mcal{B}}_0} \big[ {{\Omega}}({{\bsym{F}}}, {{\bbbm{M}}})-\rho_0 {{\widehat{\Omega}}}({{\bsym{F}}}, {{\overline{\bbbm{M}}}}) + \rho_0 J{{\bbbm{h}}}^e({\mathpalette\irchi\relax}(\mbf{X})) \cdot {{\bsym{F}}}^{-\top}{{\overline{\bbbm{M}}}} \big]{dv}_0 + \nonumber \\
 &+ \frac{{\mu_0}}{2} \int\nolimits_{\vol_0 } J {{\bsym{C}}}^{-1}{{\bbbm{H}}} \cdot {{\bbbm{H}}} {dv}_0 -\int\nolimits_{\vol_0} {{\bbbm{H}}}^e \cdot {{\bbbm{B}}} {dv}_0 \nonumber \\
& - \frac{1}{2}{{\mu_0}} \int\nolimits_{ {\mcal{B}}_0}J{{\bsym{C}}}^{-1}{{\bbbm{H}}}^s \cdot {{\bbbm{H}}}^s {dv}_0 - \frac{1}{2}{{\mu_0}} \int\nolimits_{{\mcal{B}}'_0} J{{\bsym{C}}}^{-1}{{\bbbm{H}}}^s \cdot {{\bbbm{H}}}^s {dv}_0.
\label{diff e1 e2}
 \end{align}
The first term in the second line can be rewritten in view of \eqref{eqn: h and b field decomposition} as
\begin{equation}
\frac{{\mu_0}}{2} \int\nolimits_{\vol_0 } J {{\bsym{C}}}^{-1}{{\bbbm{H}}} \cdot {{\bbbm{H}}} {dv}_0 = \frac{{\mu_0}}{2}\int\nolimits_{\vol_0 } J {{\bsym{C}}}^{-1}\bigg[ {{\bbbm{H}}}^e \cdot {{\bbbm{H}}}^e + 2 {{\bbbm{H}}}^e \cdot {{\bbbm{H}}}^s + {{\bbbm{H}}}^s \cdot {{\bbbm{H}}}^s \bigg] {dv}_0.
\end{equation}
Upon substituting the above back to \eqref{diff e1 e2}, we get
\begin{align}
 {\mathit{E}_{\mathtt{I}}} - {\mathit{E}_{\mathtt{I\hspace{-2pt}I}}} &= \int\nolimits_{{\mcal{B}}_0} \big[ {{\Omega}}({{\bsym{F}}}, {{\bbbm{M}}})- \rho_0{{\widehat{\Omega}}}({{\bsym{F}}}, {{\overline{\bbbm{M}}}}) +\rho_0 J{{\bbbm{h}}}^e({\mathpalette\irchi\relax}(\mbf{X})) \cdot {{\bsym{F}}}^{-\top}{{\overline{\bbbm{M}}}} \big]{dv}_0 \nonumber \\
 &+ {{\mu_0}} \int\nolimits_{\vol_0 } J {{\bsym{C}}}^{-1}{{\bbbm{H}}}^e \cdot {{\bbbm{H}}}^s {dv}_0 +\frac{{\mu_0}}{2}\int\nolimits_{\vol_0 } J {{\bsym{C}}}^{-1} {{\bbbm{H}}}^e \cdot {{\bbbm{H}}}^e dv_0- \int\nolimits_{\vol_0} {\bbbm{H}}^e \cdot {\bbbm{B}}^s {dv}_0\\
 &-{{\mu_0}} \int\nolimits_{\vol_0} J {{\bsym{C}}}^{-1}{\bbbm{H}}^e \cdot {\bbbm{H}}^e {dv}_0\nonumber \\
 &= \int\nolimits_{{\mcal{B}}_0} \big[ {{\Omega}}({{\bsym{F}}}, {{\bbbm{M}}})-\rho_0 {{\widehat{\Omega}}}({{\bsym{F}}}, {{\overline{\bbbm{M}}}}) + \rho_0J{{\bbbm{h}}}^e({\mathpalette\irchi\relax}(\mbf{X})) \cdot {{\bsym{F}}}^{-\top}{{\overline{\bbbm{M}}}} \big]{dv}_0 \nonumber \\
 & - \int\nolimits_{{\mcal{B}}_0} J {\bsym{C}}^{-1} {\bbbm{H}}^e \cdot {\bbbm{M}} {dv}_0-\frac{{\mu_0}}{2} \int\nolimits_{\vol_0} J {{\bsym{C}}}^{-1}{\bbbm{H}}^e \cdot {\bbbm{H}}^e {dv}_0.
 \end{align}
Since, ${\bbbm{M}} = \rho_0 {\overline{\bbbm{M}}}$ and ${\bbbm{H}}^e = {\bsym{F}}^\top {\bbbm{h}}^e$, the above can be rewritten as
\begin{align}
 {\mathit{E}_{\mathtt{I}}} - {\mathit{E}_{\mathtt{I\hspace{-2pt}I}}} &= \int\nolimits_{{\mcal{B}}_0} \big[ {{\Omega}}({{\bsym{F}}}, {{\bbbm{M}}})-\rho_0 {{\widehat{\Omega}}}({{\bsym{F}}}, {{\overline{\bbbm{M}}}}) \big]{dv}_0 -\frac{{\mu_0}}{2} \int\nolimits_{\vol_0} J {{\bsym{C}}}^{-1}{\bbbm{H}}^e \cdot {\bbbm{H}}^e {dv}_0.
\end{align}
Thus, the two potential energies differ not only by the definition of the respective stored energy density functions but also an extra term; the latter term, clearly, is a constant term though it could be infinite for ${\bbbm{H}}^e\ne0$ while the former can be made zero by naturally identifying the stored energy density functions.

From equations \eqref{eqn: potential in reference configuration E2} and \eqref{PifunctionalKR} (using \eqref{defKvec}), we get
\begin{align}
 {\mathit{E}_{\mathtt{I\hspace{-2pt}I}}}[{\mathpalette\irchi\relax}, {{\overline{\bbbm{M}}}}] - {\mathit{E}_{\mathtt{I\hspace{-2pt}I\hspace{-2pt}I}}}[{\mathpalette\irchi\relax}, {{\bbbm{K}}}] = \int\nolimits_{{\mcal{B}}_0} \rho_0 \big[ {{\widehat{\Omega}}}({{\bsym{F}}}, {{\overline{\bbbm{M}}}})- {{\widetilde{\Omega}}}({{\bsym{F}}}, {{\bbbm{K}}}) \big]{dv}_0 .
\end{align}
These two potential energies differ only by the definition of the respective stored energy density functions which can be naturally identified to achieve an equivalence.

\subsection{Comparison with the expressions provided by \cite{kankanala2004} using a modified potential energy functional}
Since ${{\bbbm{h}}}^e$ is the gradient of a potential, and in view of \eqref{eqn: b h relation vacuum}, by a direct calculation we have $\int\nolimits_{\bbm{R}^3} {{\bbbm{h}}}^e\cdot {{\bbbm{b}}}^s {dv}=0$, as a result of which we get
\begin{align}
\int\nolimits_{{\mcal{B}}_0} \rho_0 {{\bbbm{h}}}^e\cdot {{\bbbm{K}}} {dv}_0&=\int\nolimits_{{\mcal{B}}} {{\bbbm{h}}}^e\cdot {{\bbbm{m}}} {dv}
=\int\nolimits_{\bbm{R}^3} {{\bbbm{h}}}^e\cdot ({{\bbbm{b}}}^s-{{\mu_0}}{{\bbbm{h}}}^s) {dv}=
-{{\mu_0}}\int\nolimits_{\bbm{R}^3} {{\bbbm{h}}}^e\cdot {{\bbbm{h}}}^s {dv}.
\end{align}
Note that ${{\mu_0}}\int\nolimits_{\bbm{R}^3} {{\bbbm{h}}}^e\cdot {{\bbbm{h}}}^s {dv}$ is non-zero, indeed, with ${{\bbbm{h}}}^s=-\grad{\phi}^s$, ${{\mu_0}}{{\bbbm{h}}}^e={{\bbbm{b}}}^e$ and $B_r\subset{\bbm{R}^3}$ as a ball of radius $r$, we find it to be equal to
\[
-\int\nolimits_{\bbm{R}^3} {{\bbbm{b}}}^e\cdot \grad{\phi}^s {dv}=\lim_{r\to\infty}\bigg[ \int\nolimits_{B_r} \div{{\bbbm{b}}}^e {\phi}^s {dv}-\int\nolimits_{\partial B_r}{\phi}^s{{\bbbm{b}}}^e\cdot\mbf{n}_0 ds \bigg],
\]
where $\div{{\bbbm{b}}}^e=0$ in the first term but ${\phi}^s$ may not necessarily go to zero in the second term as $r=\lVert\mbf{x}\rVert\to\infty$.
Thus, an equivalent potential energy functional is
\[
{{\mathit{E}_{\mathtt{I\hspace{-2pt}I\hspace{-2pt}I}}}}({\mathpalette\irchi\relax}, {{\bbbm{K}}})+\int\nolimits_{{\mcal{B}}_0} \rho_0 {{\bbbm{h}}}^e\cdot {{\bbbm{K}}} {dv}_0+{{\mu_0}}\int\nolimits_{\bbm{R}^3} {{\bbbm{h}}}^e\cdot {{\bbbm{h}}}^s {dv},
\]
in addition to which by including the constant term $ \frac{1}{2}{{\mu_0}} \int\nolimits_{\bbm{R}^3}{{\bbbm{h}}}^e \cdot {{\bbbm{h}}}^e {dv}$ too, we get (recall \eqref{eqn: total energy 1})
\begin{align}
\overline{{\mathit{E}_{\mathtt{I\hspace{-2pt}I\hspace{-2pt}I}}}} ({\mathpalette\irchi\relax}, {{\overline{\bbbm{m}}}}) 
& =\int\nolimits_{{\mcal{B}}} \rho {{\widehat{\Omega}}}({{\bsym{F}}}, {{\overline{\bbbm{m}}}}) {dv}+\mcal{W}_M\nonumber \\
&+ \frac{1}{2}{{\mu_0}} \int\nolimits_{{\mcal{B}}}{\bbbm{h}} \cdot {\bbbm{h}} {dv}+\frac{1}{2}{{\mu_0}} \int\nolimits_{{\mcal{B}}'} {\bbbm{h}} \cdot {\bbbm{h}} {dv},
\end{align}
with its referential form (to be compared with \eqref{PifunctionalKR}) as
\begin{align}
\widehat{{\mathit{E}_{\mathtt{I\hspace{-2pt}I\hspace{-2pt}I}}}} ({\mathpalette\irchi\relax}, {{\bbbm{K}}}) 
& =\int\nolimits_{{\mcal{B}}_0} \rho_0 {{\widetilde{\Omega}}}({{\bsym{F}}}, {{\bbbm{K}}}){dv}_0 +\mcal{W}_M\nonumber \\
&+ \frac{1}{2}{{\mu_0}} \int\nolimits_{ {\mcal{B}}_0}J{{\bsym{C}}}^{-1}{{\bbbm{H}}} \cdot {{\bbbm{H}}} {dv}_0+\frac{1}{2}{{\mu_0}} \int\nolimits_{{\mcal{B}}'_0} J{{\bsym{C}}}^{-1}{{\bbbm{H}}} \cdot {{\bbbm{H}}} {dv}_0.
\end{align}
Above expression coincides with the potential energy functional of \eqref{eqn: potential energy functional for M formulation 1} except for the last term (which is absent in the present scenario as $\vol=\bbm{R}^3$) and more importantly a different measure of magnetisation; note that 
\[
{{\bbbm{K}}}(\mbf{X})={{\overline{\bbbm{m}}}}(\mbf{x})=\rho^{-1}(\mbf{x}){{\bsym{F}}}^{-\top}(\mbf{X}){{\bbbm{M}}}(\mbf{X})
\]
by \eqref{LagrangianM} and \eqref{eqn: defn mpm and mpv}.
Similar to \eqref{KTeq1term}, we have
\begin{align}
{\sdel}{{\bbbm{B}}} 
& =\Big[ [{{\bsym{F}}}^{-\top} \cdot {\sdel}{{\bsym{F}}} ] \bsym{I}-{{\bsym{C}}}^{-1} {\sdel}{{\bsym{F}}}^\top {{\bsym{F}}}\Big]{{\mu_0}}J{{\bsym{C}}}^{-1}{{\bbbm{H}}}
-{{\bsym{F}}}^{-1} {\sdel}{{\bsym{F}}} {{\bbbm{B}}}\nonumber\\
&- {{\mu_0}} J {{\bsym{C}}}^{-1} \Grad {\sdel}{\Phi}+\rho_0 {{\bsym{F}}}^{-1} {\sdel}{{\bbbm{K}}},
\label{KTeq1termnew}
\end{align}
and similar to \eqref{HBiden1}${}_2$, we have
(with $B_R\subset{\bbm{R}^3}$ as a ball of radius $R$)
\begin{align}
-\int\nolimits_{{\mcal{B}}_0'} {{\bbbm{H}}} \cdot {\sdel}{{\bbbm{B}}} {dv}_0&=
\int\nolimits_{{\mcal{B}}_0'} \Grad {\Phi}\cdot {\sdel}{{\bbbm{B}}} {dv}_0\nonumber \\
&=\lim_{R\to\infty}
(\int\nolimits_{\partial B_R} {\Phi}\mbf{n}_0 \cdot {\sdel}{{\bbbm{B}}} ds_0
-\int\nolimits_{B_R}{\Phi}\div {\sdel}{{\bbbm{B}}} {dv}_0)
-\int\nolimits_{{\partial\mcal{B}} _0} \mbf{n}_0 \cdot {\Phi} {\sdel}{{\bbbm{B}}} ds_0\nonumber\\
&=-\int\nolimits_{{\partial\mcal{B}} _0} \mbf{n}_0 \cdot {\Phi} {\sdel}{{\bbbm{B}}} ds_0,
\label{HBiden1new}
\end{align}
assuming that ${\sdel}{{\bbbm{B}}}\cdot\mbf{n}_0 $ vanishes as $R=\lVert\mbf{X}\rVert\to\infty$ in a suitable manner.
By carrying out the first variation analysis similar to that presented above in this section we get
\begin{align}
{\sdel}\widehat{{\mathit{E}_{\mathtt{I\hspace{-2pt}I\hspace{-2pt}I}}}}({\mathpalette\irchi\relax}, {{\bbbm{K}}}) 
& =\int\nolimits_{{\mcal{B}}_0} \rho_0 \big[{{\widetilde{\Omega}}}_{,{{\bsym{F}}}} \cdot {\sdel}{{\bsym{F}}}+{{\widetilde{\Omega}}}_{, {{\bbbm{K}}}} \cdot {\sdel}{{\bbbm{K}}}\big]{dv}_0 
- \int\nolimits_{{\mcal{B}}_0} \widetilde{{\mbf{f}}}^e \cdot {\sdel}{\mathpalette\irchi\relax} {dv}_0\nonumber \\
&-\int\nolimits_{{\partial\mcal{B}} _0} \widetilde{\mbf{t}}^e \cdot{\sdel}{\mathpalette\irchi\relax} ds_0+\int\nolimits_{{\mcal{B}}_0} (- {{\widehat{{\bsym{P}}}}}_m \cdot {\sdel}{{\bsym{F}}}){dv}_0+\int\nolimits_{{\mcal{B}}'_0} (- {{\widehat{{\bsym{P}}}}}_m \cdot {\sdel}{{\bsym{F}}}){dv}_0\nonumber\\
&+\int\nolimits_{{\mcal{B}}_0} {{\mathring{{\bsym{P}}}}}_m \cdot{\sdel}{{\bsym{F}}} {dv}_0 - \int\nolimits_{{\mcal{B}}_0}{{\bbbm{H}}}^s \cdot(\rho_0 {{\bsym{F}}}^{-1} {\sdel}{{\bbbm{K}}} ) {dv}_0-\int\nolimits_{{\partial\mcal{B}} _0} \mbf{n}_0 \cdot {\Phi}^s {\sdel}{{\bbbm{B}}}^s ds_0\nonumber\\
&+\int\nolimits_{{\mcal{B}}_0'} {{\mathring{{\bsym{P}}}}}_m \cdot{\sdel}{{\bsym{F}}} {dv}_0+\int\nolimits_{{\partial\mcal{B}} _0} \mbf{n}_0 \cdot {\Phi}^s {\sdel}{{\bbbm{B}}}^s ds_0,
\end{align}
where ${{\widehat{{\bsym{P}}}}}_m$ is defined by \eqref{eqn: max type tensor 1} and ${{\mathring{{\bsym{P}}}}}_m$ is defined by \eqref{eqn: max type tensor 2}.
The Euler--Lagrange equations by setting ${\sdel}\widehat{{\mathit{E}_{\mathtt{I\hspace{-2pt}I\hspace{-2pt}I}}}} =0$ are derived as
\begin{subequations}
 \begin{align}
 \Div \left( {\bsym{P}} \right) + \widetilde{{\mbf{f}}}^e =\mbf{0} & \quad \text{ in } \quad {\mcal{B}}_0, \\
 \jump{{\bsym{P}}} \mbf{n}_0 + \widetilde{\mbf{t}}^e =\mbf{0} & \quad \text{ on } \quad {\partial\mcal{B}} _0, \\
 \Div \left( {\bsym{P}} \right) =\mbf{0} & \quad \text{ in } \quad {\mcal{B}}_0', \\
 {{\bsym{F}}} {{\bbbm{H}}}^s ={{\widetilde{\Omega}}}_{, {{\bbbm{K}}}} & \quad \text{ in } \quad {\mcal{B}}_0, 
 \end{align}
 \label{eqn: sec6 EL equations 2}
\end{subequations}
by recognising that for this potential energy functional, the first Piola--Kirchhoff stress is given by
\begin{subequations}
\begin{align}
 {\bsym{P}} =\rho_0 {{\widetilde{\Omega}}}_{,{{\bsym{F}}}} + {{\mathring{{\bsym{P}}}}}_m - {{\widehat{{\bsym{P}}}}}_m & \quad \text{ in } \quad {\mcal{B}}_0, \\
 {\bsym{P}} ={{\mathring{{\bsym{P}}}}}_m - {{\widehat{{\bsym{P}}}}}_m & \quad \text{ in } \quad {\mcal{B}}_0' .
\end{align}
\end{subequations}

Upon a direct comparison of the above with \eqref{eqn: sec6 EL equations 1} and \eqref{eqn: Sec6 stress expression}, we note that due to the inclusion of extra terms with ${\bbbm{h}}^e$, the expressions for first Piola--Kirchhoff stress and the Maxwell stress in vacuum are different.
This leads to the vanishing of the equivalent of electromagnetic body force term in \eqref{eqn: sec6 EL equations 2}a and a modified constitutive equation \eqref{eqn: sec6 EL equations 2}d.

\section{Concluding remarks}
In this paper, we have presented five variational formulations of nonlinear magnetoelastostatics that differ from each other with respect to the independent field variable for the magnetic effect.
The formulations based on the magnetic field~${\bbbm{H}}$, the magnetic induction~${\bbbm{B}}$, and referential magnetization vector per unit volume~${\bbbm{M}}$ are analogous to the variational formulations of electroelastostatics presented by \cite{Saxena2020}.
Variational formulation based on referential magnetization per unit mass~${\overline{\bbbm{M}}}$ was originally postulated by \cite{Brown1965} and that based on a pull-back of the magnetization per unit mass to reference configuration~${\bbbm{K}}$ was given by \cite{kankanala2004}.
A direct equivalence between all five principles by means of Legendre transform and properties of Maxwell equations is the highlight of  Section \ref{sec: comparison} of this paper.

The principles can broadly be divided into two categories.
For the first kind based on ${\bbbm{H}}, {\bbbm{B}},$ and ${\bbbm{M}}$, the total energy is defined over a bounded domain $\vol$ with the external magnetic loading being specified by means of potential on the boundary $\bvol$.
For the second kind based on ${\overline{\bbbm{M}}}$ and ${\bbbm{K}}$, integral is defined over an infinite space and the notion of an external field becomes necessary to supply external loading. 
The choice to include this (constant) external field in the total energy can lead to a different definition of the Maxwell stress, and result in changes in the body force and traction terms.
Our analysis suggests caution with the choice of variational principle appropriate to the physical problem and control variables.

The analysis presented in this paper can be easily extended to the special case of incompressibility. 
For this purpose, see Remark 4 in the recent exposition and formulation for the electroelastic counterpart \citep{Saxena2020}.
Further extension of the present analysis to include mixed boundary conditions and discontinuities in the magnetoelastic body or free space can shed further light on the issues around correspondences between the five principles.
Inclusion of kinetic energy and the effect of time-dependent boundaries is another possible interesting area for extension of the analysis presented here.
We have restricted our analysis to nonlinear deformation and coupling. A linearised analysis to study deformation close to the reference configuration may lead to simplifications and influence the equivalence analysis presented in Section \ref{sec: comparison}.
These avenues are currently under investigation and shall appear in suitable forum elsewhere.

\section*{Acknowledgements}
Basant Lal Sharma acknowledges the support of SERB MATRICS grant MTR/2017/ 000013. 
Prashant Saxena acknowledges the support of startup funds from the James Watt School of Engineering at the University of Glasgow.

\appendix

\section{Variation of some relevant kinematic quantities}
\label{appendix: variations}
We list the first and second variations of key kinematic variables (see, for example, \citep{Saxena2020} for detailed derivations).
Upon a perturbation ${\mathpalette\irchi\relax} \to {\mathpalette\irchi\relax} + {\sdel}{\mathpalette\irchi\relax}$, we get
${{\bsym{F}}} ({\mathpalette\irchi\relax} + {\sdel}{\mathpalette\irchi\relax}) =\Grad {\mathpalette\irchi\relax} + \Grad ({\sdel}{\mathpalette\irchi\relax}) \Rightarrow {\sdel}{{\bsym{F}}} =\Grad ({\sdel}{\mathpalette\irchi\relax}), {\sddel} {{\bsym{F}}} =\mbf{0}.$
The right Cauchy--Green deformation tensor changes as
\begin{align}
{{\bsym{C}}}({\mathpalette\irchi\relax} + {\sdel}{\mathpalette\irchi\relax}) ={{\bsym{C}}} + {\sdel}{{\bsym{C}}} + {\sddel} {{\bsym{C}}} \dotsc, \mathrm{ with } {\sdel}{{\bsym{C}}} ={{\bsym{F}}}^{\top} {\sdel}{{\bsym{F}}} + [{\sdel}{{\bsym{F}}}]^{\top} {{\bsym{F}}}, {\sddel}{{\bsym{C}}} =[{\sdel}{{\bsym{F}}}]^{\top} {\sdel}{{\bsym{F}}}.
\end{align}
For the determinant $J =\mathrm{det} {{\bsym{F}}}$, we get
$J({\mathpalette\irchi\relax} + {\sdel}{\mathpalette\irchi\relax} ) =J + {\sdel}J + {\sddel} J + \dotsc$ with
\begin{equation}
{\sdel}J =J {{\bsym{F}}}^{-\top} \cdot {\sdel}{{\bsym{F}}}, {\sddel} J ={{\bsym{F}}} \cdot \mathrm{cof} ({\sdel}{{\bsym{F}}}).
\end{equation}
As ${\sdel}{{\bsym{F}}} =\Grad({\sdel}{\mathpalette\irchi\relax})$, the second of the above expressions, ${\sddel} J$, is written in component form as
${\sddel} J =\frac{1}{2} \varepsilon_{imn} \varepsilon_{jpq} F_{ij} [{\sdel}{\mathpalette\irchi\relax}_{m,p} ] [{\sdel}{\mathpalette\irchi\relax}_{n,q}].$
Here $\varepsilon_{ijk}$ is the third order permutation tensor. 
It can also be shown that 
\begin{equation}
{\sddel} J =\frac{1}{2} J[ \big[ {{\bsym{F}}}^{-\top} \cdot {\sdel}{{\bsym{F}}} \big] \big[ {{\bsym{F}}}^{-\top} \cdot {\sdel}{{\bsym{F}}} \big] - {{\bsym{F}}}^{-\top} \big[ {\sdel}{{\bsym{F}}}]^\top {{\bsym{F}}}^{-\top} \cdot {\sdel}{{\bsym{F}}}].
\label{eqn: modified expression for delta 2 J}
\end{equation}
Taylor's expansion for the inverse of determinant $J^{-1}$ is
$J^{-1} ({\mathpalette\irchi\relax} + {\sdel}{\mathpalette\irchi\relax}) =J_0 + J_1 + J_2 + \dotsc$
where $J_0 =J^{-1}, \quad J_1 =- J^{-1} {{\bsym{F}}}^{-\top} \cdot {\sdel}{{\bsym{F}}}, \quad J_2 =- J^{-2} {{\bsym{F}}} \cdot \mathrm{cof} ({\sdel}{{\bsym{F}}}) + J^{-1} \left[ {{\bsym{F}}}^{-\top} \cdot {\sdel}{{\bsym{F}}} \right]^2.$
Using the expression \eqref{eqn: modified expression for delta 2 J}, we rewrite $J_2$ as
$J_2 =({2 J})^{-1}[[ {{\bsym{F}}}^{-\top} \cdot {\sdel}{{\bsym{F}}}]^2 + {{\bsym{F}}}^{-\top} [{\sdel}{{\bsym{F}}}]^\top {{\bsym{F}}}^{-\top} \cdot {\sdel}{{\bsym{F}}}].$
For the inverse tensors,
$[{{\bsym{F}}} ({\mathpalette\irchi\relax} + {\sdel}{\mathpalette\irchi\relax}) ]^{-1} ={{\bsym{F}}}^{-1} + D_1 {{\bsym{F}}}^{-1} + D_2 {{\bsym{F}}}^{-1} + \dotsc$,
with
\begin{align}
D_1 {{\bsym{F}}}^{-1} =- {{\bsym{F}}}^{-1} [ {\sdel}{{\bsym{F}}} ] {{\bsym{F}}}^{-1}, \quad \quad D_2 {{\bsym{F}}}^{-1} ={{\bsym{F}}}^{-1} [ {\sdel}{{\bsym{F}}} ] {{\bsym{F}}}^{-1} [ {\sdel}{{\bsym{F}}} ] {{\bsym{F}}}^{-1}.
\end{align}
and
$[{{\bsym{C}}} ({\mathpalette\irchi\relax} + {\sdel}{\mathpalette\irchi\relax}) ]^{-1} ={{\bsym{C}}}^{-1} + D_1 {{\bsym{C}}}^{-1} + D_2 {{\bsym{C}}}^{-1} + \dotsc$
with
\begin{align}
D_1 {{\bsym{C}}}^{-1} &=- {{\bsym{C}}}^{-1} [{\sdel}{{\bsym{F}}}]^{\top} {{\bsym{F}}}^{-\top} - {{\bsym{F}}}^{-1} [{\sdel}{{\bsym{F}}}] {{\bsym{C}}}^{-1}, \\
D_2 {{\bsym{C}}}^{-1} &={{\bsym{C}}}^{-1} [{\sdel}{{\bsym{F}}}]^{\top} {{\bsym{F}}}^{-\top} [{\sdel}{{\bsym{F}}}]^{\top} {{\bsym{F}}}^{-\top} + {{\bsym{F}}}^{-1} [{\sdel}{{\bsym{F}}}] {{\bsym{C}}}^{-1} [{\sdel}{{\bsym{F}}}]^{\top} {{\bsym{F}}}^{-\top} \nonumber \\
&+ {{\bsym{F}}}^{-1} [ {\sdel}{{\bsym{F}}} ] {{\bsym{F}}}^{-1} [ {\sdel}{{\bsym{F}}} ] {{\bsym{C}}}^{-1}.
\end{align}

\section{Variational formulation based on magnetic induction}
\label{sec: B formulation}

Using the fact that ${{\bbbm{B}}}$ is found in terms of ${\bbbm{A}}$ by \eqref{eqn: potentials introduction}${}_1$, i.e., ${{\bbbm{B}}} =\Curl {\bbbm{A}}$,
the total potential energy of the system, i.e., the body {${\mcal{B}}_0$} and its exterior {${\mcal{B}}'_0$}, is written as a functional depending on the deformation ${\mathpalette\irchi\relax}$ and ${\bbbm{A}}$ as \citep{Dorfmann2014b}
\begin{align}
{{\mathit{E}_{\mathtt{I\hspace{-1pt}V}}}}[{\mathpalette\irchi\relax}, {\bbbm{A}}] &\defnt \int\nolimits_{{\mcal{B}}_0} {{\mathring{\Omega}}}({{\bsym{F}}}, {{\bbbm{B}}}) {dv}_0 + \frac{1}{2 {\mu_0}} \int\nolimits_{{\mcal{B}}'_0} J^{-1} [{{\bsym{F}}} {{\bbbm{B}}}] \cdot [{{\bsym{F}}} {{\bbbm{B}}}] {dv}_0\nonumber \\
& + \int\nolimits_{\bvol} \left[ {\bbbm{h}}^e \wedge {\bbbm{a}} \right]\cdot \mbf{n} ds - \int\nolimits_{{\mcal{B}}_0} \widetilde{{\mbf{f}}}^e \cdot {\mathpalette\irchi\relax} {dv}_0 - \int\nolimits_{{\partial\mcal{B}} _0} \widetilde{\mbf{t}}^e \cdot {\mathpalette\irchi\relax} ds_0, 
\label{eqn: potential energy functional for B formulation}
\end{align}
where ${{\mathring{\Omega}}}$ is the (scalar) {total} (magnetoelastic) stored energy density per unit volume, 
${\bbbm{h}}^e$ is the externally applied magnetic (vector) field whose tangential component is prescribed on $\bvol$.
The integral terms in equation \eqref{eqn: potential energy functional for B formulation} involve the reference configuration as the spatial fields are mapped to the reference {configuration},
with the exception of the third term which is written in terms of the current region $\vol$.
It assumed that the boundary (typically, {\em infinitally} far away) is fixed (i.e., it does not change in space between reference and spatial description), so that the third term in equation \eqref{eqn: potential energy functional for B formulation} is also rewritten in the reference configuration simply as
$\int\nolimits_{\bvol_0} \left[ {{\bbbm{H}}}^e \wedge {\bbbm{A}} \right] \cdot \mbf{n}_0 ds_0.$
Notice that $\mbf{n}_0$ and $\mbf{n}$ are used to denote the {respective} outward unit normals for the region $\vol_0$ and $\vol$ {(as well as ${\mcal{B}}_0$ and ${\mcal{B}}$)}.

\subsection{Equilibrium: first variation}
\label{sec: A first var}
In order to describe {the deformation} ${\mathpalette\irchi\relax}$ and {the referential magnetic vector potential} ${\bbbm{A}}$ when the body is in a state of equilibrium, the first variation of the potential energy functional should vanish, that is, using the functional \eqref{eqn: potential energy functional for B formulation},
${\sdel}{{\mathit{E}_{\mathtt{I\hspace{-1pt}V}}}}\equiv {\sdel}{{\mathit{E}_{\mathtt{I\hspace{-1pt}V}}}}[{\mathpalette\irchi\relax}, {\bbbm{A}}; ({\sdel}{\mathpalette\irchi\relax}, {\sdel}{\bbbm{A}})] =0.$
An expansion of the functional ${{\mathit{E}_{\mathtt{I\hspace{-1pt}V}}}}$ up to the first order, owing to a variation of its arguments ${\mathpalette\irchi\relax}$ and ${\bbbm{A}}$, is given by
\begin{align}
{{\mathit{E}_{\mathtt{I\hspace{-1pt}V}}}} [{\mathpalette\irchi\relax} + {\sdel}{\mathpalette\irchi\relax}, {\bbbm{A}} + {\sdel}{\bbbm{A}}] &=\int\nolimits_{{\mcal{B}}_0} {{\mathring{\Omega}}} ({{\bsym{F}}} +{\sdel}{{\bsym{F}}}, {{\bbbm{B}}} + {\sdel}{{\bbbm{B}}}) {dv}_0 \nonumber \\
&+ \frac{1}{2 {\mu_0}} \int\nolimits_{{\mcal{B}}'_0} [J + {\sdel}J]^{-1} \left[ [{{\bsym{F}}} + {\sdel}{{\bsym{F}}}] \left[{{\bbbm{B}}} + {\sdel}{{\bbbm{B}}} \right] \right] \cdot \left[ [ {{\bsym{F}}} + {\sdel}{{\bsym{F}}}] \left[{{\bbbm{B}}} + {\sdel}{{\bbbm{B}}} \right] \right] {dv}_0 \nonumber \\
&+ \int\nolimits_{\bvol_0} \left[ {{\bbbm{H}}}^e \wedge [{\bbbm{A}} + {\sdel}{\bbbm{A}}] \right] \cdot \mbf{n}_0 ds_0 \nonumber \\
& - \int\nolimits_{{\mcal{B}}_0} \widetilde{{\mbf{f}}}^e \cdot [ {\mathpalette\irchi\relax}+ {\sdel}{\mathpalette\irchi\relax} ] {dv}_0 - \int\nolimits_{{\partial\mcal{B}} _0} \widetilde{\mbf{t}}^e \cdot [ {\mathpalette\irchi\relax}+ {\sdel}{\mathpalette\irchi\relax} ] ds_0.
\end{align}
Taking a{dv}antage of the referential description, noting that
${\sdel}\mbf{D} =\Curl {\sdel}{\bbbm{A}},$
while using expressions for first order variations as derived in \citep{Saxena2020},
we simplify further the expression of ${{\mathit{E}_{\mathtt{I\hspace{-1pt}V}}}} [{\mathpalette\irchi\relax} + {\sdel}{\mathpalette\irchi\relax}, {\bbbm{A}} + {\sdel}{\bbbm{A}}] $ stated above. Thus, it is found that the first variation \eqref{eqn: first variation condition}
of ${{\mathit{E}_{\mathtt{I\hspace{-1pt}V}}}}$ is given by
\begin{align}
{\sdel}{{\mathit{E}_{\mathtt{I\hspace{-1pt}V}}}} &={{\mathit{E}_{\mathtt{I\hspace{-1pt}V}}}} [{\mathpalette\irchi\relax} + {\sdel}{\mathpalette\irchi\relax}, {\bbbm{A}} + {\sdel}{\bbbm{A}}] - {{\mathit{E}_{\mathtt{I\hspace{-1pt}V}}}} [{\mathpalette\irchi\relax}, {\bbbm{A}} ] \nonumber\\
&=\int\nolimits_{{\mcal{B}}_0} \left[ {{\mathring{\Omega}}}_{,{{\bsym{F}}}} \cdot {\sdel}{{\bsym{F}}} + {{\mathring{\Omega}}}_{, {{\bbbm{B}}}} \cdot \Curl {\sdel}{\bbbm{A}} \right] {dv}_0 \nonumber \\
& + \frac{1}{2 {\mu_0}} \int\nolimits_{{\mcal{B}}'_0} \bigg[ - J^{-1} \left[ {{\bsym{F}}}^{-\top} \cdot {\sdel}{{\bsym{F}}} \right] [{{\bsym{F}}} {{\bbbm{B}}}] \cdot [{{\bsym{F}}} {{\bbbm{B}}}] + 2 J^{-1} [[{{\bsym{F}}} {{\bbbm{B}}} ] \otimes {{\bbbm{B}}}] \cdot {\sdel}{{\bsym{F}}}\nonumber \\
& + 2 [{{\bsym{C}}} {{\bbbm{B}}}] \cdot \Curl {\sdel}{\bbbm{A}} \bigg] {dv}_0 + \int\nolimits_{\bvol_0} \left[ \mbf{n}_0 \wedge {{\bbbm{H}}}^e \right] \cdot {\sdel}{\bbbm{A}} ds_0\nonumber \\
& - \int\nolimits_{{\mcal{B}}_0} \widetilde{{\mbf{f}}}^e \cdot {\sdel}{\mathpalette\irchi\relax} {dv}_0 - \int\nolimits_{{\partial\mcal{B}} _0} \widetilde{\mbf{t}}^e \cdot {\sdel}{\mathpalette\irchi\relax} ds_0.
\end{align}
Using an elementary identity for vector fields $\mbf{u}$ and $\mbf{v}$, namely,
\begin{equation}
{ \mbf{v} \cdot \Curl \mbf{u} =\Div [\mbf{u} \wedge \mbf{v}] + [\Curl \mbf{v}] \cdot \mbf{u}, }
\label{eqn: curl identity}
\end{equation}
we expand the above expression for $ {\sdel}{{\mathit{E}_{\mathtt{I\hspace{-1pt}V}}}}$ as
\begin{align}
{\sdel}{{\mathit{E}_{\mathtt{I\hspace{-1pt}V}}}} &=\int\nolimits_{{\mcal{B}}_0} \left[ {{\mathring{\Omega}}}_{,{{\bsym{F}}}} \cdot {\sdel}{{\bsym{F}}} + [ { \Curl }{{\mathring{\Omega}}}_{, {{\bbbm{B}}}} ] \cdot {\sdel}{\bbbm{A}} \right] {dv}_0 + \int\nolimits_{{\partial\mcal{B}} _0^{-}} \mbf{n}_0 \cdot \left[ {{\mathring{\Omega}}}_{,{{\bbbm{B}}}} \wedge {\sdel}{\bbbm{A}} \right] ds_0 \nonumber \\
& - \frac{1}{{\mu_0}} \int\nolimits_{{\partial\mcal{B}} _0^+} \mbf{n}_0 \cdot \left[ {{\bsym{C}}} {{\bbbm{B}}} \wedge {\sdel}\mbf{A} \right] ds_0 \nonumber \\
& + \frac{1}{2 {\mu_0}} \int\nolimits_{{\mcal{B}}'_0} \bigg[ - J^{-1} \left[ {{\bsym{F}}}^{-\top} \cdot {\sdel}{{\bsym{F}}} \right] [{{\bsym{F}}} {{\bbbm{B}}}] \cdot [{{\bsym{F}}} {{\bbbm{B}}}] + 2 J^{-1} [[{{\bsym{F}}} {{\bbbm{B}}} ] \otimes {{\bbbm{B}}}] \cdot {\sdel}{{\bsym{F}}} \nonumber \\
& + [\Curl({{\bsym{C}}} {{\bbbm{B}}})] \cdot {\sdel}{\bbbm{A}} \bigg] {dv}_0 
+ \int\nolimits_{\bvol_0} [ \mbf{n}_0 \wedge [ {{\bbbm{H}}}^e - \frac{1}{{\mu_0}} {{\bsym{C}}} {{\bbbm{B}}} ] ] \cdot {\sdel}{\bbbm{A}} ds_0\nonumber \\
& - \int\nolimits_{{\mcal{B}}_0} \widetilde{{\mbf{f}}}^e \cdot {\sdel}{\mathpalette\irchi\relax} {dv}_0 - \int\nolimits_{{\partial\mcal{B}} _0} \widetilde{\mbf{t}}^e \cdot {\sdel}{\mathpalette\irchi\relax} ds_0.
\end{align}
Inspection of above leads to consideration of the definition of a tensor field given by \eqref{eqn: maxwell strPK}.
Using the definition \eqref{eqn: maxwell strPK}, we rewrite the first variation ${\sdel}{{\mathit{E}_{\mathtt{I\hspace{-1pt}V}}}}$ of the total potential as
\begin{align}
{\sdel}{{\mathit{E}_{\mathtt{I\hspace{-1pt}V}}}} &=\int\nolimits_{{\mcal{B}}_0} \left[ - [ \Div \left({{\mathring{\Omega}}}_{,{{\bsym{F}}}} \right) + \widetilde{{\mbf{f}}}^e ] \cdot {\sdel}{\mathpalette\irchi\relax} + [ { \Curl} {{\mathring{\Omega}}}_{, {{\bbbm{B}}}} ] \cdot {\sdel}{\bbbm{A}} \right] {dv}_0 \nonumber \\
& + \int\nolimits_{{\partial\mcal{B}} _0} [ [ \left[ {{\mathring{\Omega}}}_{,{{\bsym{F}}}}|_- - {{\bsym{P}}}_m|_+ \right] \mbf{n}_0 - \widetilde{\mbf{t}}^e ] \cdot {\sdel}{\mathpalette\irchi\relax} + [ \mbf{n}_0 \wedge [ {{\mathring{\Omega}}}_{, {{\bbbm{B}}}}|_- - \frac{1}{{\mu_0}} {{\bsym{C}}} {{\bbbm{B}}}|_+ ] ] \cdot {\sdel}{\bbbm{A}} ] ds_0 \nonumber \\
& + \int\nolimits_{{\mcal{B}}'_0} [- \Div {{\bsym{P}}}_m \cdot {\sdel}{\mathpalette\irchi\relax} 
+ \frac{1}{2 {\mu_0}} [ { \Curl } {{\bsym{C}}} {{\bbbm{B}}}] \cdot {\sdel}{\bbbm{A}}] {dv}_0 \nonumber \\ 
&+ \int\nolimits_{\bvol_0}[ {{\bsym{P}}}_m \mbf{n}_0 \cdot {\sdel}{\mathpalette\irchi\relax} + [ \mbf{n}_0 \wedge [ {{\bbbm{H}}}^e - \frac{1}{{\mu_0}} {{\bsym{C}}} {{\bbbm{B}}} ] ] \cdot {\sdel}{\bbbm{A}}] ds_0. 
\label{eqn: first variation condition1}
\end{align}
The {total} (first Piola--Kirchhoff) stress ${{\bsym{P}}}$ in the body is ${{\bsym{P}}} ={{\mathring{\Omega}}}_{,{{\bsym{F}}}}, \text{ in } {\mcal{B}}_0,$ and the (Maxwell) stress {exterior to} the body is given by \eqref{eqn: maxwell strPK}, i.e., ${{\bsym{P}}} ={{\bsym{P}}}_m, \text{ in } {\mcal{B}}_0'.$

Upon applying the condition \eqref{eqn: first variation condition} to the first variation \eqref{eqn: first variation condition1} calculated above, the coefficients of arbitrary variations ${\sdel}{\mathpalette\irchi\relax}$ and ${\sdel}{\bbbm{A}}$ should vanish for ${\sdel}{{\mathit{E}_{\mathtt{I\hspace{-1pt}V}}}}$ to vanish.
As a consequence of the vanishing of the coefficients of ${\sdel}{\mathpalette\irchi\relax}$ results in the following equations
\begin{subequations}
\begin{align}
\Div {{\bsym{P}}} + \widetilde{{\mbf{f}}}^e =\mbf{0} &\text{ in } {\mcal{B}}_0,
\label{eqn: gov 1 B formulation} \\
\Div {{\bsym{P}}} =\mbf{0} &\text{ in } {\mcal{B}}_0',\\
\jump{{{\bsym{P}}}} \mbf{n}_0 + \widetilde{\mbf{t}}^e=\mbf{0} &\text{ on } {\partial\mcal{B}} _0, \\
{{\bsym{P}}} \mbf{n}_0 =\mbf{0} &\text{ on } \bvol_0.
\end{align}
\label{eqn: Euler Eq B form}
\end{subequations}
We {thus obtain} the magnetic field ${{\bbbm{H}}}$ in the body as
\begin{equation}
{{\bbbm{H}}} ={{\mathring{\Omega}}}_{, {{\bbbm{B}}}} =\frac{1}{{\mu_0}} \left[ J^{-1} {{\bsym{C}}} {{\bbbm{B}}} - {{\bbbm{M}}} \right] \text{ in } {\mcal{B}}_0,
\label{eqn: H =omegaB}
\end{equation}
and {exterior to} the body as
\begin{equation}
{{\bbbm{H}}} =\frac{1}{{\mu_0} }J^{-1} {{\bsym{C}}} {{\bbbm{B}}} \text{ in } {\mcal{B}}_0',
\end{equation}
because the magnetisation ${{\bbbm{M}}}$ vanishes in ${\mcal{B}}_0'$ and use has been made of the constitutive relation \eqref{eqn: constitutive B H M}.
Since the body ${\mcal{B}}_0$ and normal to the boundary $\mbf{n}_0$ can be chosen arbitrarily, we get the following relations from the vanishing of the coefficients of ${\sdel}{\bbbm{A}}$ 
\begin{subequations}
\begin{align}
\Curl({{\bbbm{H}}} ) =\mbf{0} &\text{ in } {\mcal{B}}_0 \cup {\mcal{B}}_0', \\
\mbf{n}_0 \wedge \jump{{\bbbm{H}}} =\mbf{0} &\text{ on } {\partial\mcal{B}} _0, \\
\mbf{n}_0 \wedge \left[ {{\bbbm{H}}}^e - {{\bbbm{H}}} \right] =\mbf{0} &\text{ on } \bvol_0.
\label{eqn: gov last B formulation}
\end{align}
\label{eqn: Euler Eq 2 B form}
\end{subequations}

\begin{sidenote}
We note that in this formulation based on the magnetic induction vector, we have apriori assumed that the equation \eqref{eqn: maxwell lagrangian}$_1$ is satisfied by ${{\bbbm{B}}}$ and have recovered the equation \eqref{eqn: maxwell lagrangian}$_2$ for the magnetic field ${{\bbbm{H}}}$ as the Euler-Lagrange equation for the variational (potential energy minimisation) problem.
This procedure implies the constitutive assumption ${{\bbbm{H}}} ={{\mathring{\Omega}}}_{,{{\bbbm{B}}}}$.
\label{note1}
\end{sidenote}

\subsection{Critical point: second variation}
\label{second variation 1}
For the analysis of critical point $({\mathpalette\irchi\relax}, {\bbbm{A}})$, we need to find the functions ${\bdel}{\mathpalette\irchi\relax}$ and ${\bdel}{\bbbm{A}}$ such that the bilinear functional defined below vanishes at the critical point, that is
${\sddel} {{\mathit{E}_{\mathtt{I\hspace{-1pt}V}}}}\equiv {\sddel} {{\mathit{E}_{\mathtt{I\hspace{-1pt}V}}}}[{\mathpalette\irchi\relax}, {\bbbm{A}}; ({\sdel}{\mathpalette\irchi\relax}, {\sdel}{\bbbm{A}}), ({\bdel}{\mathpalette\irchi\relax}, {\bdel}{\bbbm{A}})]
=0. $
Upon using the expressions derived in \citep{Saxena2020}, the bilinear functional associated with the second variation 
of ${{\mathit{E}_{\mathtt{I\hspace{-1pt}V}}}}$ is expanded into the form
\begin{align}
{\sddel} {{\mathit{E}_{\mathtt{I\hspace{-1pt}V}}}}& =\int\nolimits_{{\mcal{B}}_0}[[ {{\mathring{\Omega}}}_{, {{\bsym{F}}} {{\bsym{F}}}} {\bdel}{{\bsym{F}}} + \frac{1}{2} {{\mathring{\Omega}}}_{, {{\bsym{F}}} {{\bbbm{B}}}} {\bdel}{{\bbbm{B}}} + \frac{1}{2} \wst{{\mathring{\Omega}}}_{ {{\bsym{F}}} {{\bbbm{B}}}} {\bdel}{{\bbbm{B}}}] \cdot {\sdel}{{\bsym{F}}} \nonumber \\
& +[ {{\mathring{\Omega}}}_{, {{\bbbm{B}}} {{\bbbm{B}}}} {\bdel}{{\bbbm{B}}} + \frac{1}{2} {{\mathring{\Omega}}}_{, {{\bbbm{B}}} {{\bsym{F}}}} {\bdel}{{\bsym{F}}} + \frac{1}{2} \wst{{\mathring{\Omega}}}_{ {{\bbbm{B}}} {{\bsym{F}}}} {\bdel}{{\bsym{F}}}] \cdot {\sdel}{{\bbbm{B}}}] {dv}_0 \nonumber \\
& + \frac{1}{2 {\mu_0}} \int\nolimits_{{\mcal{B}}_0'} J^{-1}[ [{{\bsym{F}}} {{\bbbm{B}}}] \cdot [{{\bsym{F}}} {{\bbbm{B}}}][[ {{\bsym{F}}}^{-\top} \cdot {\bdel}{{\bsym{F}}}][ {{\bsym{F}}}^{-\top} \cdot {\sdel}{{\bsym{F}}}] + {{\bsym{F}}}^{-\top} [{\bdel}{{\bsym{F}}}]^\top {{\bsym{F}}}^{-\top} \cdot {\sdel}{{\bsym{F}}} \nonumber \\
& - 2 [[ {\bdel}{{\bsym{F}}} {{\bbbm{B}}}] \cdot[ {{\bsym{F}}} {{\bbbm{B}}}] +[ {{\bsym{F}}} {\bdel}{{\bbbm{B}}}] \cdot[ {{\bsym{F}}} {{\bbbm{B}}}]] {{\bsym{F}}}^{-\top} \cdot {\sdel}{{\bsym{F}}} \nonumber \\
& - 2[[ {\sdel}{{\bsym{F}}} {{\bbbm{B}}}] \cdot[ {{\bsym{F}}} {{\bbbm{B}}}] +[ {{\bsym{F}}} {\sdel}{{\bbbm{B}}}] \cdot[ {{\bsym{F}}} {{\bbbm{B}}}]] {{\bsym{F}}}^{-\top} \cdot {\bdel}{{\bsym{F}}} \nonumber \\
& + 2[ {\sdel}{{\bsym{F}}} {\bdel}{{\bbbm{B}}} + {\bdel}{{\bsym{F}}} {\sdel}{{\bbbm{B}}}] \cdot[ {{\bsym{F}}} {{\bbbm{B}}}] + 2 {\sdel}{{\bsym{F}}} {{\bbbm{B}}} \cdot {{\bsym{F}}} {\bdel}{{\bbbm{B}}} + 2 {\bdel}{{\bsym{F}}} {{\bbbm{B}}} \cdot {{\bsym{F}}} {\sdel}{{\bbbm{B}}} \nonumber \\
& + 2[ {\bdel}{{\bsym{F}}} {{\bbbm{B}}}] \cdot[ {\sdel}{{\bsym{F}}} {{\bbbm{B}}}] + 2[ {{\bsym{F}}} {\bdel}{{\bbbm{B}}}] \cdot[ {{\bsym{F}}} {\sdel}{{\bbbm{B}}}]] {dv}_0.
\label{eqn: expanded d2E B formulation}
\end{align}
In the expression stated above we have defined the third order tensors $\wst{{\mathring{\Omega}}}_{{{\bsym{F}}} {{\bbbm{B}}}}$ and $\wst{{\mathring{\Omega}}}_{{{\bbbm{B}}} {{\bsym{F}}} }$ according to the following property
\begin{equation}
[ \wst{{\mathring{\Omega}}}_{{{\bsym{F}}} {{\bbbm{B}}}} \mbf{u} ] \cdot \bsym{U} =[ {{\mathring{\Omega}}}_{, {{\bbbm{B}}} {{\bsym{F}}}} \bsym{U}] \cdot \mbf{u}, \quad [ \wst{{\mathring{\Omega}}}_{ {{\bbbm{B}}} {{\bsym{F}}}} \bsym{U}] \cdot \mbf{u} =\left[ {{\mathring{\Omega}}}_{, {{\bsym{F}}} {{\bbbm{B}}}} \mbf{u} \right] \cdot \bsym{U},
\end{equation}
which holds for arbitrary $\mbf{u}$ and $\bsym{U}$, while $\mbf{u}$ is a vector and $\bsym{U}$ is a second order tensor.
Using the expression \eqref{eqn: expanded d2E B formulation} of ${\sddel} {{\mathit{E}_{\mathtt{I\hspace{-1pt}V}}}}$, in the region ${\mcal{B}}_0'$ the terms containing ${\sdel}{{\bbbm{B}}}$ can be written in the form $\mbf{v}_0 \cdot {\sdel}{{\bbbm{B}}}$, where the vector field $\mbf{v}_0$ is defined by
\begin{align}
\mbf{v}_0 \defnt \frac{1}{ {\mu_0} J}[ -[ {{\bsym{F}}}^{-\top} \cdot {\bdel}{{\bsym{F}}}] {{\bsym{F}}}^\top {{\bsym{F}}} {{\bbbm{B}}} + [{\bdel}{{\bsym{F}}}]^\top {{\bsym{F}}} {{\bbbm{B}}} + {{\bsym{F}}}^\top {\bdel}{{\bsym{F}}} {{\bbbm{B}}} + {{\bsym{F}}}^\top {{\bsym{F}}} {\bdel}{{\bbbm{B}}}].
\end{align}
Since equation \eqref{eqn: constitutive B H M} gives ${{\bbbm{H}}} =J^{-1} {\mu_0}^{-1} {{\bsym{C}}} {{\bbbm{B}}}$ in ${\mcal{B}}_0'$, it is easy to see that 
$\mbf{v}_0 ={\bdel}{{\bbbm{H}}}.$
Also, in the expression \eqref{eqn: expanded d2E B formulation} of ${\sddel} {{\mathit{E}_{\mathtt{I\hspace{-1pt}V}}}}$, in the region ${\mcal{B}}_0'$ the terms containing ${\sdel}{{\bsym{F}}}$ can be written in the form $\bsym{T} \cdot {\sdel}{{\bsym{F}}}$ where the second order tensor $\bsym{T}$ is defined by
\begin{align}
\bsym{T} & \defnt \frac{1}{2 {\mu_0} J} \Bigg[ [{{\bsym{F}}} {{\bbbm{B}}}] \cdot [{{\bsym{F}}} {{\bbbm{B}}}] \bigg[ \big[ {{\bsym{F}}}^{-\top} \cdot {\bdel}{{\bsym{F}}} \big] {{\bsym{F}}}^{-\top} + {{\bsym{F}}}^{-\top} [{\bdel}{{\bsym{F}}}]^\top {{\bsym{F}}}^{-\top} \bigg] \nonumber \\
& - 2 \bigg[ \Big[ {\bdel}{{\bsym{F}}} {{\bbbm{B}}} \Big] \cdot \big[ {{\bsym{F}}} {{\bbbm{B}}} \big] + \Big[ {{\bsym{F}}} {\bdel}{{\bbbm{B}}} \Big] \cdot \big[ {{\bsym{F}}} {{\bbbm{B}}} \big] \bigg] {{\bsym{F}}}^{-\top} - 2 \big[ {{\bsym{F}}}^{-\top} \cdot {\bdel}{{\bsym{F}}} \big] [{{\bsym{F}}} {{\bbbm{B}}}] \otimes {{\bbbm{B}}} \nonumber \\
& + 2 [{{\bsym{F}}} {{\bbbm{B}}}] \otimes {\bdel}{{\bbbm{B}}} + 2 [{{\bsym{F}}} {\bdel}{{\bbbm{B}}}] \otimes {{\bbbm{B}}} + 2 [{\bdel}{{\bsym{F}}} {{\bbbm{B}}}] \otimes {{\bbbm{B}}} \Bigg].
\label{eqn: delta Pm expression 1}
\end{align}
By expanding the expression stated in equation \eqref{eqn: maxwell strPK}, to first order perturbation, it is seen that 
$\bsym{T} ={\bdel}{{\bsym{P}}}_m.$
{Based on a repeated }application of the triple product identity involving the curl operator \eqref{eqn: curl identity} and the divergence theorem, while observing that the variations ${\sdel}{\mathpalette\irchi\relax}$ and ${\sdel}{\bbbm{A}}$ are arbitrary, the equation ${\sddel} {{\mathit{E}_{\mathtt{I\hspace{-1pt}V}}}}=0$ \eqref{eqn: expanded d2E B formulation} finally leads to the following partial differential equations
\begin{align}
\Div({{\mathring{\Omega}}}_{, {{\bsym{F}}} {{\bsym{F}}}} {\bdel}{{\bsym{F}}} + \frac{1}{2}[ {{\mathring{\Omega}}}_{, {{\bsym{F}}} {{\bbbm{B}}}} + \wst{{\mathring{\Omega}}}_{ {{\bsym{F}}} {{\bbbm{B}}}}] {\bdel}{{\bbbm{B}}}) &=0 \text{ in } {\mcal{B}}_0, \\
\Curl({{\mathring{\Omega}}}_{, {{\bbbm{B}}} {{\bbbm{B}}}} {\bdel}{{\bbbm{B}}} + \frac{1}{2}[ {{\mathring{\Omega}}}_{, {{\bbbm{B}}} {{\bsym{F}}}} + \wst{{\mathring{\Omega}}}_{{{\bbbm{B}}} {{\bsym{F}}} }] {\bdel}{{\bsym{F}}}) &=0 \text{ in } {\mcal{B}}_0, \\
[ [ {{\mathring{\Omega}}}_{, {{\bsym{F}}} {{\bsym{F}}}} {\bdel}{{\bsym{F}}} + \frac{1}{2}[ {{\mathring{\Omega}}}_{, {{\bsym{F}}} {{\bbbm{B}}}} + \wst{{\mathring{\Omega}}}_{ {{\bsym{F}}} {{\bbbm{B}}}} ] {\bdel}{{\bbbm{B}}}]|_- - \bsym{T}|_+] \mbf{n}_0 &=0 \text{ on } {\partial\mcal{B}} _0,\\
[ {{\mathring{\Omega}}}_{, {{\bbbm{B}}} {{\bbbm{B}}}} {\bdel}{{\bbbm{B}}} + \frac{1}{2} [ {{\mathring{\Omega}}}_{, {{\bbbm{B}}} {{\bsym{F}}}} + \wst{{\mathring{\Omega}}}_{{{\bbbm{B}}} {{\bsym{F}}} }] {\bdel}{{\bsym{F}}}|_- - \mbf{v}_0|_+] \wedge \mbf{n}_0 &=0 \text{ on } {\partial\mcal{B}} _0,\\
\Div \bsym{T} &=0 \text{ in } {\mcal{B}}_0',\\
\Curl\mbf{v}_0 &=0 \text{ in } {\mcal{B}}_0',\\
\bsym{T} \mbf{n}_0 &=0 \text{ on } \bvol_0,\\
\mbf{v}_0 \wedge \mbf{n}_0 &=0 \text{ on } \bvol_0.
\label{eqn: second variation based PDE}
\end{align}

\begin{sidenote}
Note that since we have proved $\bsym{T} ={\bdel}{{\bsym{P}}}_m$ and $\mbf{v}_0 ={\bdel}{{\bbbm{H}}}$, it also follows that the above set of equations for the variations ${\bdel}{{\bbbm{B}}}$ and ${\bdel}{{\bsym{F}}}$ {in ${\mcal{B}}_0'$} can be alternatively obtained by {perturbing} the corresponding equations of equilibrium \eqref{eqn: gov 1 B formulation}--\eqref{eqn: gov last B formulation}.
{However, perturbation of the equilibrium equations in ${\mcal{B}}_0$ do not result in the above equations due to presence of the $ \frac{1}{2}[{{\mathring{\Omega}}}_{, {{\bsym{F}}} {{\bbbm{B}}}} + \wst{{\mathring{\Omega}}}_{ {{\bsym{F}}} {{\bbbm{B}}}}]$ and $\frac{1}{2}[{{\mathring{\Omega}}}_{, {{\bbbm{B}}} {{\bsym{F}}}} + \wst{{\mathring{\Omega}}}_{{{\bbbm{B}}} {{\bsym{F}}} }]$ terms. 
This general argument can be relaxed in cases when the energy density function ${{\mathring{\Omega}}}$ is assumed to be sufficiently continuous as has been considered, for example, by \cite{Bustamante2012}. }
\label{perturbeqlb1}
\end{sidenote}

\section{Variational formulation based on magnetic field}
\label{sec: H formulation}
Noting that ${{\bbbm{H}}} =- \Grad {\Phi}$, the total potential energy of the system is written as \citep{Dorfmann2014b}
\begin{align}
{\mathit{E}_{\mathtt{V}}}[{\mathpalette\irchi\relax}, {\Phi}] &\defnt \int\nolimits_{{\mcal{B}}_0} {{\check{\Omega}}} ({{\bsym{F}}}, {{\bbbm{H}}}) {dv}_0 - \frac{1}{2 }{\mu_0} \int\nolimits_{{\mcal{B}}'_0} J \left[ {{\bsym{F}}}^{-\top} {{\bbbm{H}}} \right] \cdot \left[ {{\bsym{F}}}^{-\top} {{\bbbm{H}}} \right] {dv}_0 - \int\nolimits_{\bvol} {\phi} {\bbbm{b}}^e \cdot \mbf{n} ds \nonumber \\
& - \int\nolimits_{{\mcal{B}}_0} \widetilde{{\mbf{f}}}^e \cdot {\mathpalette\irchi\relax} {dv}_0 - \int\nolimits_{{\partial\mcal{B}} _0} \widetilde{\mbf{t}}^e \cdot {\mathpalette\irchi\relax} ds_0, 
\label{eqn: potential energy functional for H formulation}
\end{align}
where ${\check{\Omega}}$ is the stored energy density per unit volume that depends on the deformation gradient ${{\bsym{F}}}$ and the referential magnetic field vector ${{\bbbm{H}}}$.
The third term in equation \eqref{eqn: potential energy functional for H formulation} is in the current configuration but the same argument as that following \eqref{eqn: potential energy functional for B formulation}
allows it to be rewritten in the reference configuration as $-\int\nolimits_{\bvol_0} {\Phi} {{\bbbm{B}}}^e \cdot \mbf{n}_0 ds_0.$

\subsection{Equilibrium: first variation}
At an state of equilibrium, ${\mathpalette\irchi\relax}$ and ${\Phi}$ are such that the first variation of the potential energy functional vanishes satisfying an analogue of equation \eqref{eqn: first variation condition}, i.e.,
${\sdel}{\mathit{E}_{\mathtt{V}}}\equiv {\sdel}{\mathit{E}_{\mathtt{V}}}[{\mathpalette\irchi\relax}, {\Phi}; ({\sdel}{\mathpalette\irchi\relax}, {\sdel}{\Phi})] =0.$
The variation of the functional ${\mathit{E}_{\mathtt{V}}}$ up to the first order in $({\sdel}{\mathpalette\irchi\relax}, {\sdel}{\Phi})$ is given by
\begin{align}
{\sdel}{\mathit{E}_{\mathtt{V}}} &={\mathit{E}_{\mathtt{V}}} [{\mathpalette\irchi\relax} + {\sdel}{\mathpalette\irchi\relax}, {\Phi} + {\sdel}{\Phi}] - {\mathit{E}_{\mathtt{V}}} [{\mathpalette\irchi\relax}, {\Phi} ] =\int\nolimits_{{\mcal{B}}_0} \left[ {{\check{\Omega}}}_{,{{\bsym{F}}}} \cdot {\sdel}{{\bsym{F}}} - {{\check{\Omega}}}_{, {{\bbbm{H}}}} \cdot \Grad {\sdel}{\Phi} \right] {dv}_0 \nonumber \\ 
& - \frac{1}{2 }{\mu_0} \int\nolimits_{{\mcal{B}}'_0} \bigg[ J {{\bsym{F}}}^{-\top} \cdot {\sdel}{{\bsym{F}}} [ {{\bsym{F}}}^{-\top} {{\bbbm{H}}} ] \cdot [ {{\bsym{F}}}^{-\top} {{\bbbm{H}}} ] -2 J \left[ {{\bsym{F}}}^{-\top} [{\sdel}{{\bsym{F}}}]^{\top} {{\bsym{F}}}^{-\top} {{\bbbm{H}}} \right] \cdot [ {{\bsym{F}}}^{-\top} {{\bbbm{H}}} ] \nonumber \\
& + 2 J [ {{\bsym{F}}}^{-\top} {{\bbbm{H}}} ] \cdot [ {{\bsym{F}}}^{-\top} {\sdel}{{\bbbm{H}}} ] \bigg] {dv}_0 - \int\nolimits_{\bvol_0} {\sdel}{\Phi} {{\bbbm{B}}}^e \cdot \mbf{n}_0 ds_0\nonumber \\
& - \int\nolimits_{{\mcal{B}}_0} \widetilde{{\mbf{f}}}^e \cdot {\sdel}{\mathpalette\irchi\relax} {dv}_0 - \int\nolimits_{{\partial\mcal{B}} _0} \widetilde{\mbf{t}}^e \cdot {\sdel}{\mathpalette\irchi\relax} ds_0. 
\label{eqn: E first var 1}
\end{align}
We define the first Piola--Kirchhoff stress ${\bsym{P}}$ and magnetic induction ${{\bbbm{B}}}$ in the body as
\begin{equation}
{\bsym{P}} ={{\check{\Omega}}}_{,{{\bsym{F}}}}, \quad \quad {{\bbbm{B}}} =- {{\check{\Omega}}}_{, {{\bbbm{H}}}} \quad \quad \text{ in } \quad {\mcal{B}}_0,
\label{eq: PK H formulation}
\end{equation}
the (Maxwell) stress ${{\bsym{P}}}_m$ {exterior to} the body as {stated} earlier in equation \eqref{eqn: maxwell strPK}
and recall the relation $J^{-1} {{\bsym{F}}} {{\bbbm{B}}} ={\mu_0} {{\bsym{F}}}^{-\top} {{\bbbm{H}}}$ in vacuum from equation \eqref{eqn: constitutive B H M}. 
Using the above relations \eqref{eq: PK H formulation}, we rewrite the first variation \eqref{eqn: E first var 1} as
\begin{align}
{\sdel}{\mathit{E}_{\mathtt{V}}}& =\int\nolimits_{{\mcal{B}}_0} \big[ \Div \left({\bsym{P}}^{\top} {\sdel}{\mathpalette\irchi\relax} \right) - [ \Div {\bsym{P}} + \widetilde{{\mbf{f}}}^e ] \cdot {\sdel}{\mathpalette\irchi\relax} + \Div \left({\sdel}{\Phi} {{\bbbm{B}}} \right) - {\sdel}{\Phi} \Div {{\bbbm{B}}} \big] {dv}_0 \nonumber \\
& + \int\nolimits_{{\mcal{B}}'_0} \big[ \Div \left({\bsym{P}}^{\top}_m {\sdel}{\mathpalette\irchi\relax} \right) - \left[ \Div {{\bsym{P}}}_m \right] \cdot {\sdel}{\mathpalette\irchi\relax} + \Div \left({\sdel}{\Phi} {{\bbbm{B}}} \right) - {\sdel}{\Phi} \Div {{\bbbm{B}}} \big] {dv}_0 \nonumber \\
& - \int\nolimits_{\bvol_0} {\sdel}{\Phi} {{\bbbm{B}}}^e \cdot \mbf{n}_0 ds_0 - \int\nolimits_{{\partial\mcal{B}} _0} \widetilde{\mbf{t}}^e \cdot {\sdel}{\mathpalette\irchi\relax} ds_0.
\label{eqn: E first var 12}
\end{align}
After an application of divergence theorem to \eqref{eqn: E first var 12}, we get
\begin{align}
{\sdel}{\mathit{E}_{\mathtt{V}}} &=\int\nolimits_{{\mcal{B}}_0}[ -[ \Div ({\bsym{P}}) + \widetilde{{\mbf{f}}}^e] \cdot {\sdel}{\mathpalette\irchi\relax} - {\sdel}{\Phi} \Div {{\bbbm{B}}}] {dv}_0 \nonumber \\
& + \int\nolimits_{{\partial\mcal{B}} _0}[[[ {\bsym{P}}|_- - {{\bsym{P}}}_m|_+] \mbf{n}_0 - \widetilde{\mbf{t}}^e] \cdot {\sdel}{\mathpalette\irchi\relax} + {\sdel}{\Phi}[ {{\bbbm{B}}}|_- - {{\bbbm{B}}}|_+] \cdot \mbf{n}_0] ds_0 \nonumber \\
& + \int\nolimits_{{\mcal{B}}_0'}[ -[ \Div {{\bsym{P}}}_m] \cdot {\sdel}{\mathpalette\irchi\relax} - {\sdel}{\Phi} \Div {{\bbbm{B}}}] {dv}_0\nonumber \\
& + \int\nolimits_{\bvol_0}[ {{\bsym{P}}}_m \mbf{n}_0 \cdot {\sdel}{\mathpalette\irchi\relax} + {\sdel}{\Phi}[ {{\bbbm{B}}} - {{\bbbm{B}}}^e] \cdot \mbf{n}_0 ] {dv}_0.
\label{eqn: E first var 13}
\end{align}
Since the two variations ${\sdel}{\mathpalette\irchi\relax}$ and ${\sdel}{\Phi}$ are arbitrary, their coefficients in each of the integrals must vanish.
Accordingly, using the coefficient of ${\sdel}{\mathpalette\irchi\relax}$ in \eqref{eqn: E first var 13}, we get the equations
\begin{subequations}
\begin{align}
\Div {{\bsym{P}}} + \widetilde{{\mbf{f}}}^e =\mbf{0} &\text{ in } {\mcal{B}}_0,
\label{eqn: gov 1 H formulation}\\
\Div {{\bsym{P}}} =\mbf{0} &\text{ in } {\mcal{B}}_0',\\
\jump{{{\bsym{P}}}} \mbf{n}_0 + \widetilde{\mbf{t}}^e=\mbf{0} &\text{ on } {\partial\mcal{B}} _0, \\
{{\bsym{P}}} \mbf{n}_0 =\mbf{0} &\text{ on } \bvol_0,
\end{align}
\label{eqn: Euler based E}
\end{subequations}
while the coefficient of ${\sdel}{\Phi}$ in \eqref{eqn: E first var 13} leads to the equations
\begin{subequations}
\begin{align}
\Div {{\bbbm{B}}} =0 &\text{ in } {\mcal{B}}_0,\\
\Div {{\bbbm{B}}} =0 &\text{ in } {\mcal{B}}_0',\\
\jump{{\bbbm{B}}} \cdot \mbf{n}_0 =0 &\text{ on } {\partial\mcal{B}} _0, \\
\jump{{\bbbm{B}}} \cdot \mbf{n}_0 =0 &\text{ on } \bvol_0.
\label{eqn: gov last H formulation}
\end{align}
\label{eqn: Euler 2 based E}
\end{subequations}

\begin{sidenote}
Parallel to the remark \ref{note1} at the end of Section \ref{sec: A first var},
we note that in this formulation based on the magnetic field (equivalently, the magnetic scalar potential), we have apriori assumed the equation \eqref{eqn: maxwell lagrangian}$_2$ that ${{\bbbm{H}}}$ should satisfy and have recovered the equation \eqref{eqn: maxwell lagrangian}$_1$ for the magnetic induction ${{\bbbm{B}}}$ as an Euler-Lagrange equation of this minimisation problem.
This procedure too implies the constitutive assumption ${{\bbbm{B}}} =- {{\check{\Omega}}}_{,{{\bbbm{H}}}}$ while it has been also independently derived earlier \citep{dorfmann2004}.
\label{note2}
\end{sidenote}

\subsection{Critical point: second variation}
For the analysis of critical point $({\mathpalette\irchi\relax}, {\Phi})$, we need to find ${\bdel}{\mathpalette\irchi\relax}$ and ${\bdel}{\Phi}$ such that certain bilinear functional based on the second variation vanishes at the critical point, that is
${\sddel} {\mathit{E}_{\mathtt{V}}}\equiv {\sddel} {\mathit{E}_{\mathtt{V}}}[{\mathpalette\irchi\relax}, {\Phi}; ({\sdel}{\mathpalette\irchi\relax}, {\sdel}{\Phi}), ({\bdel}{\mathpalette\irchi\relax}, {\bdel}{\Phi})]
=0. $
{The second variation of the functional in \eqref{eqn: potential energy functional for H formulation} based on the magnetic field ${{\bbbm{H}}}$ is given by}
\begin{align}
{\sddel} {\mathit{E}_{\mathtt{V}}} &=\int\nolimits_{{\mcal{B}}_0} \bigg[ \Div \left(\big[ {{\check{\Omega}}}_{, {{\bsym{F}}} {{\bsym{F}}}} {\bdel}{{\bsym{F}}} 
+ \frac{1}{2} {{\check{\Omega}}}_{, {{\bsym{F}}} {{\bbbm{H}}}} {\bdel}{{\bbbm{H}}} 
+ \frac{1}{2} \wst{{\check{\Omega}}}_{ {{\bsym{F}}} {{\bbbm{H}}} } {\bdel}{{\bbbm{H}}} \big] ^\top {\sdel}{\mathpalette\irchi\relax} \right) \nonumber \\
& - \Div \left({{\check{\Omega}}}_{, {{\bsym{F}}} {{\bsym{F}}}} {\bdel}{{\bsym{F}}} 
+ \frac{1}{2} {{\check{\Omega}}}_{, {{\bsym{F}}} {{\bbbm{H}}}} {\bdel}{{\bbbm{H}}} 
+ \frac{1}{2} \wst{{\check{\Omega}}}_{ {{\bsym{F}}} {{\bbbm{H}}} } {\bdel}{{\bbbm{H}}} \right) \cdot {\sdel}{\mathpalette\irchi\relax} \nonumber \\
& - \Div \left(\big[ \frac{1}{2} \wst{{\check{\Omega}}}_{{{\bbbm{H}}} {{\bsym{F}}}} {\bdel}{{\bsym{F}}}
+ \frac{1}{2} {{\check{\Omega}}}_{, {{\bbbm{H}}} {{\bsym{F}}}} {\bdel}{{\bsym{F}}} 
+ {{\check{\Omega}}}_{, {{\bbbm{H}}} {{\bbbm{H}}}} {\bdel}{{\bbbm{H}}}
\big] {\sdel}{\Phi} \right) \nonumber \\
& + \Div \left(\frac{1}{2} \wst{{\check{\Omega}}}_{{{\bbbm{H}}} {{\bsym{F}}}} {\bdel}{{\bsym{F}}}
+ \frac{1}{2} {{\check{\Omega}}}_{, {{\bbbm{H}}} {{\bsym{F}}}} {\bdel}{{\bsym{F}}} 
+ {{\check{\Omega}}}_{, {{\bbbm{H}}} {{\bbbm{H}}}} {\bdel}{{\bbbm{H}}} \right) {\sdel}{\Phi} \bigg] {dv}_0 \nonumber \\
& + \int\nolimits_{{\mcal{B}}_0'} \big[ \Div (\widetilde{\bsym{T}}^\top {\sdel}{\mathpalette\irchi\relax} ) - \Div \widetilde{\bsym{T}} \cdot {\sdel}{\mathpalette\irchi\relax} + \Div \left(\widetilde{\mbf{v}}_0 {\sdel}{\Phi} \right) - \Div \widetilde{\mbf{v}}_0 {\sdel}{\Phi} \big] {dv}_0
\label{eqn: second variation condition E 1}, 
\end{align}
where we have introduced the tensor $\widetilde{\bsym{T}}$ and the vector $\widetilde{\mbf{v}}_0$ as
\begin{align}
\widetilde{\bsym{T}} \defnt & {J}{{\mu_0}} \Bigg[ {{\bsym{F}}}^{-\top} [{\bdel}{{\bsym{F}}}]^\top {{\bsym{F}}}^{-\top} {{\bbbm{H}}} \otimes {{\bsym{F}}}^{-1} {{\bsym{F}}}^{-\top} {{\bbbm{H}}} + {{\bsym{F}}}^{-\top} {{\bbbm{H}}} \otimes {{\bsym{F}}}^{-1} {\bdel}{{\bsym{F}}} {{\bsym{F}}}^{-1} {{\bsym{F}}}^{-\top} {{\bbbm{H}}} \nonumber \\
&- {{\bsym{F}}}^{-\top} {\bdel}{{\bbbm{H}}} \otimes {{\bsym{F}}}^{-1} {{\bsym{F}}}^{-\top} {{\bbbm{H}}} - {{\bsym{F}}}^{-\top} {{\bbbm{H}}} \otimes {{\bsym{F}}}^{-1} {{\bsym{F}}}^{-\top} {\bdel}{{\bbbm{H}}} \nonumber \\
& + {{\bsym{F}}}^{-\top} {{\bbbm{H}}} \otimes {{\bsym{F}}}^{-1} {{\bsym{F}}}^{-\top} [{\bdel}{{\bsym{F}}}]^\top {{\bsym{F}}}^{-\top} {{\bbbm{H}}} - \big[ {{\bsym{F}}}^{-\top} \cdot {\bdel}{{\bsym{F}}} \big] {{\bsym{F}}}^{-\top} {{\bbbm{H}}} \otimes {{\bsym{F}}}^{-1} {{\bsym{F}}}^{-\top} {{\bbbm{H}}}
\nonumber \\
& + \bigg[ - \big[ {{\bsym{F}}}^{-\top} \left[ {\bdel}{{\bsym{F}}} \right]^{\top} {{\bsym{F}}}^{-\top} {{\bbbm{H}}} \big] \cdot \big[ {{\bsym{F}}}^{-\top} {{\bbbm{H}}} \big] 
+ \big[ {{\bsym{F}}}^{-\top} {{\bbbm{H}}} \big] \cdot {{\bsym{F}}}^{-\top} \big[ {\bdel}{{\bbbm{H}}} \big] \bigg]{{\bsym{F}}}^{-\top} \nonumber \\
& - \frac{1}{2} \big[ {{\bsym{F}}}^{-\top} {{\bbbm{H}}} \big] \cdot \big[ {{\bsym{F}}}^{-\top} {{\bbbm{H}}} \big] \bigg[ \big[ {{\bsym{F}}}^{-\top} \cdot {\bdel}{{\bsym{F}}} \big] {{\bsym{F}}}^{-\top} - {{\bsym{F}}}^{-\top} [{\bdel}{{\bsym{F}}}]^\top {{\bsym{F}}}^{-\top} \bigg ], \\
\widetilde{\mbf{v}}_0 \defnt & {J}{{\mu_0}} \bigg[ {{\bsym{F}}}^{-1} {\bdel}{{\bsym{F}}} {{\bsym{F}}}^{-1} {{\bsym{F}}}^{-\top} + {{\bsym{F}}}^{-1} {{\bsym{F}}}^{-\top} \big[ {\bdel}{{\bsym{F}}} \big]^\top {{\bsym{F}}}^{-\top} - \left[ {{\bsym{F}}}^{-\top} \cdot {\bdel}{{\bsym{F}}} \right] {{\bsym{F}}}^{-1} {{\bsym{F}}}^{-\top} \bigg] {{\bbbm{H}}} \nonumber \\
& - J {\mu_0} {{\bsym{F}}}^{-1} {{\bsym{F}}}^{-\top} {\bdel}{{\bbbm{H}}},
\end{align}
while we have also utilized the definitions of two third order tensors $\wst{{\check{\Omega}}}_{ {{\bsym{F}}} {{\bbbm{H}}}}$ and $\wst{{\check{\Omega}}}_{ {{\bbbm{H}}} {{\bsym{F}}} }$, according to the relations
\begin{equation}
[ \wst{{\check{\Omega}}}_{{{\bsym{F}}} {{\bbbm{H}}}} \mbf{u}] \cdot \bsym{U} =[ { {{\check{\Omega}}}_{, {{\bbbm{H}}} {{\bsym{F}}}}} \bsym{U} ] \cdot \mbf{u}, \quad [\wst{{\check{\Omega}}}_{ {{\bbbm{H}}} {{\bsym{F}}}} \bsym{U}] \cdot \mbf{u} =[ {{{\check{\Omega}}}_{, {{\bsym{F}}} {{\bbbm{H}}}} } \mbf{u} ] \cdot \bsym{U},
\end{equation}
where $\mbf{u}$ and $\bsym{U}$ are arbitrary vector and arbitrary second order tensor, respectively.

An application of divergence theorem to \eqref{eqn: second variation condition E 1} gives
\begin{align}
{\sddel} {\mathit{E}_{\mathtt{V}}} &=\int\nolimits_{{\mcal{B}}_0} \Bigg[ - \Div ({{\check{\Omega}}}_{, {{\bsym{F}}} {{\bsym{F}}}} {\bdel}{{\bsym{F}}} 
+ \frac{1}{2} {{\check{\Omega}}}_{, {{\bsym{F}}} {{\bbbm{H}}}} {\bdel}{{\bbbm{H}}} 
+ \frac{1}{2} \wst{{\check{\Omega}}}_{ {{\bsym{F}}} {{\bbbm{H}}} } {\bdel}{{\bbbm{H}}} ) \cdot {\sdel}{\mathpalette\irchi\relax} \nonumber \\
& + \Div (\frac{1}{2} \wst{{\check{\Omega}}}_{{{\bbbm{H}}} {{\bsym{F}}}} {\bdel}{{\bsym{F}}}
+ \frac{1}{2} {{\check{\Omega}}}_{, {{\bbbm{H}}} {{\bsym{F}}}} {\bdel}{{\bsym{F}}} 
+ {{\check{\Omega}}}_{, {{\bbbm{H}}} {{\bbbm{H}}}} {\bdel}{{\bbbm{H}}} ) {\sdel}{\Phi} \Bigg] {dv}_0 \nonumber \\
& + \int\nolimits_{{\partial\mcal{B}} _0} \Bigg[ \bigg[ {{\check{\Omega}}}_{, {{\bsym{F}}} {{\bsym{F}}}} {\bdel}{{\bsym{F}}} 
+ \frac{1}{2} {{\check{\Omega}}}_{, {{\bsym{F}}} {{\bbbm{H}}}} {\bdel}{{\bbbm{H}}} 
+ \frac{1}{2} \wst{{\check{\Omega}}}_{ {{\bsym{F}}} {{\bbbm{H}}} } {\bdel}{{\bbbm{H}}} \bigg]|_- - \widetilde{\bsym{T}}|_+ \Bigg] \mbf{n}_0 \cdot {\sdel}{\mathpalette\irchi\relax} ds_0 \nonumber \\
& - \int\nolimits_{{\partial\mcal{B}} _0} \Bigg[ \bigg[ \frac{1}{2} \wst{{\check{\Omega}}}_{{{\bbbm{H}}} {{\bsym{F}}}} {\bdel}{{\bsym{F}}}
+ \frac{1}{2} {{\check{\Omega}}}_{, {{\bbbm{H}}} {{\bsym{F}}}} {\bdel}{{\bsym{F}}} 
+ {{\check{\Omega}}}_{, {{\bbbm{H}}} {{\bbbm{H}}}} {\bdel}{{\bbbm{H}}} \bigg]|_- - \widetilde{\mbf{v}}_0|_+ \Bigg] \cdot \mbf{n}_0 {\sdel}{\Phi} ds_0 \nonumber \\
& + \int\nolimits_{{\mcal{B}}_0'} \bigg[ - \Div \widetilde{\bsym{T}} \cdot {\sdel}{\mathpalette\irchi\relax} - \Div \widetilde{\mbf{v}}_0 {\sdel}{\Phi} \bigg] {dv}_0 + \int\nolimits_{\bvol_0} \bigg[ \widetilde{\bsym{T}} \mbf{n}_0 \cdot {\sdel}{\mathpalette\irchi\relax} + \widetilde{\mbf{v}}_0 \cdot \mbf{n}_0 {\sdel}{\Phi} \bigg] ds_0.
\end{align}
Since the variations ${\sdel}{\mathpalette\irchi\relax}$ and ${\sdel}{\Phi}$ are arbitrary, we arrive at the following equations for the unknown functions $({\bdel}{\mathpalette\irchi\relax}, {\bdel}{\Phi})$
\begin{align}
\Div ({{\check{\Omega}}}_{, {{\bsym{F}}} {{\bsym{F}}}} {\bdel}{{\bsym{F}}} 
+ \frac{1}{2} {{\check{\Omega}}}_{, {{\bsym{F}}} {{\bbbm{H}}}} {\bdel}{{\bbbm{H}}} 
+ \frac{1}{2} \wst{{\check{\Omega}}}_{ {{\bsym{F}}} {{\bbbm{H}}} } {\bdel}{{\bbbm{H}}} ) & =\mbf{0} \text{ in } {\mcal{B}}_0,\\
\Div (\frac{1}{2} \wst{{\check{\Omega}}}_{{{\bbbm{H}}} {{\bsym{F}}}} {\bdel}{{\bsym{F}}}
+ \frac{1}{2} {{\check{\Omega}}}_{, {{\bbbm{H}}} {{\bsym{F}}}} {\bdel}{{\bsym{F}}} 
+ {{\check{\Omega}}}_{, {{\bbbm{H}}} {{\bbbm{H}}}} {\bdel}{{\bbbm{H}}} ) & =0 \text{ in } {\mcal{B}}_0,\\
[ \big[ {{\check{\Omega}}}_{, {{\bsym{F}}} {{\bsym{F}}}} {\bdel}{{\bsym{F}}} 
+ \frac{1}{2} {{\check{\Omega}}}_{, {{\bsym{F}}} {{\bbbm{H}}}} {\bdel}{{\bbbm{H}}} 
+ \frac{1}{2} \wst{{\check{\Omega}}}_{ {{\bsym{F}}} {{\bbbm{H}}} } {\bdel}{{\bbbm{H}}} \big]|_- - \widetilde{\bsym{T}}|_+ ] \mbf{n}_0 & =\mbf{0} \text{ on } {\partial\mcal{B}} _0,\\
[[ \frac{1}{2} \wst{{\check{\Omega}}}_{{{\bbbm{H}}} {{\bsym{F}}}} {\bdel}{{\bsym{F}}}
+ \frac{1}{2} {{\check{\Omega}}}_{, {{\bbbm{H}}} {{\bsym{F}}}} {\bdel}{{\bsym{F}}} 
+ {{\check{\Omega}}}_{, {{\bbbm{H}}} {{\bbbm{H}}}} {\bdel}{{\bbbm{H}}}]|_- - \widetilde{\mbf{v}}_0|_+] \cdot \mbf{n}_0 & =0 \text{ on } {\partial\mcal{B}} _0,\\
\Div\widetilde{\bsym{T}} & =\mbf{0} \text{ in } {\mcal{B}}_0',\\
\Div\widetilde{\mbf{v}}_0 & ={0} \text{ in } {\mcal{B}}_0',\\
\widetilde{\bsym{T}} \mbf{n}_0 & =\mbf{0} \text{ on } \bvol_0,\\
\widetilde{\mbf{v}}_0 \cdot \mbf{n}_0 & ={0} \text{ on } \bvol_0,
\label{eqn: Perturb based E}
\end{align}
describing the onset of bifurcation.

\begin{sidenote}
Note that a variation of the relation $ {{\bbbm{B}}} =J {\mu_0} {{\bsym{C}}}^{-1} {{\bbbm{H}}} $ from equation \eqref{eqn: constitutive B H M} gives
${\bdel}{{\bbbm{B}}}=\widetilde{\mbf{v}}_0,$
since
\begin{align}
{\bdel}{{\bbbm{B}}}&=J {\mu_0} \big[ {{\bsym{F}}}^{-1} {{\bsym{F}}}^{-\top} {\bdel}{{\bbbm{H}}} - {{\bsym{F}}}^{-1} {\bdel}{{\bsym{F}}} {{\bsym{F}}}^{-1} {{\bsym{F}}}^{-\top} {{\bbbm{H}}} - {{\bsym{F}}}^{-1} {{\bsym{F}}}^{-\top} [{\bdel}{{\bsym{F}}}]^\top {{\bsym{F}}}^{- \top} {{\bbbm{H}}} \nonumber \\&
+ \big[ {{\bsym{F}}}^{-\top} \cdot {\bdel}{{\bsym{F}}} \big] {{\bsym{F}}}^{-1} {{\bsym{F}}}^{-\top} \big].
\end{align}
A variation of the Maxwell stress \eqref{eqn: maxwell strPK} (after writing it in terms of ${{\bbbm{H}}}$ using the relation \eqref{eqn: constitutive B H M}) gives
${\bdel}{{\bsym{P}}}_m=\widetilde{\bsym{T}},$
since
\begin{align}
{\bdel}{{\bsym{P}}}_m & =J {\mu_0} \Bigg[ {{\bsym{F}}}^{-\top} {\bdel}{{\bbbm{H}}} \otimes {{\bsym{F}}}^{-1} {{\bsym{F}}}^{-\top} {{\bbbm{H}}} + {{\bsym{F}}}^{-\top} {{\bbbm{H}}} \otimes {{\bsym{F}}}^{-1} {{\bsym{F}}}^{-\top} {\bdel}{{\bbbm{H}}} \nonumber \\
& + \big[ {{\bsym{F}}}^{-\top} \cdot {\bdel}{{\bsym{F}}} \big] {{\bsym{F}}}^{-\top} {{\bbbm{H}}} \otimes {{\bsym{F}}}^{-1} {{\bsym{F}}}^{-\top} {{\bbbm{H}}} - {{\bsym{F}}}^{-\top} [{\bdel}{{\bsym{F}}}]^{ -\top} {{\bbbm{H}}} \otimes {{\bsym{F}}}^{- 1} {{\bsym{F}}}^{-\top} {{\bbbm{H}}} \nonumber \\
& - {{\bsym{F}}}^{-\top} {{\bbbm{H}}} \otimes {{\bsym{F}}}^{-1} {{\bsym{F}}}^{-\top} [{\bdel}{{\bsym{F}}}]^\top {{\bsym{F}}}^{-\top} {{\bbbm{H}}} - {{\bsym{F}}}^{-\top} {{\bbbm{H}}} \otimes {{\bsym{F}}}^{-1} [{\bdel}{{\bsym{F}}} ] {{\bsym{F}}}^{-1 } {{\bsym{F}}}^{-\top} {{\bbbm{H}}} \nonumber \\
& + \frac{1}{2} [{{\bsym{F}}}^{-\top} {{\bbbm{H}}}] \cdot [{{\bsym{F}}}^{-\top} {{\bbbm{H}}}][ {{\bsym{F}}}^{-\top} [{\bdel}{{\bsym{F}}}]^\top {{\bsym{F}}}^{-\top} - [{{\bsym{F}}}^{-\top} \cdot {\bdel}{{\bsym{F}}}] {{\bsym{F}}}^{-\top}] \nonumber \\
& + \big[ -[ {{\bsym{F}}}^{-\top}[ {\bdel}{{\bsym{F}}}]^{\top} {{\bsym{F}}}^{-\top} {{\bbbm{H}}}] \cdot[ {{\bsym{F}}}^{-\top} {{\bbbm{H}}}] 
+[ {{\bsym{F}}}^{-\top} {{\bbbm{H}}}] \cdot {{\bsym{F}}}^{-\top}[ {\bdel}{{\bbbm{H}}}] \big] {{\bsym{F}}}^{-\top} \Bigg].
\label{eqn: delta Pm expression 2}
\end{align}
Alternative to the statements $\widetilde{\mbf{v}}_0 ={\bdel}{{\bbbm{H}}}$ 
and $\widetilde{\bsym{T}} ={\bdel}{{\bsym{P}}}_m$,
it can be also shown that the above set of equations for the perturbations ${\bdel}{{\bbbm{H}}}$ and ${\bdel}{{\bsym{F}}}$ can be obtained by linearising the equations of equilibrium \eqref{eqn: gov 1 H formulation}--\eqref{eqn: gov last H formulation}.
\label{perturbeqlb2}
\end{sidenote}

\bibliographystyle{./author-year-prashant}
\bibliography{./references_used2}
\end{document}